\documentclass[11pt,a4paper]{report}
\usepackage{psfig}

\newcommand{\myblstr}{1.5}
\renewcommand{\baselinestretch}{\myblstr}

\newcommand{\be}{\begin{equation}}
\newcommand{\ee}{\end{equation}}
\newcommand{\bea}{\begin{eqnarray}}
\newcommand{\eea}{\end{eqnarray}}
\newcommand{\bearr}[1]{\begin{equation} \begin{array}{{#1}}}
\newcommand{\eearr}{\end{array} \end{equation}}
\newcommand{\bref}[1]{(\ref{#1})}

\renewcommand{\th}{\theta}
\newcommand{\la}{\lambda}
\newcommand{\eps}{\epsilon}
\newcommand{\epb}{\bar{\epsilon}}
\newcommand{\vp}{\varphi}
\newcommand{\thb}{\bar{\theta}}
\newcommand{\dr}{\partial_r}
\newcommand{\dth}{\partial_\th}
\newcommand{\cconj}{(\mbox{c.\ c.})}
\newcommand{\hconj}{(\mbox{h.\ c.})}

\newcommand{\gen}[1]{\mbox{$\tau\!\left[{\scriptstyle {#1}}\right]$}}
\newcommand{\gena}{\mbox{$\tau^a$}}
\newcommand{\ph}[1]{\Phi_{#1}}
\newcommand{\phvac}[1]{\ph{{#1}}^{\rm vac}}
\newcommand{\pharrow}[1]{\stackrel
	{ \rule[-2pt]{0.0 in}{2 pt} \scriptstyle \ph{{#1}} }{\longrightarrow}}
\newcommand{\sprod}[2]{({#1} \times {#2})_S}
\newcommand{\sig}[1]{\sigma^{#1}}

\newcommand{\ga}{\alpha}
\newcommand{\gb}{\beta}
\newcommand{\Mab}{{M_{\ga \gb}}}
\newcommand{\Cab}{{C_{\ga \gb}}}

\newcommand{\sbu}{u}
\newcommand{\sbd}{d}
\newcommand{\sbe}{e}
\newcommand{\sbG}{{\scriptscriptstyle \rm G}}
\newcommand{\sbE}{{\scriptscriptstyle \rm E}}
\newcommand{\sbud}{{\sbu,\sbd}}
\newcommand{\sbdu}{{\sbd,\sbu}}

\newcommand{\etag}{\eta_\sbG}
\newcommand{\etau}{\eta_\sbu}
\newcommand{\etad}{\eta_\sbd}
\newcommand{\etaud}{\eta_\sbud}
\newcommand{\etae}{\eta_\sbE}
\newcommand{\etamed}{\eta_{\scriptscriptstyle \rm M}}
\newcommand{\yukg}{g_\sbG}
\newcommand{\yuke}{g_\sbE}

\newcommand{\fl}{\Psi_L}
\newcommand{\flc}{\Psi_L^c}
\newcommand{\flb}{\bar{\Psi}_L}

\newcommand{\fr}{\Psi_R}

\newcommand{\WL}[1]{W_{\! L}^{{#1}}}
\newcommand{\WLpm}{\mbox{$\WL{\pm}$}}
\newcommand{\X}[2]{X_{{#1}}^{{#2}}}	
\newcommand{\Xpm}{\mbox{$\X{i}{\pm}$}}
\newcommand{\Y}[2]{Y_{{#1}}^{{#2}}}	
\newcommand{\Ypm}{\mbox{$\Y{i}{\pm}$}}
\newcommand{\WR}[1]{W_{\! R}^{{#1}}}
\newcommand{\WRpm}{\mbox{$\WR{\pm}$}}
\newcommand{\Xs}[2]{X_{S{#1}}^{{#2}}}	
\newcommand{\Xspm}{\mbox{$\Xs{i}{\pm}$}}
\newcommand{\Xg}[2]{{X_{{#1}}'}^{{#2}}}	
\newcommand{\Xgpm}{\mbox{$\Xg{i}{\pm}$}}
\newcommand{\Yg}[2]{{Y_{{#1}}'}^{{#2}}}	
\newcommand{\Ygpm}{\mbox{$\Yg{i}{\pm}$}}

\newcommand{\Nr}{{\rm e}_1}
\newcommand{\Hu}[1]{H_\sbu^{{#1}}}
\newcommand{\Hd}[1]{H_\sbd^{{#1}}}
\newcommand{\Hud}[1]{H_\sbud^{{#1}}}
\newcommand{\Hdu}[1]{H_\sbdu^{{#1}}}

\newcommand{\ts}{T_{\rm s}}
\newcommand{\tew}{T_{\rm ew}}
\newcommand{\rs}{r_{\rm s}}
\newcommand{\rew}{r_{\rm ew}}

\newcommand{\Lag}{{\cal L}}
\newcommand{\Lagf}{{\cal L}_{\rm fermions}}
\newcommand{\mass}[1]{m_{\rm #1}}

\newcommand{\spinU}{\mbox{\scriptsize 
		$\left(\begin{array}{@{}c@{}}1\\0\end{array}\right)$}}

\newcommand{\bmat}[1]{\left(\begin{array}{{#1}}}
\newcommand{\emat}{\end{array}\right)}

\newlength{\myfigwidth}
\newlength{\struthgt}
\newcommand{\mystrut}{\rule{0.0 in}{\struthgt}}
\newcommand{\setstrut}[1]{\setlength{\struthgt}{#1}}

\begin{document}


\thispagestyle{empty}
\ 

\vspace{0.5in}

\centerline{\Huge \bf Microphysics of} 

\vspace{0.1in}

\centerline{\Huge \bf Cosmic Strings in} 

\vspace{0.1in}

\centerline{\Huge \bf Supersymmetric and}

\vspace{0.1in}

\centerline{\Huge \bf Grand Unified Theories}

\vspace{2.0in}

\centerline{\LARGE \bf Stephen Christopher Davis}

\vspace{0.2in}

\centerline{\large Trinity College, Cambridge}

\vspace{1.5in}

\begin{center}{\large
A dissertation submitted for the degree of Doctor of Philosophy \\
University of Cambridge \\ September 1998}
\end{center}

\newpage

\renewcommand{\thepage}{\roman{page}}
\setcounter{page}{1}

\centerline{\bf \LARGE Acknowledgements}
\vspace{.2in}

I wish to thank my supervisor, Dr.\ Anne C. Davis, for all her advice and
support. I am also grateful to Dr.\ Mark Trodden and Dr.\ Warren
B. Perkins for useful discussions.

\vspace{1in}
\centerline{\bf \LARGE Declaration}
\vspace{.2in}

Apart from the introductory sections, including
chapter~\ref{Ch:Intro}, and where indicated in the text, the work
presented is my own. The material in chapters 2--5 is published as follows:
Parts of chapters 2 and 3 appeared in Physical Review D 55 (1997)
1879--1895 (with Dr.\ A. C. Davis), most of chapter 3 appeared in
Physics Letters B 408 (1997) 81--90
(with Dr.\ A. C. Davis and Dr.\ W. B. Perkins),
chapter 4 was published in
Physics Letters B 405 (1997) 257--264 
(with Dr.\ A. C. Davis and Dr.\ M. Trodden),
and chapter 5 was published in Physical Review D 57 (1998) 5184--5188
(with Dr.\ A. C. Davis and Dr.\ M. Trodden). 
The work contained in chapter 6 is in a preliminary form.
Whilst the papers have been published in collaboration, I have only
included my own contributions in this dissertation. 

To the best of my knowledge, the work presented as my own is original.

\tableofcontents

\listoffigures

\newpage

\centerline{\bf \LARGE Abstract}
\vspace{.2in}

In this thesis we investigate the microphysics of cosmic strings in
non-minimal quantum field theories. In particular we consider theories
in which fermion fields couple to the strings, and those with larger
symmetry groups, such as grand unified and supersymmetric
theories. By considering these extensions to the minimal
model, we obtain a more realistic picture of the properties of cosmic strings.

In considering grand unified
theories, which have multiple phase transitions, we show that a
cosmic string formed at one phase transition can cause the creation of
another string-like solution at a later transition. This string-like
solution will have many of the properties and implications of a normal cosmic
string. We consider this effect for a general string solution, and
illustrate it with a realistic $SO(10)$ unified theory. As
well as the usual abelian strings, this theory also contains
more exotic string solutions. We consider both types of cosmic string. 
Separately, we examine the form of cosmic string solutions in
supersymmetric theories, and the effect of soft supersymmetry breaking on them.

We investigate the existence of conserved fermion currents in a variety of
cosmic string models. We show that supersymmetry may be used to find
the form of some solutions analytically. We also derive an expression
for the number and type of massless fermion currents in a general
model. The existence of conserved currents can conflict with
observations, so these results may be used to constrain models. We
find the number of massless currents in the $SO(10)$ and supersymmetric
theories mentioned above. We show that currents present on a string
can be destabilised by later phase transitions or supersymmetry
breaking. This may allow any conflict that the current's existence has
with observations to be avoided. We also examine massive fermion
currents in a simple model, and determine the spectrum of such states.

\newcommand{\vac}{\left| 0 \right>}
\newcommand{\vacG}{\vac_{\scriptstyle \rm GUT}}
\newcommand{\vman}{{\cal M}}
\newcommand{\pd}[1]{\partial_{{#1}}}
\renewcommand{\r}[1]{r_{\rm #1}}

\chapter{Introduction}
\label{Ch:Intro}

\setcounter{page}{1}
\renewcommand{\thepage}{\arabic{page}}

\section{Particle Cosmology and Cosmic Strings}
\label{In Cosmo}
In the last twenty years there have been many significant developments
in our understanding and approach to cosmology. In the past, the
evolution of the universe was mainly studied through Einstein's
equations of General Relativity. Modern cosmology also makes use of
quantum field theory. This is particularly important when considering
the early universe. According to the big bang model, the universe was once
very small and at a very high temperature. Since classical
approximations will certainly break down under these conditions, the 
investigation of quantum effects is vital.

The combination of General Relativity and Particle Physics has led to
a very accurate model known as the Standard Cosmology. This
successfully predicts many phenomena, such as the expansion of the
universe, light element abundances, and the uniformity of the cosmic
microwave background. Despite its successes, there are still many
unanswered questions. It is not clear how structures such as galaxies
formed in a universe which seems to have been homogeneous and
isotropic at very early times, nor how the universe became so uniform
in the first place.  There are also the questions of the origin of
galactic magnetic fields, and why there is
more matter than anti-matter in the universe.

The Standard Model of particle physics does not provide answers to
these and other questions. It is therefore necessary to go beyond the
Standard Model and consider more speculative ideas in quantum field
theory, such as unification of the fundamental forces, or
supersymmetry. Unfortunately, testing some of these ideas is beyond
current particle accelerators. However, as well as predicting
cosmology from quantum field theory, it is possible to use
cosmological observations to make predictions about high energy
quantum field theory. Study of the early universe can thus give
insight into particle physics as well as cosmology.

Topological defects (such as cosmic strings) are one example of
quantum field theory's contribution to cosmology~\cite{string book,
CSreport}. They provide possible explanations for structure
formation~\cite{structure}, baryogenesis~\cite{baryogenesis1}, cosmic
microwave background anisotropies~\cite{CMB}, and the origin of high
energy cosmic rays~\cite{rays1,current-rays}. Once formed, a defect
will not decay (unless it collides with another similar defect). Thus
even though any defect formation will be restricted to the early
universe, their effects will continue to the present.

Topological defects form at phase transitions~\cite{Kibble}. 
It is believed that the universe passed through several phase
transitions shortly after it came into existence.
These transitions reduced the symmetry of the field theory and gave
mass to certain fundamental particles. This causes their interactions to be
suppressed, and explains why they are not observed at everyday
temperatures. Inside a defect the transition does not occur, so the
laws of physics there resemble those of the
universe before the phase transition at which they formed. This allows
interactions to occur inside the defect which are heavily suppressed in
today's universe. For example, if baryon violating
processes were unsuppressed proton decay would occur more
frequently~\cite{proton}. This property of defects could provide an
important window into the physics of the very early universe. 

In the past few years it has been realised that cosmic strings may have
considerably richer microstructure than previously
thought~\cite{wbp&acd}. In particular, the presence of conserved
currents in the spectrum of a cosmic string has profound implications
for the cosmology of the defects. If the particles making up the
current carry electric charge, and the string was formed at a mass
scale of $10^{16}$GeV (a reasonable value for grand unification) the
string could carry an electric current of up to
$10^{20}$A~\cite{Witten}. There are many candidates for such a
current, one class of which is introduced in section~\ref{In FermZM}.

Another important consequence of cosmic strings is their gravitational
effects. The mass per unit length of a string formed at an energy
scale around $10^{16}$GeV would be $10^{22}$g/cm. The gravitational
field of the string is such that although matter is not attracted to a
stationary string, it is attracted to the wake of a moving string. The
evolution of a network of cosmic strings produced at high energy
scales thus provides a possible origin for the seed density
perturbations which became the large scale structure of the 
observed universe~\cite{structure}. The gravitational effects of the
string could also explain anisotropies in the cosmic microwave
background~\cite{CMB}.

The occurrence of phase transitions and defect formation is not
restricted to quantum field theory. They also occur in condensed
matter systems such as superconductors, ${}^4$He and ${}^3$He
superfluids, and nematic liquid crystals~\cite{condmatter}. Defects
have actually been observed in these cases, and have been used to gain
insight into the evolution of cosmological defects.

\section{Grand Unification and Symmetry Breaking}
\label{In PT}
In the Standard Model the weak and electromagnetic forces are unified
at energies of order $10^2$GeV. This idea can be extended to give a Grand
Unified Theory (GUT) in which the strong force is also unified with the
electroweak force. If this is the case, then it is believed that
unification will occur at an energy scale of order $10^{16}$GeV. The
unified theory will be based on a simple continuous Lie group, $G$. At
high energies the vacuum state of the theory respects the full
symmetry of the Lagrangian. Such a situation occurs in the early
universe, when its temperature is extremely high. As it cools, the
gauge theory undergoes a series of spontaneous symmetry breakings,
until it becomes $SU(3)_c \times U(1)_Q$, which are the gauge groups
of QCD and QED. This can be represented by
\be
G \rightarrow H \rightarrow \cdots \rightarrow
SU(3)_c \times SU(2)_L \times U(1)_Y \rightarrow SU(3)_c \times U(1)_Q \ ,
\ee
where the first breaking occurs around $10^{16}$GeV.
Unfortunately modern particle accelerators cannot hope to access
energies above 1TeV. The only place where suitable energies are
likely to be reached is the early universe, so cosmology may provide
evidence for unification.

The spontaneous symmetry breaking mechanism works as follows: At high
temperatures the vacuum state, $\vacG$, is invariant under $G$, so
$g(x)\vacG = \vacG$ for all $g(x) \in G$. At a lower temperature the theory's
potential changes, and a different vacuum state, $\vac$, becomes energetically
favourable. This new state is only invariant under $H$, a subgroup of
$G$, so the symmetry has been broken from $G$ to $H$. 

Although the vacuum state is no longer invariant under $G$, the theory
itself must still be. Thus if $\vac$ is a minimal energy vacuum state,
so is $g\vac$, hence the vacuum is degenerate. The vacuum manifold
then consists of all the distinct states of the form $g\vac$. Since
$g_1\vac = g_2\vac$ if $g_1^{-1}g_2 \in H$, the vacuum manifold 
$\vman$ is equal to the coset space $G/H$.

\begin{figure}
\centerline{\psfig{file=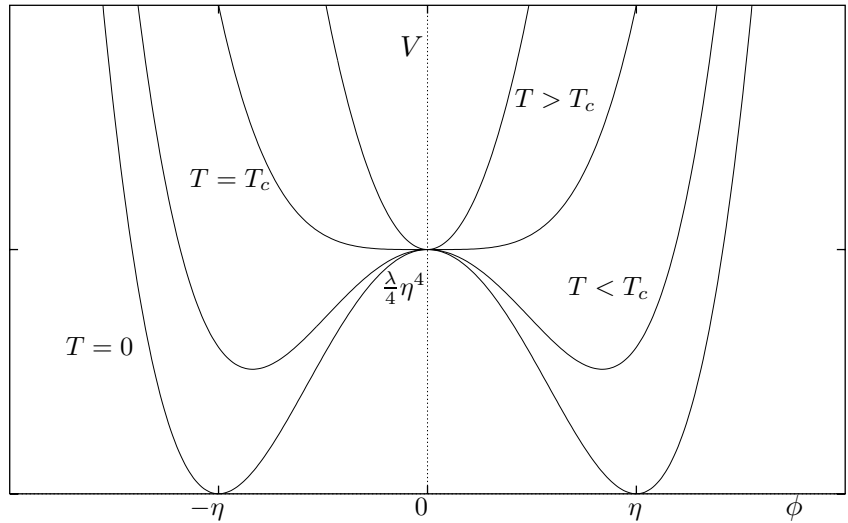,width=\myfigwidth}}
\vskip -2.6 in \hskip 3.0 in $V$
{\small 
\vskip 0.1 in \hskip 3.6 in $T > T_c$
\vskip 0.25 in \hskip 1.9 in $T = T_c$
\vskip 0.4 in \hskip 2.9 in $\frac{\lambda}{4}\eta^4$
\hskip 0.7 in $T < T_c$
\vskip 0.15 in \hskip 1.25 in $T = 0$
\vskip 0.65 in \hskip 1.9 in $-\eta$ \hskip 0.95 in  $0$
\hskip 1.0 in $\eta$} 
\hskip 0.7 in $\phi$
\caption{Finite temperature effective potential.}
\label{Vplot fig}
\end{figure}

The different vacuum states can be labelled by the expectation value of
a scalar Higgs field. In a general GUT there will be several Higgs
fields, each associated with a subsequent phase transition at which
symmetry is broken. As a simple example consider a model with $G=U(1)$
and $H=I$. It has the Lagrangian 
\be
\Lag = (D_\mu \phi)^\ast(D^\mu \phi) 
		- \frac{1}{4}F_{\mu \nu} F^{\mu \nu} - V(\phi) \ ,
\label{In U1Model}
\ee
where $\phi$ is the (complex) Higgs field, $D_\mu\phi = \left(\pd{\mu} -
ieA_\mu\right)\phi$ and 
$F_{\mu \nu} = \partial_\mu A_\nu - \partial_\nu A_\mu$.
$V$ is the potential energy of $\phi$. It is equal to 
$(\la/4) (|\phi|^2 - \eta^2)^2$ at zero temperature. When thermal
corrections are added, its minimum varies with temperature. Above
some critical temperature $T_c$ the vacuum energy is minimised by
$\phi=0$. Below this temperature the minimum has $\phi \neq 0$. $T_c$
is the temperature at which the phase transition occurs. If $T$ is the
temperature of the universe, a suitable $V$ is
\be
V=\frac{\lambda}{4}\left(|\phi|^4 - 
2\eta^2\left[1 - \frac{T^2}{T_c^2}\right] |\phi|^2 + \eta^4\right) \ .
\ee
Thus the vacuum states have $\phi=0$ for $T > T_c$ and 
$\phi = \eta e^{i\alpha} \sqrt{1 - T^2/T_c^2} =$ constant for
$T < T_c$ (see figure~\ref{Vplot fig}). $\phi = \eta e^{i\alpha}$ when $T=0$.
The above potential gives a second order phase transition.

\renewcommand{\baselinestretch}{1}
\renewcommand{\baselinestretch}{\myblstr}

The gauge and fermion fields of a GUT will couple to the Higgs
fields. When a Higgs field gains a non-zero vacuum expectation
value (VEV), it gives masses to some of the gauge and fermion fields
via this coupling. The masses from a GUT scale phase transition will
be enormous, and so the corresponding particles and their interactions
will not be observed. This explains why we do not observe
significant proton decay in today's universe. However, such forces
must have been unsuppressed at some time in the early universe, since
the universe could not have reached its current state without
them. Experimental measurements of the variation of gauge couplings
suggest that they converge at high energies~\cite{Ross}. This is further
circumstantial evidence for unification.

In a grand unified theory there will be one unified force, and one
unified fermion field. The idea of supersymmetry (SUSY) is a natural
extension of this, in which the gauge and Higgs fields are unified
with the fermion fields~\cite{susybook}. If the strengths of the
fundamental forces are carefully extrapolated to high energies, they
do not quite converge, as would be required in a GUT. This problem is
solved if SUSY is added to the theory, suggesting it does have physical
relevance. Furthermore, supersymmetry and grand unification form a
central part of superstring theory, which is the most promising theory
to describe gravity on the quantum scale. Since any full theory of
everything includes quantum gravity, it seems likely that SUSY must
play some role in a realistic quantum field theory. Another motivation for
supersymmetry is its solution to the hierarchy
problem~\cite{softSSbreak}. We expect the high energy scale GUT fields
to couple to the low energy scale Higgs field of the Standard
Model. The mass of this Higgs field receives quadratically divergent
radiative corrections from these couplings, which will generally lead
to conflict with experiment. The symmetry of a supersymmetric theory
ensures that the contributions from the fermion and boson fields
cancel each other exactly, avoiding the problem.

\section{Topological Defects}
\label{In TopDefects}

Phase transitions would have occured in the early
universe as it cooled down. Because information can travel no faster
than the speed of light, $\phi$ would gain different vacuum
expectation values in parts of the universe that were not in
causal contact~\cite{Kibble}. It is
energetically favourable for these variations to disappear. However if
$\vman$ has any non-trivial homotopy groups it is
possible that topologically stable configurations with $\phi$ not
constant will form. These are called topological defects. The type of
defect depends on the non-trivial homotopy group.

If $\pi_0(\vman) \neq I$ then $\vman$ is disconnected. Suppose that
$\phi \in \vman_A$ at $x=\infty$ (cartesian coordinates) and
$\phi \in \vman_B$ at $x=-\infty$, where $\vman_A$ and $\vman_B$ are
disconnected components of $\vman$. By continuity there must be a region where 
$\phi \not\in \vman$. This region is called a domain wall. In three
dimensions it will be a two dimensional surface. 

In the model given by \bref{In U1Model}, $\pi_1(G/H) = Z \neq I$, so $\vman$
contains non-contractible loops. In two dimensions, a configuration
with $\phi = \eta e^{in\th}$ at infinity (with $n$ any integer) cannot be
continuously deformed to $\phi =$ constant, and so is topologically
stable. Regularity will force $\phi$ to leave $\vman$ in some
region (see next section). Such defects are called cosmic strings,
as in three dimensions they are one dimensional. For topological
reasons it is not possible for cosmic strings to have ends, so they
must either be infinitely long, or a closed loop. Suppose a cosmic
string did have an end. Consider the variation of the Higgs field
along a closed path around a string. This variation corresponds to a
non-contractible loop in $\vman$. By sliding it off the string we can
continously change this path to one which corresponds to a
contractible loop in $\vman$. This is equivalent to continously moving
between the disconnected parts of $H$, which is not possible. Hence
strings cannot have ends.

In three dimensions, if $\pi_2(G/H) \neq I$ the corresponding
point-like defects are called monopoles. In this case there are mappings
of $S^2$ to itself which cannot be continously deformed to the identity.
Thus if $\vman$ is equal to $S^2$, and $\phi$ at
$r=\infty$ is such a map, the resulting configuration will be
topologically stable. As with strings, there will be a region in which
$\phi \not\in \vman$.

The size of a defect can be estimated by balancing the potential and
kinetic terms of its Lagrangian \bref{In U1Model}. The potential
energy is of order $\la \eta^4$ inside the defect. If $\delta$ is the
width of the defect, the kinetic terms $(\partial \phi)^2$ are of
order $(\eta/\delta)^2$. Equating these terms gives
$\delta \sim 1/(\sqrt{\la} \eta) \sim \mass{s}^{-1}$, where $\mass{s}$ is
the mass of the Higgs field.

The cosmological consequences of domain walls and monopoles strongly
conflict with observations. Domain walls will come to dominate the
energy density of the universe~\cite{domwall}, and the predicted
monopole density is unacceptablely high~\cite{monopoleabun}. Thus any GUT which
predicts them must also have a mechanism which ensures most of them
are destroyed. The properties of cosmic strings do not conflict with
observations (see section~\ref{In network}). Furthermore they provide
explanations of many phenomena, so cosmic strings are the most
cosmologically significant defect.

\section{The Abelian Cosmic String Model}
\label{In AbString}

As it stands, the string solution described in the previous
section has a non-vanishing covariant derivative which gives an
infinite contribution to the energy. This is avoided by having a
non-zero gauge term. The resulting solution is a cosmic string. More
precisely (taking $T=0$), it has the form
\bea
\phi &=& \eta f(r) e^{in\th} \ , \label{In U1Ansatza} \\
A_\mu &=& n\frac{a(r)}{er}\delta^\th_\mu \ .
\label{In U1Ansatzb}
\eea
In order for the solution to be regular at the origin
$f(0)=a(0)=0$. If the solution is to have finite energy, $f(r)$ and $a(r)$
must tend to 1 as $r \rightarrow \infty$. Substituting \bref{In U1Ansatza} 
and \bref{In U1Ansatzb} into the theory's field equations gives
\be
f'' + \frac{f'}{r} - n^2 \frac{(1-a)^2}{r^2}f 
= \frac{\la}{2}\eta^2 (f^2 - 1)f \ ,
\label{In U1f}
\ee
\be
a'' - \frac{a'}{r} = -2 e^2\eta^2 (1-a) f^2 \ .
\label{In U1a}
\ee
The resulting string is the well known Nielsen-Olesen vortex~\cite{Nielsen}.
It turns out that $f$ and $a$ take their asymptotic values everywhere
outside of a small region around the string. Thus $|\phi|$ is constant
and $A_\mu$ is pure gauge away from the string. The sizes of the
regions in which the magnetic field is non-zero and $|\phi|$ is not
constant ($\r{v}$ and $\r{s}$) are roughly the inverses of the masses
of the corresponding particles. Thus if $\mass{s}$ and $\mass{v}$ are the
Higgs and gauge field masses,
\be
\r{s}^{-1} \approx \mass{s} = \sqrt{\la}\eta \ , \ \ 
\r{v}^{-1} \approx \mass{v} = \sqrt{2}e\eta \ .
\ee
By rescaling $r$, the $e$ and $\eta$ dependence can be removed from
\bref{In U1f} and \bref{In U1a}. The form of the solutions then only
depends on only one parameter: $\beta = (\mass{s}/\mass{v})^2$.

The above solution is a point-like defect in two dimensions. If it is
extended to three dimensions it will take the form of an infinite
line, or a closed loop. The defect is now one dimensional and string
shaped, hence its name. There is a non-zero magnetic field inside the
string. Its flux is
\be
\Phi_B = \int_{r=\infty} A_\mu dx^\mu = \frac{2\pi n}{e} \ .
\ee

Many of the observational consequences of cosmic strings arise from
their gravitational effects. Defining $\mu$ to be the mass per unit
length of the string, the size of these effects is proportional
to the dimensionless quantity $G\mu \sim (T_c/m_{\rm pl})^2$, where
$G$ is Newton's constant and $m_{\rm pl}$ is the Planck mass.
We find that $\mu \sim \eta^2$, so for a grand unification scale
string $G\mu \sim 10^{-6}$. This is the right magnitude to explain the
fluctuations in the cosmic microwave background and the matter
distribution of the observed universe. 

The spacetime around a string is approximately conical. It resembles a
flat space with a wedge of angular size $8\pi G\mu$~\cite{deficitangle}
removed, and the two faces of the wedge identified. If two objects
travelling on parallel paths pass each side of a string, they will
begin moving towards each other despite having experienced no
force. The result of this effect is that as a string moves about,
it increases the density of the regions through which it passes. This is an
example of how cosmic strings provide a mechanism for structure
formation.

At the centre of an abelian cosmic string $\phi$ is zero. Since the
breaking of $G$ to $H$ is caused by $\phi$ being non-zero, this means
that $G$ is not broken inside the string. The result is also true
for other types of defects (monopoles and domain walls). Thus the
grand unified theory is restored inside the string, even though the
universe has cooled to temperatures that would usually break it.

The above ansatz (\ref{In U1Ansatza},\ref{In U1Ansatzb}) describes an
infinite number of string solutions since the winding number, $n$, can
be any non-zero integer. Although they are all topologically stable
with respect to the vacuum and other strings, it is possible for
$|n|>1$ strings to decay by splitting into several strings with
smaller winding numbers~\cite{string book}.

If $\beta > 1$ (where $\beta = (\mass{s}/\mass{v})^2$) this does
happen, since the force between the strings is repulsive. The reverse
is true when $\beta < 1$. In this case strings will tend to combine to
produce a single string with a larger winding number. The situation is
similar in superconductors. In this case $\beta<1$ and $\beta>1$
correspond to type I and type II superconductors respectively. 

When $\beta = 1$ there is no force between strings. It is then
possible to reduce \bref{In U1f} and \bref{In U1a} to first order equations:
\be
f' = |n| \frac{f}{r}(1-a) \ ,
\ee
\be
|n| \frac{a'}{r} = \frac{\lambda}{2} \eta^2 (1 - f^2) \ .
\ee

\section{Determination of the String Field Profiles}

Although full analytic solutions of the string field equations have
not been found, their asymptotic forms are easily determined. For
large $r$, $f(r) \approx 1$ and \bref{In U1a} can be reduced to the
equation for a modified Bessel function. The solution is
\be
 1 - a(r) \sim r K_1 (\mass{v}r) \ .
\label{In largear}
\ee
When $\beta > 4$ the gauge term in \bref{In U1f} determines the
behaviour of $f(r)$~\cite{asympNO}. Otherwise it can be dropped
and
\be
 1 - f(r) \sim K_0 (\mass{s}r) \ .
\label{In largefr}
\ee
The forms of $f(r)$ and $a(r)$ at small $r$ are
\be
f(r) \sim r^{|n|}  \ , \ \  a(r) \sim r^2 \ .
\label{In smallr}
\ee

\renewcommand{\baselinestretch}{1}
\begin{figure}
\centerline{\psfig{file=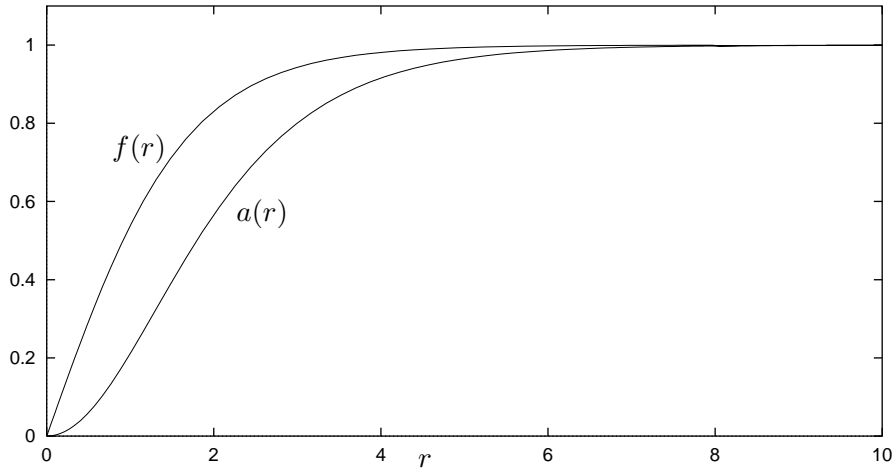,width=\myfigwidth}}
\vskip -1.8 in \hskip 1.3 in $f(r)$
\vskip 0.15 in \hskip 1.95 in $a(r)$
\vskip 1.1 in \hskip 2.9 in $r$
\caption{Plot of abelian string field profiles with $\mass{s}=\mass{v}=1$.}
\label{fa fig}
\end{figure}
\renewcommand{\baselinestretch}{\myblstr}

It is possible to find solutions of the string field equations
numerically. One suitable method is shooting. In this method, values
for the free parameters in the small $r$ solution \bref{In smallr} are
guessed, and the equations are then integrated out to large $r$. The
values of $f(r)$ and $a(r)$ are then compared with the required
values. This is repeated several times, with the choice of free parameters
being adjusted each time to get closer to the required solution at large $r$.

Rather that trying to match boundary conditions at infinity, a large
finite value of $r$ will do. In fact, as well as the required
solution, \bref{In U1f} and \bref{In U1a} have an exponentially
increasing large $r$ solution. Thus the accuracy of the computer
will severely limit the range of $r$ which can be sensibly
used. Alternatively it is possible to shoot from both small and large $r$,
using (\ref{In largear},\ref{In largefr}), and require that the
solutions meet at some point in the middle. 

A better, but more complex and memory intensive approach is
relaxation. The idea is to start with an approximate solution for all
$r$, and then see how closely it satisfies the field equations. The
difference from the required values of the field equations is then
used to calculate an adjustment to the guessed solution. This is
repeated several times until the modified guess satisfies the field
equations closely enough. Unlike shooting, the whole solution is
stored in the computer's memory, so this approach uses more
resources. In the case of cosmic string field equations, it is very
reliable. An example of solutions obtained by relaxation is shown in
figure~\ref{fa fig}.

\section{String Superconductivity}
\label{In FermZM}

In the core of a cosmic string the Higgs and gauge fields do not take their
usual values. It is possible that other fields will also have
different VEVs, resulting in the string having additional
properties. The first example of this was discussed by
Witten~\cite{Witten}. He considered an abelian string theory with
an extra scalar field, $\sigma$, whose usual VEV is zero. For a
suitable choice of potential this scalar field gains a non-zero VEV
in the core of a string (due to the variation of $\phi$). Since
$\sigma$ is charged, its non-zero VEV will break electromagnetism. If $z$ and
$t$ dependence are added to $\sigma$, the resulting solution is a
superconducting current, which is conserved. Since $\sigma$ is electrically
charged, the current will have long range effects.

It is also possible to have currents made up of fermions or gauge
bosons. Gauge boson superconductivity only occurs in more complex
nonabelian theories, and will not be discussed here. Fermion
superconductivity can occur in simpler models, such as the abelian
model discussed in section \ref{In AbString}. The fermion
superconductivity arises from Yukawa couplings, and unlike scalar boson
superconductivity, it is not necessary to use a specific potential,
so fermion superconductivity is more generic. Strictly, cosmic strings
with fermion currents are not superconductors since electromagnetism
is not broken inside the string. The strings act as perfect
conductors, since the current flows without resistance. However, the term
`superconductivity' is often used for both cases.

If charged currents can exist on a cosmic string, they will be
generated when the string passes through an electric or magnetic
field. Currents can also be generated by interaction with the plasma,
particularly when the string forms, or by string collisions. It has
been suggested that internal phase transitions occuring on the string
could generate currents too~\cite{intPTcurrent}. The presence of a
current will alter the evolution of a string network. Decaying
currents may explain observed high energy cosmic
rays~\cite{current-rays,Witten}, and could also provide a mechanism for
baryogenesis~\cite{currbaryo}. Charged currents on a fast moving string
can create shockwaves in the plasma, which has implications for
structure formation~\cite{OTW}. It is also possible that currents
could stabilise loops of cosmic string~\cite{vorton}. Such
configurations are called `vortons'. If they form at high energy
scales they can have dramatic consequences (see next section). Vortons
formed at low energies may provide a dark matter candidate~\cite{string book}.

The maximum fermion current on a string is restricted by the mass of the
corresponding particle off the string. If the current's momentum
exceeds this it can escape from the string. The ease with which the
current then escapes will depend on the curvature of the cosmic
string. For a string with radius of curvature $R$, the following 
bound on the current is obtained~\cite{maxcurrent}
\be
J_{\rm max} \sim e m^2 R \ .
\ee
In fact the maximum value the current will reach is likely to be less
than this, and will depend on the details of the model.
 
Consider an extension of \bref{In U1Model} to include a two-component
fermion, with charge $1/2$. The extra terms in the Lagrangian will then be
\be
\Lagf = \bar{\psi} i\sig{\mu} D_\mu \psi 
- \frac{1}{2} \left[ig_Y \bar{\psi} \phi \psi^c + \hconj \right] \ , 
\label{In Lagferm}
\ee
where $\sig{\mu} = (-I,\sig{i})$, $D_\mu\psi = \left(\partial_\mu -
\frac{1}{2}ieA_\mu\right)\psi$, and $\psi^c = i\sig{2} (\bar{\psi})^T$ 
is the charge conjugate of $\psi$. This gives the field equations
\be
\left(\begin{array}{cc} 
-e^{i\th}\left[\dr +\frac{i}{r}\dth +n\frac{a(r)}{2r} \right] &
\pd{z} + \pd{t} \\ \pd{z} - \pd{t} & 
e^{-i\th}\left[\dr -\frac{i}{r}\dth -n\frac{a(r)}{2r} \right] 
\end{array}\right) \psi  - \mass{f} f(r) e^{in\th} \psi^\ast = 0 \ ,
\label{In U1Ferm} \ee
where the expressions (\ref{In U1Ansatza},\ref{In U1Ansatzb}) have
been substituted for $\phi$ and $A_\mu$, and $\mass{f}=g_Y \eta$. 

Non-trivial solutions of \bref{In U1Ferm} with only $r$ and $\th$
dependence exist. It has been shown that there are $|n|$ such
solutions which are normalisable, in the sense that 
$\int |\psi|^2 d^2 x$ is finite~\cite{Jackiw}. They have zero energy,
and are referred to as zero modes. If $n=1$ the single solution can be
found analytically,
\be
\psi(r,\th) = \spinU \exp\left(
-\int^r_0 \mass{f} f(s) + \frac{a(s)}{2s} ds\right) \ .
\label{In aU1FermSol}
\ee
For higher $|n|$ it is not generally possible to determine all such
solutions analytically, although some of their properties can be
found. Like \bref{In aU1FermSol}, they are all eigenstates of
$\sig{3}$. Their eigenvalues are $+1$ if $n>0$, and $-1$ if
$n<0$. All the solutions decay exponentially outside the string, and
so are confined to it. They can be regarded as fermions trapped on the
string.

The solutions can be extended to include $z$ and $t$ dependence. This
is achieved by multiplying $\psi$ by $\alpha(z,t)$, which satisfies
$(\pd{z} \mp \pd{t})\alpha = 0$, depending on whether $\sigma^3\psi=\pm\psi$.
Thus the trapped fermions move at the speed of light, in the $\pm z$
direction. Other fermions which couple to $\phi^\ast$ can be added to
the theory. $\phi^\ast$ has the opposite winding number to $\phi$, and the
fermion currents flow in the opposite direction. Considering
both sets of fermions, currents can flow in both directions.

Although $\psi(r,\th)$ was easy to find for $n=1$, this is not
generally the case for other winding numbers, or for more complex
theories with several fermion fields (but see chapter~\ref{Ch:SUSY}). 
More realistic grand unified theories will have several Higgs fields,
and possibly more complex string solutions than the abelian case
discussed in section~\ref{In AbString}. It is still possible to determine
the existence of fermion zero modes in such cases. This is done for a
general theory in chapter~\ref{Ch:Index}. A specific grand unified
theory is also considered there. 

In supersymmetry the fermion and boson fields are related by a
symmetry. This means that the properties of a cosmic string (which is
a bosonic object) can be used to make predictions about the fermion
solutions on the string. This is a useful idea, and it is
investigated in chapters \ref{Ch:SUSY} and \ref{Ch:SUSY2}.

\section{Evolution of Cosmic String Networks}
\label{In network}

Although cosmic strings form in the early universe, many of their
observable effects will take place in the later universe. It is
therefore important to know how a network of cosmic strings will evolve.

When considering a network of cosmic string it is useful to define a
characteristic length scale $\xi$. There are various possible choices of
$\xi$, for instance $\xi^3$ can be the volume of space which
contains cosmic string of average length $\xi$. Initially $\xi$ will
grow faster than the size of the universe~\cite{CSreport}. This
cannot continue for long since causality implies that $\xi < t$.

Logically, $\xi$ can do one of two things after this point. It can
either approach a scaling solution where $\xi/t$ is constant, or it
can grow less quickly, so $\xi/t$ decreases. If $\xi$ grows more
slowly, the strings will come to dominate the energy density of the
universe. This is sometimes referred to as `overclosing' the
universe. This would substantially alter its evolution. Such evolution
is strongly ruled out by observations.

The most obvious way for a string network to evolve is to just stretch
with the expansion of the universe. Unfortunately this implies 
$\xi \propto \sqrt{t}$, and so the strings dominate the energy
density. To avoid this, there must be some mechanism for transferring
energy away from the string network.

Strings can lose energy by radiating particles. Unfortunately strings
are not usually charged, and their interactions with other fields are
likely to be weak. Strings formed at high energy scales will
emit significant amounts of gravitational radiation, although not
enough to solve the problem with the network evolution.

For topological reasons, strings are either infinite or closed loops.
String loops provide a solution to the energy density problem. When
a string intersects itself, a loop will break off. The loop will then
start losing energy by gravitational radiation. It will begin to contract, and
will eventually disappear. Unless this loop rejoins the string
network, its contribution to the network energy density is
lost. Smaller loops are unlikely to do this, and loop formation
provides a substantial enough energy loss mechanism to ensure the
string network has a scaling solution. The decay of string loops can
also explain the observed baryon asymmetry~\cite{asymmetry}.

If conserved currents exist on the strings, such as those discussed
in section~\ref{In FermZM}, the above solution fails. It is possible
for string loops to be stabilised by the angular momentum of the
trapped charge carriers~\cite{vorton}. Such stable loops are called
vortons. The loops do not decay, so they will continue to
contribute to the energy density of the cosmic string network. Since
it no longer has a sufficiently strong mechanism of energy loss, the
string network will no longer be able to reach a scaling solution. The
universe will then become dominated by vortons. The possibility of
this happening allows the underlying particle physics to be cosmologically
constrained~\cite{vortbounds}. The study of fermion (and other)
currents will give insight into the possibility of vorton formation.

When strings collide they will intercommute. It is because of this
that a loop is formed when a string intersects itself. Some
strings, such as type I ($\mass{s}^2/\mass{v}^2 < 1$) abelian strings,
do not intercommute~\cite{abnoncomm}. In order for such strings to be
cosmologically viable, they must either form late in the universe (and
so not have time to dominate it), or possess a new mechanism for
energy loss. Because of this, type II strings are likely to be more
physically significant than type I strings.

Similar arguments can be applied to domain walls and monopoles. There
is no equivalent of loop formation for domain walls. These
defects would be sure to dominate the energy density of the universe, hence
they are ruled out~\cite{domwall}. The monopole density evolves the same way
as the matter density. However, monopoles have long range effects
which strongly conflict with observations~\cite{monopoleabun}. Thus
they are ruled out too.

\section{Overview}

Although a simple $U(1)$ model is useful for illustrating the
existence of strings and their basic properties, a more realistic theory
is needed to get accurate phenomenological results. Furthermore,
strings in such theories will have a richer microstructure, which can
give rise to additional properties and phenomena. It may also alter, or even
remove, some of the properties suggested by the simplest models. In
this thesis we will consider cosmic strings in grand unified and
supersymmetric theories. Perhaps the most natural place
for cosmic strings to arise is in unified theories, so it is important to
consider the implications of their more complex symmetry breaking for
strings. Many credible extensions of the Standard Model involve
supersymmetry, and so its effects should also be examined.

In chapter~\ref{Ch:SO10} a grand unified theory which has cosmic
strings is outlined. Its unifying symmetry group is $SO(10)$. There is
a wider range of string solutions in this theory than the $U(1)$
model. Many of these have a more complex internal structure. Instead
of just one winding number, their Higgs fields will have several. We
show that a cosmic string can affect subsequent phase transitions,
forcing other Higgs fields to take string-like solutions. We
investigate this effect in the $SO(10)$ theory. Generalisation of the
results to other related theories is also discussed.

As was mentioned in section~\ref{In FermZM}, precise determination of
fermion zero modes is rarely possible. Their existence can be
determined by examining the field equations, but this is
time-consuming for more complex theories. In chapter~\ref{Ch:Index} an
index theorem is found which gives the number of zero modes for a general
theory. Whilst index theorems have been found previously, they only
give the difference between the number of left and right moving
currents. Ours is more general, and gives the total number of massless
currents. By applying the theorem before and after a phase transition we
can investigate the fate of fermion zero modes during sequences of phase
transitions in a variety of models. Depending on the couplings that
the breaking introduces, the zero modes may be destroyed and the
superconductivity of the string removed. Vortons will then dissipate,
relaxing the constraints on the theory. We discuss the features of the
theory that are required to produce this behaviour and consider the
implications of spectral flow. We apply the theorem to the unified
theory of chapter~\ref{Ch:SO10}. It is applied to other
theories in sections \ref{SS ZeroModesI} and \ref{SS2 ZeroModes}. 

In chapter~\ref{Ch:SUSY}, the microphysics of supersymmetric cosmic
strings is discussed. For simplicity two $N=1$ supersymmetric abelian
Higgs models are considered. The vortex solutions are found, and it is
shown that the two simplest supersymmetric cosmic string models admit
fermionic conductivity. In a SUSY theory, fermion and boson fields can be
transformed into each other. This allows string solutions with non-zero
fermion fields to be found analytically. These solutions are fermion
zero modes, and are found to first order explicitly. We note that this
constrains all supersymmetric grand unified theories with abelian strings.

In chapter~\ref{Ch:SUSY2} we extend the results of the previous chapter
to a model with abelian and nonabelian strings. The strings in this
model have some resemblance to those of the $SO(10)$ model discussed
in chapter~\ref{Ch:SO10}, and so may give
some insight into the properties of SUSY GUT strings. We give the string
solutions, and find analytic fermion zero mode solutions using SUSY
transformations. We consider the effects of soft supersymmetry
breaking on these cosmic string solutions, as well as those in the
previous chapter. We also examine the implications of SUSY breaking for
the fermion zero modes. Soft SUSY breaking terms are those
which break SUSY without giving quadratically divergent corrections to
the electroweak Higgs field.

So far we have only considered massless fermion currents, arising from
zero modes. In chapter~\ref{Ch:Bound} we look for massive currents
arising from bound states. We show that
there are no space-like fermion currents in any model. In contrast to
the null (or light-like) currents, it is difficult to determine the
existence of the time-like currents analytically, so we use numerical
methods instead. We determine the spectrum of fermion bound states and
currents in the abelian string model. We also speculate about similar
states in other theories.

Finally, in chapter~\ref{Ch:Summary}, the various results are
summarised, and possible future work is discussed.

\newcommand{\pzero}[1]{\ph{{#1}}^{(0)}}

\newcommand{\lamg}{\lambda_\sbG}
\newcommand{\lame}{\lambda_\sbE}
\newcommand{\lamx}{\lambda_\times}
\newcommand{\lammed}{\lambda_{\scriptscriptstyle \rm M}}

\newcommand{\hhu}[1]{h^\sbu_{{#1}}}
\newcommand{\hhd}[1]{h^\sbd_{{#1}}}
\newcommand{\hhud}[1]{h^\sbud_{{#1}}}

\newcommand{\evalu}[1]{\lambda^\sbu_{#1}}
\newcommand{\evald}[1]{\lambda^\sbd_{#1}}
\newcommand{\evalud}[1]{\lambda^\sbud_{#1}}
\newcommand{\evecu}[1]{\phi^\sbu_{#1}}
\newcommand{\evecd}[1]{\phi^\sbd_{#1}}
\newcommand{\evecud}[1]{\phi^\sbud_{#1}}

\chapter{Microphysics of SO(10) Cosmic Strings}
\label{Ch:SO10}

\section{Introduction}
As was discussed in chapter~\ref{Ch:Intro}, many significant
cosmological properties of cosmic strings arise from their microstructure.
In particular, at subsequent phase transitions the core of the cosmic
string acquires additional features. For example, the string can cause
electroweak symmetry restoration in a much larger region around it,
proportional to the electroweak scale itself~\cite{wbp&acd,
Goodband}. This microphysical structure has been used to provide a new
scenario for electroweak baryogenesis~\cite{defectbaryogenesis}, and to
investigate the current-carrying properties of cosmic
strings~\cite{acd&wbp,currenta}.

Previous work considered the simplest extension to the Standard
Model that would allow the formation of strings. A $U(1)$ symmetry,
whose breaking produced an abelian string, was added to the usual
Standard Model symmetries. The resulting theory had two coupling
constants, of arbitrary ratio. It was shown that if the ratio was
large enough, the electroweak Higgs field would not only be zero at
its centre, but would also wind like a
string. Whether this is likely to happen with phenomenological strings
can be found by considering a realistic grand unified theory, where
there is less arbitrariness. 

By using a larger gauge group it is also possible to consider the effects
of nonabelian strings, which could not occur in the theories
considered in refs.~\cite{wbp&acd,Goodband}. Nonabelian strings have
significantly different behaviour to abelian strings, since the
associated string generators do not all commute with the Standard
Model fields, or the other gauge fields. It is thus necessary to approach
them in a slightly different way.

In this chapter we examine these issues in detail for strings formed in
a realistic grand unified theory (GUT) based on $SO(10)$. In
section~\ref{SO10 GUT} the theory to be used is outlined. The possible
strings that form in it prior to the electroweak phase transition are
discussed in section~\ref{SO10 GUTstrings}. The effects
of a cosmic string on the subsequent phase transitions of a general
theory are considered in section~\ref{SO10 multiHiggs}. In 
section~\ref{SO10 EWSymBrk} we consider these effects in greater
detail for the electroweak phase transition of the $SO(10)$ GUT, and
examine the various electroweak string solutions in
section~\ref{SO10 EWstrings}. The form of the electroweak fields, and
numerical solutions of the corresponding field equations, are found. In
section~\ref{SO10 OtherSymmRest} some other, simpler symmetry
restorations occuring in the theory are discussed, in particular that
of the intermediate $SU(5)$ symmetry. Although one specific theory is
considered, many of the results generalise to other theories. The
implications of our results for such theories are discussed in
section~\ref{SO10 OtherGUT}. In section~\ref{SO10 Conc} we summarise
our results and discuss the conclusions.

\section{An SO(10) Grand Unified Theory}
\label{SO10 GUT}

A realistic GUT which has a symmetry breaking pattern which produces
strings is $SO(10)$. Its properties have a reasonable agreement
with physical results. Consider the symmetry breaking
\bea
SO(10) & \pharrow{126} & SU(5) \times Z_2 \nonumber \\ 
& \pharrow{45} & SU(3)_c \times SU(2)_L \times U(1)_Y \times Z_2 \nonumber \\
& \pharrow{10} & SU(3)_c \times U(1)_Q \times Z_2 \ ,
\label{symbreak}
\eea
where $\ph{\rm N}$ transforms under the {\bf N} representation of $SO(10)$.
The actual grand unified gauge group is {\it Spin}$(10)$, the covering group of
$SO(10)$, but for simplicity the symmetry breaking is shown in
terms of the Lie algebras. 
The discrete $Z_2$ symmetry left by the $\ph{126}$ Higgs field
leads to the formation of a variety of cosmic strings.  Comparison
of the effects of the various symmetry breakings is simplified by
expressing everything in terms of the same representation. Since
${\bf 126} + {\bf 10} = \sprod{\bf 16}{\bf 16}$ this is possible. 
Conveniently, {\bf 16} is also the representation that acts on the
fermions. The $SO(10)$ fermions consist of the usual standard model
fermions, plus a right handed neutrino. The fermionic part of the
theory is then expressed in terms of the left-handed fermions and the
charge conjugates of the right-handed fermions. The fields of the theory are
discussed in more detail in appendix~\ref{Ch:app}.

The maximal subgroup of $SO(10)$ is actually $SU(5) \times U(1)_P$, and 
$P$ can be used to decompose $SO(10)$ into representations of $SU(5)$
\be
{\bf 16} \longrightarrow {\bf 1}_5 + {\bf 10}_1 + {\bf \bar{5}}_{-3} \ ,
\ee
where the subscripts are the eigenvalues of $P$. ${\bf 126}$ and ${\bf
10}$ can be similarly decomposed by considering symmetric products of
{\bf 16}. 
\be
{\bf 126} \longrightarrow {\bf 1}_{10} + \ldots  \ ,\hspace{.6in}
{\bf 10} \longrightarrow {\bf 5}_{-2} + {\bf \bar{5}}_2 \ .
\ee
$P$ can also be used to describe the non-trivial element of
the discrete symmetry of (\ref{symbreak}), which is $d = \exp(2\pi i P/10)$.

Defining $\phvac{\rm N}$ to be the usual constant vacuum expectation value
of $\ph{\rm N}$, $\phvac{126}$ has a magnitude of
$\etag$, which is of order $10^{16}$GeV. It is in the
${\bf 1}_{10}$ component of {\bf 126}, and so must be equal to
$\etag (\Nr \times \Nr)$, where $\Nr$ is in the ${\bf 1}_5$
component of the {\bf 16} representation (the corresponding field in
the fermion representation is the charge conjugate of the right-handed
neutrino). $\phvac{10}$ is made up of both the chargeless
components of {\bf 10}. If $\Hd{0}$ and $\Hu{0}$ are the chargeless
components of ${\bf 5}_{-2}$ and ${\bf \bar{5}}_2$ respectively,
$\phvac{10} = \etad\Hd{0} + \etau\Hu{0}$. Since {\bf 10}
is contained in $\sprod{\bf 16}{\bf 16}$, $\ph{10}$ can be
expressed as a sum of symmetric products of components of {\bf
16}s. The {\bf 45} is contained in ${\bf 16} \times {\bf
\bar{16}}$. $\ph{45}$'s effect on the form of the cosmic string
solutions is far less significant than the other Higgs fields, so we
will ignore it for now.
 
Strings can form at the first $SO(10) \longrightarrow SU(5) \times Z_2$
symmetry breaking. In this case $\ph{126}$ is not constant, and takes
the form $e^{i \th \ts}\pzero{126}(r)$. $\pzero{126}$ is independent
of $\th$, and satisfies the boundary condition 
$\pzero{126}(\infty) = \phvac{126}$. $\ts$ is made up of the broken
generators of $SO(10)$, and must give a single-valued $\ph{126}$.  
If $e^{2 \pi i \ts} = U \times d$ (for some $U$ in $SU(5)$), then while 
$\ph{126}$ will be single valued, it will not be topologically equivalent to
$\ph{126} =$ constant, and so the string will be topologically
stable. If $e^{2 \pi i \ts} = U \times I$ the string is not
topologically stable, but may have a very long lifetime, and so still
be physically significant~\cite{Witten}.

The Lagrangian of the system is 
\bea
\Lag &=& (D_\mu \ph{126})^{\ast}(D^\mu \ph{126})
 +(D_\mu \ph{10})^{\ast}(D^\mu \ph{10})
 + (D_\mu \ph{45})^{\ast}(D^\mu \ph{45}) \nonumber \\
 &&\hspace{.2in} {}- \frac{1}{4} F^a_{\mu \nu}F^{\mu \nu a} 
- V(\ph{126},\ph{45},\ph{10})+ \Lagf \ ,
\label{lagrangian}
\eea
where $D_\mu = \partial_{\mu} - \frac{1}{2}igA_{\mu}$ and $F_{\mu \nu} =
\partial_{\mu}A_{\nu} - \partial_{\nu}A_{\mu} - 
\frac{1}{2}ig[A_{\mu},A_{\nu}]$. 
There are 45 gauge fields in all, most of which acquire superheavy
masses and so are not observed at everyday temperatures. They consist
of the usual standard model fields, with the $W$-bosons
denoted by $\WL{i}$; $\WR{i}$, which are right-handed versions of the
$\WL{i}$, coupling right handed neutrinos to electrons; some
leptoquark bosons: \Ypm, \Xpm, \Xspm, where \Ypm\ and \Xpm\ are $SU(5)$
gauge fields;  
some more general gauge fields \Xgpm\ and \Ygpm,
which couple quarks to leptons and different coloured quarks;
 and a fifth uncharged field, $B'$.
The index $i$ takes the values 1, 2, 3, and is related to colour. Two uncharged
diagonal fields, $Z'$ and $B$, are
made up of orthogonal linear combinations of $\WR{3}$ and
$B'$. Linear combinations of $B$ and $\WL{3}$ produce the $Z$ boson and
the photon, $A$.
At the first symmetry breaking $Z'$, \WRpm, \Xgpm, \Ygpm, and \Xspm\ are
all given superheavy masses. The second stage gives high masses to \Xpm, \Ypm,
and additional masses to \WRpm, \Xgpm, \Ygpm\ and \Xspm. Finally the third
symmetry breaking gives masses to \WLpm and $Z$, with further
masses being given to the $Z'$, \WRpm, \Ypm\ and \Ygpm\ fields.

\section{GUT Strings}
\label{SO10 GUTstrings}

Neglecting fermions, the Euler-Lagrange equations obtained from
\bref{lagrangian} are
\bea
D_\mu D^\mu \ph{i} &=& - \frac{\partial V}{\partial \ph{i}^\ast} \ ,
\label{phiequ} \\
(D_\mu F^{\mu \nu})^a 
&=& -g\mbox{Im} \sum_i (D^\nu \ph{i})^\ast(\gena \ph{i}) \ ,
\label{SO10 gaugeequ}
\eea
where $D_\mu F^{\mu \nu} = \partial_\mu F^{\mu \nu}
- i\frac{g}{2} \left[A_\mu,F ^{\mu \nu}\right]$.
At high temperatures $V$ is such that $\ph{126}$ is the only non-zero
Higgs field. \bref{phiequ} and \bref{SO10 gaugeequ} have various cosmic string
solutions. The different solutions correspond to different choices of $\ts$. 
In the $SO(10) \longrightarrow SU(5) \times Z_2$ symmetry breaking,
21 of $SO(10)$'s 45 generators are broken, and $\ts$ will be a linear
combination of them. One of them, $P$,  
corresponds to the $U(1)$ symmetry not embedded in $SU(5)$. The
corresponding string is abelian, and has the solution
\bea
\ph{126} &=& f(r)e^{in\th \ts}\phvac{126} \ , \nonumber \\
A_{\th} &=& n \frac{2a(r)}{gr} \ts \ , \hspace{0.8in}
A_{\mu} = 0 \mbox{ \ otherwise} \ ,
\label{SO10 abansatz}
\eea
where $\ts$, the string generator, equals $P/10$, and $n$ is
an integer. The non-zero gauge field is required to give a zero
covariant derivative, and hence zero energy, at $r=\infty$. It
corresponds to a non-zero $Z'$ field. \bref{SO10 abansatz} can be
simplified using $P\Nr = 5\Nr$, to give 
$e^{in\th \ts}\phvac{126} = e^{in\th}\phvac{126}$. Substituting
\bref{SO10 abansatz} into \bref{phiequ} and \bref{SO10 gaugeequ} gives the
Nielsen-Olesen vortex equations, as would be expected. Regularity at
the centre of the string, and finite energy due to a vanishing
covariant derivative and potential at infinity, imply the boundary
conditions $f(0)=a(0)=0$ and $f(\infty)=a(\infty)=1$.

The situation for the other generators is more complicated. For a
general string generator $\ts$, the left and right hand sides of \bref{phiequ}
are proportional to $\ts^2 \pzero{126}$ and $\pzero{126}$ respectively,
which in general are not proportional~\cite{Ma}. Thus the solution
\bref{SO10 abansatz} will not work. This is resolved by expressing 
$\phvac{126}$ in terms of the eigenstates of $\ts^2$, to give 
$\phvac{126} = \sum_m \phi_m$, 
where $\ts^{2} \phi_m = m^2 \phi_m$. Since $\ts$ is Hermitian, $m^2$
will be positive and real. A suitable string solution can now be
constructed
\bea
\ph{126} &=& e^{in\th \ts}\sum_m f_m(r)\phi_m \ , \nonumber \\ 
A_{\th} &=& n\frac{2a(r)}{gr} \ts \ , \hspace{0.8in}
A_{\mu} = 0 \mbox{ \ otherwise} \ .
\label{SO10 nabansatz}
\eea
In order for $\ph{126}$ to be single valued the various $m$ must all be
integers, and $\ts\phi_{0}$ must be zero. The boundary conditions on
$a$ and $f_m$ will be the same as those for \bref{SO10 abansatz},
except that $f_0$ need not be zero at $r=0$. The simplest examples
of such solutions occur when $\ts^{2}\Nr=\frac{1}{4}\Nr$, in which case
\bea
\phi_1 &=& \frac{\etag}{2}\left[ 
(\Nr \times \Nr) + 4 (\ts \Nr \times \ts \Nr) \right]
\ , \\
\phi_0 &=& \frac{\etag}{2}\left[ 
(\Nr \times \Nr) - 4 (\ts \Nr \times \ts \Nr) \right]
\ . \label{nabphis}
\eea
$\ts \Nr$ and $\Nr$ are orthogonal, so $\phi_0$ and $\phi_1$ are
orthogonal. In this case only part of the Higgs
field winds around the string. This type of string was first suggested
by Aryal and Everett~\cite{AryalE}, and has been examined in detail by
Ma~\cite{Ma}. It turns out to have lower (about half as much) energy
than the abelian string \bref{SO10 abansatz}. This is because the Higgs
field is not forced to be zero at the string's centre, which reduces
the contribution to the energy from the potential terms. Also since
$\ph{126}$ varies less, the covariant derivative terms are smaller.

Of course, such vortex-like solutions are only topological strings if
$e^{2\pi i\ts}$ is not contained in $SU(5)$. If $n$ is even this is
not the case, and the solution is topologically equivalent to the
vacuum.  Similarly, odd values of $n$ are all topologically equivalent
to each other, so there is only one topologically distinct type of
string of this form. Strings with higher $n$ can unwind into
strings with lower $n$. The same is true of the abelian string.
However it is possible that the lifetime of an $n>1$ string will be
very long, so in a general theory all values of $n$ should be considered. 

\renewcommand{\baselinestretch}{1}
\begin{figure}
\centerline{\psfig{file=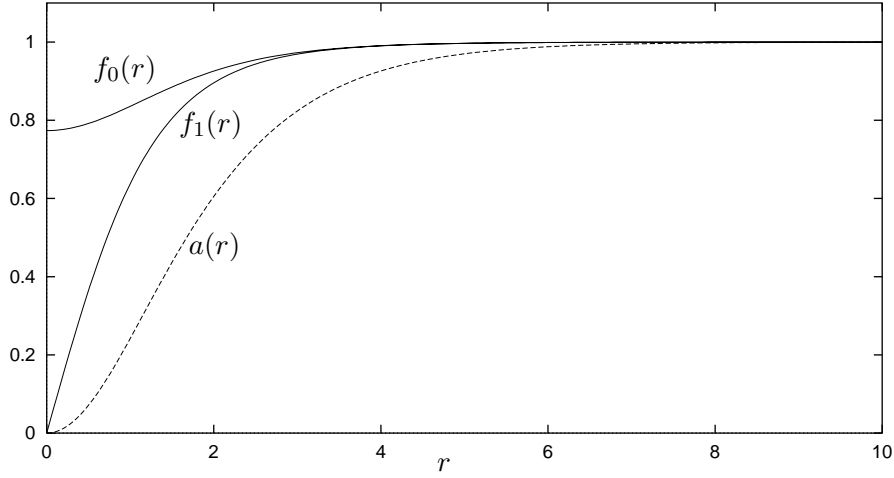,width=\myfigwidth}}
\vskip -2.2 in \hskip 1.2 in $f_0(r)$ 
\vskip 0.1 in \hskip 1.65 in $f_1(r)$
\vskip 0.45 in \hskip 1.7 in $a(r)$
\vskip 0.95 in \hskip 3.0 in $r$
\caption{$SU(2)$ string field profiles, 
with $\etag=\lamg=g=1$ and $\lamg'=5$.}
\label{nabstr fig}
\end{figure}
\renewcommand{\baselinestretch}{\myblstr}

As shown in~\cite{Ma} the most general potential reduces to a
different form to that of the abelian case, and leads to these
equations for $a$ and the $f_{m}$'s   
\be 
f_0'' + \frac{f_0'}{r} = \frac{\etag^2}{4} \left[\lamg
\left(f_1^2 + f_0^2 - 2\right) - \lamg'
\left(f_1^2 - f_0^2\right)\right]f_0 \ ,
\label{nabequa}
\ee
\be
f_1'' + \frac{f_1'}{r} - n^2\frac{(1-a)^2}{r^2}f_1 = 
\frac{\etag^2}{4} \left[\lamg\left(f_1^2 + f_0^2 -
2\right) + \lamg' \left(f_1^2 - f_0^2\right)\right]f_1 \ ,
\label{nabequb}
\ee
\be
a'' - \frac{a'}{r} = -g^2 \etag^2 (1-a)f_1^2 \ . 
\label{nabequc}
\ee
$\lamg$ and $\lamg'$ are such that $f_1 (\infty)$ and 
$f_0 (\infty)$ will both be 1. A numerical solution of the above
equations is shown in figure~\ref{nabstr fig}. Since $f_0 \neq 0$ at
the string's centre, symmetry is still broken there (although the VEV
of $\ph{126}$ is lower). The value of $f_0(0)$ depends on the ratio of
$\lamg$ and $\lamg'$. If $\lamg'>\lamg$, $f_0(0)<1$ while $\lamg'<\lamg$
gives $f_0(0)>1$. If $\lamg'=\lamg$, $f_0 =1$ everywhere, and
(\ref{nabequb},\ref{nabequc}) reduce to abelian string field equations.

The corresponding equations for the abelian string ($\ts = P/10$) are
\be
f'' + \frac{f'}{r} - n^2 \frac{(1-a)^2}{r^2}f =
				 \frac{1}{2}\etag^2 \lamg (f^2 - 1)f \ ,
\label{SO10 abGUTphi}
\ee
\be
a'' - \frac{a'}{r} = -5 g^2 \etag^2 (1-a)f^2 \ . 
\label{SO10 abGUTgauge}
\ee
The above nonabelian strings are in fact all $SU(2)$ strings. There
are other more complicated possibilities, for which $\ts^2 \Nr$ is not
proportional to $\Nr$, but none of these are topologically stable. We
shall only consider topologically stable strings, and the closely
related solutions with higher winding numbers. We will consider the
strings corresponding to each of the broken generators.
 These are all equivalent under
$SU(5)$, but not under $SU(3)_c \times U(1)_Q$, so they will be
distinct after the electroweak symmetry breaking. Apart from the
abelian string the four cases correspond to non-zero \WRpm, \Xspm,
\Xgpm\ and \Ygpm\ fields. Under $SU(3)_c \times U(1)_Q$
any linear combination of \Xspm\ generators can be gauge transformed
into any other combination, thus they are equivalent. The same is true
for the other generators, so there are just 5 distinct types
of string at low temperatures. The 4 nonabelian strings can be labelled
by their gauge fields. Under $U(1)_Q$, nonabelian strings
with winding number $-n$ are gauge equivalent to ones with winding
number $n$ (for any choice of $\ts$), so it is sufficient to
consider only $n > 0$ strings. 

Since all the nonabelian strings are gauge equivalent under $SU(5)$,
they have the same energy, which is about half that of the abelian
string~\cite{Ma}. Later phase transitions will alter the string
solutions and remove this degeneracy.

\section{Cosmic Strings and Multiple Higgs Fields}
\label{SO10 multiHiggs}

Before examining the effect of the other phase transitions on the
$SO(10)$ cosmic strings, we will consider a general GUT. Suppose it
has the symmetry breaking
\be
G_0 \pharrow{0} G_1 \times D \pharrow{1} G_2 \times D \cdots
G_{n-1} \times D \pharrow{n-1} G_n \times D 
\ee
where $D$ is a discrete group, such as $Z_N$. Define $\phvac{i}$
to be the usual constant vacuum expectation value of $\ph{i}$, and 
$D_\mu = \partial_\mu - i\frac{g}{2} A_\mu$ to be the covariant
derivative. Because of the discrete group cosmic strings can form at
the first phase transition. Away from the string core the solution
will take the form
\be
\ph{0} = R_0(\th)\phvac{0} \ , \ \ 
A_\th = \frac{2}{igr} R_0^{-1}(\th) \dth R_0(\th) \ ,
\ee
where $R_0(\th) \in G_0$. We will choose $R_0(0) = I$. Since $\ph{0}$
is single valued, $R$ must obey $R_0(2\pi)\phvac{0} = \phvac{0}$, thus
$R_0(2\pi) = U_1 \times d$, where  $U_1 \in G_1$ and $d \in D$. 
If the string is to be topologically stable $d$ must not be equal to $I$.
The gauge field ensures a vanishing covariant derivative, and thus
finite energy.

At the next phase transition $\ph{1}$ gains a VEV. If its covariant
derivative is also to vanish, $\ph{1}$ must wind like $\ph{0}$, so
\be
\ph{1} = R_0(\th)\phvac{1} \ .
\ee
Unfortunately if $R_0(2\pi)\phvac{1} \neq \phvac{1}$, then $\ph{1}$
is not single valued. The same is true of gauge and fermion fields who
get their masses from $\ph{1}$. Therefore this ansatz is unacceptable.

If the product $G_1 \times D$ is a direct product, it is possible
to alter the above ansatz to give a single valued $\ph{1}$, without
changing $\ph{0}$. This is achieved by applying an extra (singular)
gauge transformation that has no effect on $\ph{0}$, but ensures the
values of $\ph{1}$ at 0 and $2\pi$ match up. A corresponding gauge
term is then added so the covariant derivative vanishes.  
\bea
\ph{1} &=&  R_0(\th)R_1(\th)\phvac{1} \ , \nonumber \\
A_\th &=& \frac{2}{igr} \left\{ R_0^{-1}(\th) \dth R_0(\th) 
	+ R_0(\th)[R_1^{-1}(\th) \dth R_1(\th)] R_0^{-1}(\th) \right\} \ ,
\eea
where $R_1(\th)$ is chosen so that $R_1(\th) \in G_1$, 
$R_1(0) = I$ and $R_0(2\pi)R_1(2\pi) = U_2 \times d$, with $U_2 \in G_2$. 
All the fields are now single valued, and since $R_1(\th)$
annihilates $\phvac{0}$, the form of the original part of the string
solution is unaffected.

The same arguments apply to the subsequent symmetry breakings, leading
to the ansatz
\bea
\ph{i} &=& \prod^i_{j=0} R_j(\th) \phvac{i} \ , \nonumber \\
A_\th &=& \frac{2}{igr} \sum^{n}_{i=0} \left\{ \prod^{i-1}_{j=0} R_j(\th) 
  [R_i^{-1}(\th) \dth R_i(\th)] \prod^0_{j=i-1} R_j^{-1}(\th) \right\} \ ,
\label{SO10 multiAnsatz}
\eea
where the gauge transformations $R_i$, $i=0\ldots n-1$, satisfy
$R_i(\th) \in G_i$, $R_i(0) = I$ and 
$\prod^i_{j=0} R_j(2\pi) = U_{i+1} \times d$ for some $U_{i+1} \in G_{i+1}$.

If we define $G_{n+1}$ to be $I$, and choose an $R_n$ which satisfies
the above conditions, the theory's gauge fields can be defined as
\be
A_\mu(\th) = \prod^{n}_{j=0} R_j(\th) A_\mu(0)
	\prod^0_{j=n} R_j^{-1}(\th) \ .
\ee
Such a definition is single valued for all $A_\mu \in L(G_1)$, and so the
resulting cosmic string is not an Alice string.

Although the conditions after \bref{SO10 multiAnsatz} restrict the choice of
$R_i$, they do not generally determine it uniquely. Thus it is
necessary to look at the string field equations in detail.

In the above discussion we have only considered the form of the string
solution away from the string core. In some string solutions different
parts of the Higgs fields have different radial and angular dependence
(see section~\ref{SO10 GUTstrings}). In these cases the angular
dependence of each part must be such that it is single valued. Also, since
the solution must regular, all the winding parts of the Higgs fields
must be zero at the string's centre.

We will now apply the above arguments to the $SO(10)$ cosmic strings
discussed in section~\ref{SO10 GUTstrings}. These have 
$R_0(\th) = e^{in\ts\th}$. For all choices of $\ts$,
$e^{2\pi in\ts}\phvac{45} = \phvac{45}$, so we can put $R_1(\th) = I$,
which is the minimal energy choice. However 
$e^{2\pi in\ts}\phvac{10} \neq \phvac{10}$ for some $\ts$, so a non-trivial
choice of $R_2(\th)$ may be required. To determine the most favourable
choice we have to look at the form of the possible solutions close to the
cosmic string.

\section{The Electroweak Symmetry Breaking} 
\label{SO10 EWSymBrk}

Topological strings only form in a symmetry breaking $G \longrightarrow H$
if $\pi_1 (G/H) \neq I$. This is not the case at the electroweak ($\ph{10}$)
symmetry breaking, so such strings do not form there. It is still possible
for $\ph{10}$ to wind, and for string-like solutions to
appear~\cite{Vachaspati}.
However, since it is energetically favourable for the Higgs field to
unwind, they are not completely stable (although they could take a
long time to decay). 

The situation is different in the presence of a topological string,
formed at a previous symmetry breaking. As we discussed in
section~\ref{SO10 multiHiggs} the string gauge field may force a
Higgs field to wind and take a string-like solution. Unlike the
electroweak strings considered by Vachaspati~\cite{Vachaspati},
such solutions would be stable. 

As we discussed in the previous section, we need to add an extra
singular gauge transformation to the cosmic string ansatz in order to
give a single valued electroweak Higgs field ($\ph{10}$). We will
define it to be $e^{i\tew\th}$. Thus at large $r$
\bea
\ph{10} &=& e^{in\ts\th}e^{i\tew\th} (\etau \Hu{0} + \etad \Hd{0}) 
\ , \\
A_\th &=& \frac{2n}{gr}\ts + 
	\frac{2}{gr} e^{in\ts\th} \tew e^{-in\ts\th} \ .
\eea 

Before we consider the radial dependence of the solution, we need to
express $\ph{10}$ and $\tew$ in terms of eigenstates of the string
generators (as we did with $\ph{126}$ in section~\ref{SO10 GUTstrings}).
Thus we set $\etaud\Hud{0} = \sum_{j,k} \evecud{jk}$, with 
$\ts \evecud{jk} = j\evecud{jk}$, $\tew \evecud{jk} = k\evecud{jk}$,
and $\tew = \sum_j T_j$, with $[\ts,T_j] = jT_j$. A suitable ansatz is
then
\bea
\ph{10} &=& e^{in\ts\th}e^{i\tew\th} \sum_{j,k} \left(\evecu{jk}
\hhu{jk}(r) + \evecd{jk} \hhd{jk}(r) \right) 
\nonumber \\
&=& \sum_{j,k} e^{i(nj+k)\th} \left(\evecu{jk} \hhu{jk}(r) 
 + \evecd{jk} \hhd{jk}(r) \right)\ , \\
A_\th &=& \frac{2na(r)}{gr}\ts 
	  + \sum_j \frac{2b_j(r)}{gr} e^{in\ts\th} T_j e^{-in\ts\th} \ .
\eea
If $[\ts,\tew] = 0$ the second gauge term reduces to $2b_0(r)\tew / (gr)$. 
If $[\ts,\tew] \neq 0$ the $r$ component of the gauge field equations
\bref{SO10 gaugeequ} becomes non-trivial and we also need to introduce
a non-zero $A_r$ field, made up of $e^{in\ts\th} jT_j e^{-in\ts\th}$ terms.
At $r=0$ the functions $b_j(r)$ and $\hhud{jk}(r)$ take values
which give a regular solution, and at $r=\infty$ they are all equal to 1.
Since $\ph{10}$ is single valued, $nj+k$ must be an integer whenever
$\evecud{jk} \neq 0$. Any $A_r$ field must be zero at $r=\infty$.

We can now construct an approximate solution and use this to get an
estimate of the energy. Defining $\rs$ to be the radius of the string
(outside of which $|\ph{126}|$ takes its usual VEV), and $\rew$ to be
the radius of the region in which $|\ph{10}|$ does not take its usual
VEV, the solution can be approximated separately in three regions. If
all the coupling constants are of order 1, $\rs$ will be of order
$|\phvac{126}|^{-1} = \etag^{-1}$ and $\rew$ will be of order
$|\phvac{10}|^{-1} = \etae^{-1}$, with $\etae^2 = \etau^2 + \etad^2$.
$\ph{10}$ takes the role of the Weinberg-Salam Higgs field and so
$\etae \sim 10^2$GeV.

We will set all fields to their asymptotic values for $r>\rew$.
For $\rs < r < \rew$ we set $a(r)$ and $\ph{126}$ to their asymptotic
values, and assume all other fields to be small. To first order the
field equations are
\be
(\hhud{jk})'' + \frac{(\hhud{jk})'}{r} -
 \frac{k^2}{r^2}\hhud{jk} = 0 \ \ , \ \ 
b''_j - \frac{b'_j}{r} = 0 \ .
\ee
These are solved by $\hhud{jk}(r) \sim r^{\pm|k|}$ and 
$b_j(r) \sim r^2$ or $b_j(r)=$ constant. For $r < \rs$ we take all
fields to be small, and the field equations reduce to
\be
(\hhud{jk})'' + \frac{(\hhud{jk})'}{r} -
 \frac{(nj+k)^2}{r^2}\hhud{jk} = 0 \ \ , \ \ 
b''_j - \frac{b'_j}{r} = 0 \ .
\ee
These are solved by $\hhud{jk}(r) \sim r^{|nj+k|}$ and $b_j(r) \sim
r^2$ (the other solutions are not regular at $r=0$). Requiring
continuity of the solution and its first derivative at $r = \rs$,
reveals that the $\hhud{jk}(r) \sim r^{-|k|}$ and $b_j(r)=$ constant
solutions can be neglected outside the string core. Requiring
continuity at $r=\rew$ gives the trial solution
\be
\hhud{jk}(r) = h_k(r) = \left\{ \begin{array}{c}
 \left(\frac{r}{\rew}\right)^{|k|} \\  1 \end{array} \right. \hspace{.2in}
b_j(r) = b(r) = \left\{ \begin{array}{cl}
 \left(\frac{r}{\rew}\right)^2 & \rs < r < \rew \\
		  1 & r > \rew \end{array} \right. \ .
\label{SO10 trial}
\ee
We assume any $A_r$ field is negligible.
An estimate of the energy of this string-like solution can now be found by
substituting the trial solution into the Lagrangian. The contribution
from the region $r<\rs$ is suppressed by powers of $\rs/\rew \sim
10^{-14}$ and can be neglected. All the contributions are zero for 
$r > \rew$, thus
\bea
E &=& 2\pi\int_{0}^{\infty} r dr \left\{
  \left|\dr \ph{10}\right|^2 + \left|D_\th\ph{10}\right|^2
+ \frac{1}{2}\left(\dr A_{\theta}^{a}\right)^2 + V \right\} \nonumber \\
  &\approx& 2\pi\int_{\rs}^{\rew} r dr  \sum_k
 \left( h'^2_k + \frac{k^2}{r^2}(1-b)^2 h_k^2 \right) 
|\evecu{k} + \evecd{k}|^2
+ \frac{1}{2}\left(\frac{2b'}{gr}\right)^2 {\rm Tr}(\tew^2) + V \nonumber \\
  &\approx&  2\pi   \left(\frac{|k|}{2} 
	+ \frac{|k|}{2}  - \frac{2k^2}{2+2|k|} + \frac{k^2}{4+2|k|} 
	\right) |\evecu{k} + \evecd{k}|^2 
 \nonumber \\ & & \hspace{1in} {}+
\frac{4}{g^2 \rew^2} {\rm Tr}(\tew^2) +
 2\pi\int_{\rs}^{\rew} V r dr \ .
\label{SO10 trialEnergy}
\eea
where $\evecud{k} = \sum_j \evecud{jk}$.

The generator $\tew$ should annihilate $\phvac{126}$, since any change
to the GUT string will give a large increase in energy. It is therefore a
combination of electroweak generators. Gluon and photon generators
do not affect $\phvac{10}$, so the most energetically favourable
choice will be of the form
\be
\tew = z \gen{Z} + \frac{1}{\sqrt{2}} 
	\left(w\gen{\WL{+}} + w^\ast \gen{\WL{-}}\right) \ ,
\label{SO10 tew}
\ee
where \gen{Z} is the generator of the Z boson field, etc.
$z$ (which is real) and $w$ (complex) are parameters to be determined.
We will now find the eigenstates ($\evecud{k}$) of $\tew$ which make up
$\Hu{0}$ and $\Hd{0}$, and the corresponding eigenvalues ($\evalud{k}$).

Defining $z' = \sqrt{\frac{5}{8}} z$, their eigenvalues are
\be
\evalu{\pm} = \frac{3}{5} z' \pm \sqrt{z'^2 +|w|^2} \ \ , \ 
\evald{\pm} = -\frac{3}{5} z' \pm \sqrt{z'^2 +|w|^2} \ ,
\label{SO10 EWeval}
\ee
and the eigenstates are
\bea
\evecu{\pm} &=& \left(\left[z' \pm \sqrt{z'^2 +|w|^2}\right] \Hu{0} 
	+ w \Hu{-} \right) \frac{\etau}{2\sqrt{z'^2 + |w|^2}} \ , \nonumber \\
\evecd{\pm} &=& \left(\left[-z' \pm \sqrt{z'^2 +|w|^2}\right] \Hd{0} 
	+ w^\ast \Hd{+} \right) \frac{\etad}{2\sqrt{z'^2 + |w|^2}} \ .
\label{SO10 EWevec}
\eea
Thus
\be
\ph{10} = e^{in\ts} \sum_k \left(e^{i\evalu{k}\th} \evecu{k} \hhu{k}(r) 
 + e^{i\evald{k}\th} \evecd{k} \hhd{k}(r) \right) 
\ee
outside the string core.

Substituting these results into \bref{SO10 trialEnergy}, and taking
$g\rew/2 = \etae^{-1}$ gives
\bea
E &\approx& 2\pi \left[ \sum_\pm \left(|\evalu{\pm}| |\evecu{\pm}|^2
+ |\evald{\pm}| |\evecd{\pm}|^2 \right) + \etae^2 {\rm Tr}(\tew^2) \right] 
\nonumber \\
&=& 10\pi\etae^2 (z^2 + |w|^2) \ .
\label{SO10 EWenergy}
\eea
Thus the minimal energy solution will have $z$ and $|w|$ as small as
possible, subject to $e^{in\ts} \evecud{\pm}$ being single valued. If
$\evecud{\pm}$ is made up of eigenstates of $\ts$ with eigenvalues
$\mu^\sbud_{\pm,j}$ then this is true if
$\evalud{\pm} + \mu^\sbud_{\pm,j}$ are all integers.

If $|w|=0$ the situation is slightly different. In this case
$\phvac{10}$ is composed of 2 eigenstates of $\tew$, and not 4. They
are $\Hu{0}$ and $\Hd{0}$, and have eigenvalues
\be
\evalu{} = \frac{8}{5} z' \ \ , \ 
\evald{} = -\frac{8}{5} z' \ .
\label{SO10 EWeval0}
\ee
If $\Hu{0}$ and $\Hd{0}$ are made up of eigenstates of $\ts$ with
eigenvalues $\mu^\sbud_{j}$, then it is only necessary for 
$\evalud{} + \mu^\sbud_{j}$ to all be integers if $\ph{10}$ is
to be single valued.

\section{Electroweak Cosmic String Solutions} 
\label{SO10 EWstrings}

Using the results of section~\ref{SO10 EWSymBrk} we can now find the
form of the electroweak Higgs and gauge fields around the various
$SO(10)$ cosmic string solutions.

\subsection{The Abelian U(1) String}
\label{SO10 EWZsect}
 
With the abelian string, $\ts \Hu{0,-} = \frac{1}{5}\Hu{0,-}$ and
$\ts \Hd{0,+} = -\frac{1}{5}\Hd{0,+}$. Thus 
$e^{i n\ts\th}\phvac{10} 
= \etau \Hu{0} e^{in\th/5} + \etad \Hd{0} e^{-in\th/5}$ 
is not generally single valued, so a non-zero $\tew$ is needed.

The most general suitable $\tew$ is given by \bref{SO10 tew}. We can
see that the eigenstates of $\tew$ which make up $\phvac{10}$ (given
by \bref{SO10 EWevec}) are also eigenstates of $\ts$. The minimal
energy choice of $\tew$ must minimise ${\rm Tr}(\tew^2)$, while
satisfying the conditions after \bref{SO10 EWenergy} (or after
\bref{SO10 EWeval0} if $|w|=0$). For this string the conditions are that 
\bea
\evald{\pm} - n/5 \ \  &\mbox{and} \ \ \evalu{\pm} + n/5  
\ \ \ &(\mbox{if} \ \ w \neq 0) \ , \\
\evald{} - n/5 \ \ &\mbox{and} \ \ \evalu{} + n/5  
\ \ \ &(\mbox{if} \ \ w = 0) \ ,
\eea
are all integers. This occurs when $w=0$ and $\sqrt{8/5} z + n/5$ is
an integer, or when $3z/(2\sqrt{10}) \pm \sqrt{5z^2 /8 + |w|^2} + n/5$
are integers. The choice which minimises \bref{SO10 EWenergy} is $w=0$,
$z=-\sqrt{5/8}\{n/5\}$, where $\{x\}$ is defined as $x$ minus the nearest
integer. Thus if $n$ is a multiple of 5, $\tew=0$.

Since $[\ts,\tew]=0$ the electroweak string solution can be written in
the form
\bea
\ph{10} &=& \left(\etau \Hu{0} e^{im\th} h_\sbu (r)
+ \etad \Hd{0} e^{-im\th} h_\sbd (r)\right) 
\ , \nonumber \\
A_\th &=& \frac{2na(r)}{gr}\ts + \frac{2b(r)}{gr} \tew
\hspace{.6in} A_\mu = 0 \mbox{ \ otherwise} \ ,
\label{EWabansatz}
\eea
where $m = n/5 - \{n/5\}$. The solution has the boundary conditions 
$h_\sbud(\infty)=b(\infty)=1$, $b(0)=0$, and if $m \neq 0$ then
$h_\sbud (0)=0$ too. If $m=0$, which occurs when $|n| < 3$, $\ph{10}$ need
not be zero a the string's centre. We can get an estimate of it value
using a trial solution like that in section~\ref{SO10 EWSymBrk}.  The
solution has $h_\sbud (r) \approx (r/\rew)^{|\{n/5\}|}$ for $\rs < r
<\rew$ and $h_\sbud (r)$ constant for $r < \rs$. By continuity at $r=\rs$,
$h_\sbud (0) \approx (\rs/\rew)^{|\{n/5\}|} \sim 10^{-14|\{n/5\}|}$. Thus
electroweak symmetry is almost fully restored for abelian
strings even when $\ph{10}$ does not wind. Although symmetry is restored,
$\ph{10}$ will only wind like a string if the GUT string has a high
winding number ($|n| > 2$). This is in agreement with earlier work by
Alford and Wilczek~\cite{Alford}.

Taking $\ph{45}$ to be constant, the $\ph{10}$ terms of the potential
will reduce to
\bea
V &=& \frac{\lame}{4} \left(|\ph{10}|^2 - \etae^2\right)^2
  + \frac{\lame'}{4} \left|\ph{10}\widetilde{C}\ph{10} - 2\etau\etad\right|^2 
\nonumber \\ &&
  {}+ \lamx \left(|\ph{126}|^2 - \etag^2\right)
	\left(|\ph{10}|^2 - \etae^2\right)
\eea
for the above string solution. $\widetilde{C}$ is a conjugation matrix which
maps $\Hud{\alpha}$ to $\Hdu{\alpha}$. In general the full potential
will contain various cross terms such as $|\ph{10}\cdot\ph{45}|^2$. With
a suitable choice of parameters these terms will ensure that $\ph{10}$
breaks $SU(2)_L$ rather than $SU(3)_c$~\cite{Ross}. The resulting
field equations are
\bea
\lefteqn{h''_\sbud + \frac{h'_\sbud}{r} - \frac{\left[\{\frac{n}{5}\}(1-b)
- \frac{n}{5}(1-a)\right]^2}{r^2} h_\sbud = 
\lamx \etag^2 (f^2 - 1) h_\sbud} \hspace{1.2 in} \nonumber \\ &&
{}+ \frac{\lame}{2} (\etau^2 h_\sbu^2 + \etad^2 h_\sbd^2 - \etae^2) h_\sbud
+ \lame' \eta_\sbdu^2 (h_\sbu h_\sbd - 1) h_\sbdu \ ,
\label{abEWphi} 
\eea
\be
b'' - \frac{b'}{r} = -\frac{4}{5}g^2
\left[(1-b)-\frac{n/5}{\{n/5\}}(1-a)\right]
(\etau^2 h_\sbu^2 + \etad^2 h_\sbd^2) \ .
\label{abEWgauge}
\ee

The electroweak string will also contribute terms to the GUT string
field equations. There will be an additional 
$\lamx (\etau^2 h_\sbu^2 + \etad^2 h_\sbd^2 - \etae^2)f$
on the right hand side of \bref{SO10 abGUTphi} and a 
$-g^2 \left[(1-a)/5 -(1-b) \{\frac{n}{5}\}/n \right] 
(\etau^2 h_\sbu^2 + \etad^2 h_\sbd^2)$ term on the
right hand side of \bref{SO10 abGUTgauge}. These extra terms are far smaller
than those already present in (\ref{SO10 abGUTphi},\ref{SO10 abGUTgauge}), 
so the back reaction from the electroweak string will be negligible.

\renewcommand{\baselinestretch}{1}
\begin{figure}
\centerline{\psfig{file=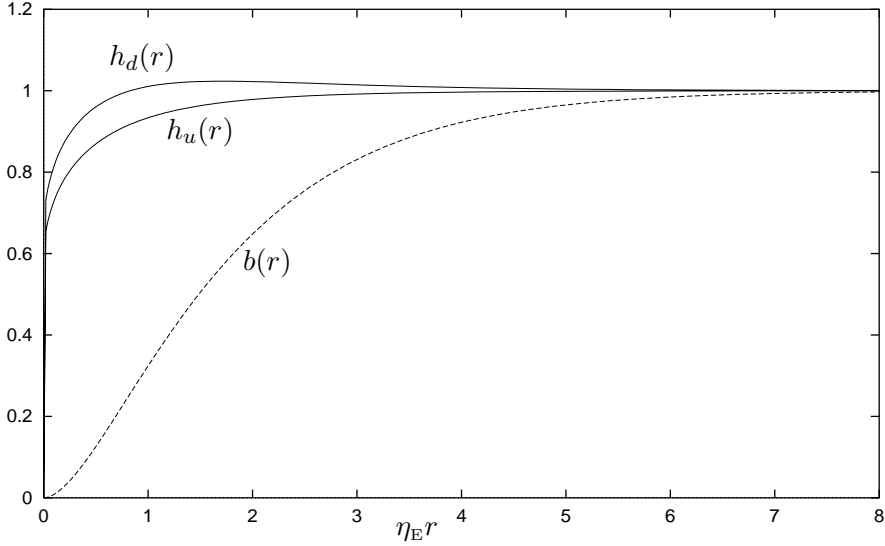,width=\myfigwidth}}
\vskip -2.6 in \hskip 1.3 in $h_\sbd(r)$ 
\vskip 0.2 in \hskip 1.6 in $h_\sbu(r)$
\vskip 0.5 in \hskip 2.0 in $b(r)$
\vskip 1.2 in \hskip 2.8 in $\etae r$
\caption{Plot of electroweak field profiles for abelian string.}
\label{ewzstr fig}
\end{figure}
\renewcommand{\baselinestretch}{\myblstr}

The field equations (\ref{abEWphi},\ref{abEWgauge}) can be solved
numerically. The electroweak fields are shown in figure~\ref{ewzstr fig}.  
The parameters used were $\etad/\etag = 10^{-14.5}$, $\etau = 3\etad$,
$g=\lame=\lame'=\lamg=1$ and $\lamx=0$. As expected, the effect of the
electroweak fields on the GUT string was negligible.

In general the size of the region of electroweak symmetry restoration is far
greater than the GUT symmetry restoration, since $\rew/\rs \sim 10^{14}$.
If $n$ is a multiple of 5 then $\tew=0$ (as would be expected). In
this case $|\ph{10}|$ will take its usual constant value for $r >\rs$,
and while the electroweak symmetry is still restored, it is over a
much smaller region. When $m=0$ the electroweak Higgs field is
non-zero at the string's centre. Its value can be evaluated
numerically, and figure~\ref{ewh0 fig} shows its
variation with respect to $\etag/\etae$. It can be seen that
$h_\sbud (0) \sim \etag/\etae$ when $\etag \gg \etae$, which is less
than the value suggested by the trial solution.

\subsection{$\Xs{}{}$ and $\Xg{}{}$ SU(2) Strings}

Although at the GUT scale all the $SU(2)$ strings have the same
properties, this is not true at low temperatures, since they affect
$\ph{10}$ differently. The generators corresponding to the \Xspm\ and
\Xgpm\ both annihilate the usual vacuum expectation value of
$\ph{10}$, and so the string gauge fields have no effect on this
symmetry breaking, and $\ph{10}$ can be constant  everywhere. There is
still the possibility of symmetry restoration from potential terms,
which is due to variation of the $\ph{126}$ and $\ph{45}$ fields (see
section~\ref{NgaugeSymRest}), although this is less significant. The
other generators do have a non-trivial effect.

\subsection{$\WR{}$ SU(2) Strings}
\label{SO10 EWWsect}

Different combinations of \WRpm\ generators give strings like
those in section~\ref{SO10 GUTstrings}, but they are all gauge equivalent
under $U(1)_Q$. Without loss of generality we can take 
$\ts = \left(\gen{\WR{+}} + \gen{\WR{-}}\right)/(2\sqrt{2})$ and
$n > 0$. For such a string,
\bea
\ts (\Hu{0} \mp \Hd{+}) = \pm \frac{1}{2}(\Hu{0} \mp \Hd{+}) \ , \nonumber \\
\ts (\Hd{0} \pm \Hu{-}) = \pm \frac{1}{2}(\Hd{0} \pm \Hu{-}) \ .
\label{SO10 tsWevec}
\eea
Thus $e^{2\pi in\ts}\phvac{10} = (-1)^n \phvac{10}$, and so 
$e^{in\ts\th}\phvac{10}$ is not single valued for odd $n$. As with the
$U(1)$ string we need a non-zero $\tew$. Again the most energetically
favourable choice will minimise ${\rm Tr}(\tew^2)$, while ensuring
that $\ph{10}$ is single valued. If we split the eigenstates of $\tew$
\bref{SO10 EWevec} up into eigenstates of $\ts$, then $\ph{10}$ will
be single valued if the sum of the $\ts$ and $\tew$ eigenvalues for
each eigenstate are integers (see section~\ref{SO10 EWSymBrk}). 

From \bref{SO10 tsWevec} we can see that the $\ts$ eigenvalues are
$\pm 1/2$. Thus we require that
\bea
\evalud{\pm} - n/2 \ \  &\mbox{and} \ \ \evalud{\pm} + n/2  
\ \ \ &(\mbox{if} \ \ w \neq 0) \ , \\
\evalud{} - n/2 \ \ &\mbox{and} \ \ \evalud{} + n/2  
\ \ \ &(\mbox{if} \ \ w = 0) \ ,
\eea
are all integers. The $\tew$ eigenvalues are $\evalud{\pm}$ or
$\evalud{}$ and are given by (\ref{SO10 EWeval},\ref{SO10 EWeval0}).
These conditions are satisfied by $\tew=0$ if $n$ is even (as
expected). If $n$ is odd then $\evalud{\pm} + 1/2$ (or just
$\evalud{} + 1/2$ if $w=0$) must be integers. The choice which
minimises ${\rm Tr}(\tew^2)$ is $z=0$, $|w|=1/2$.

Using the trial solution \bref{SO10 trial} does not tell us what phase
of $w$ will give the lowest energy. We can find this by considering the
contribution to the energy from the region $r < \rs$. If $|w|=1/2$, the
electroweak Higgs field is given by
\be
\ph{10} = \sum_{j,k=\pm1} \phi^j_k h_{jk}(r) e^{i\frac{nj+k}{2}\th} \ ,
\ee
where
\be
\phi^j_k = \frac{1}{4}(\etau - 2jk w \etad)
	\left[\Hu{0} - 2kw^\ast \Hu{-} - 2jkw^\ast\Hd{0} - j \Hd{+}\right] \ .
\ee
Matching solutions at $r=\rs$ gives 
$h_{jk}(r) \approx \sqrt{\rs/\rew}(r/\rs)^{|nj+k|/2}$. Its contribution
to the energy in the string core is then
\bea
E&\approx&2\pi\int_{0}^{\rs} r dr
  \left|\dr \ph{10}\right|^2 + \left|D_\th\ph{10}\right|^2 \nonumber \\
&\approx&\pi\sqrt{\frac{\rs}{\rew}} \left[|n+1||\phi^+_+ +\phi^-_- |^2 +
  |n-1||\phi^+_-+\phi^-_+ |^2\right] \nonumber \\
&=&\pi\sqrt{\frac{\rs}{\rew}}\left[(\etau^2 + \etad^2)n 
	- 2\etau\etad(w + w^\ast)\right] \ .
\eea
This is minimised by $w=1/2$. For a more general choice of $\ts$,
this implies the gauge fields are related by 
$\WR{+}/\WR{-} = \WL{+}/\WL{-}$. Combining the above results, and
noting that $[\ts,\tew]=0$, leads to the ansatz
\bea
\ph{10} &=& \frac{1}{\sqrt{2}} e^{i(n\ts + \tew)\th} \left[ 
	 (\Hu{0} - \Hd{0}) \eta_+ h_+(r) 
	+ (\Hu{0} + \Hd{0}) \eta_- h_-(r)\right] \nonumber \\
	&=& \left(e^{im'\th}\phi^+_+ +
	          e^{-im'\th}\phi^-_-\right) h_+(r) 
	+ \left(e^{im\th}\phi^+_- 
	+ 	e^{-im\th}\phi^-_+\right) h_-(r) 	
\ , \nonumber \\
A_\th &=& \frac{2na(r)}{gr}\ts + \frac{2b(r)}{gr} \tew
\hspace{.6in} A_\mu = 0 \mbox{ \ otherwise} \ ,
\label{EWWansatz}
\eea
where $\eta_\pm = (\etau \mp \etad)/\sqrt{2}$, $m = (n-1)/2$ and
$m' = (n+1)/2$. For the solution to have the correct asymptotic form
and to be regular at $r=0$ the required boundary conditions are
$h_\pm(\infty) = b(\infty) = 1$, $h_+(0)=b(0)=0$, and also $h_-(0)=0$
if $n \neq 1$. Ignoring any potential terms which couple $\ph{10}$ to
any of the other Higgs fields, the field equations for odd $n$ are
\bea
\lefteqn{h_\pm''+ \frac{h_\pm'}{r} -
 \frac{\left[(1-b) \pm n(1-a)\right]^2}{4r^2}h_\pm =} \hspace{1in}
\nonumber \\ && \hspace{-.1in} 
\left[ \frac{\lame}{2} (\eta_+^2 h_+^2 + \eta_-^2 h_-^2 - \etae^2)
\pm \frac{\lame'}{2}(\eta_+^2 h_+^2 - \eta_-^2 h_-^2 + 2\etau\etad) 
\right] h_\pm \ ,
\label{EWWphi}
\eea
\be
b'' - \frac{b'}{r} = -\frac{1}{2} g^2 \left\{ 
\left[(1-b)+n(1-a)\right] \eta_+^2 h_+^2
 + \left[(1-b)-n(1-a)\right] \eta_-^2 h_-^2 \right\} \ .
\label{EWWgauge}
\ee

\renewcommand{\baselinestretch}{1}
\begin{figure}
\centerline{\psfig{file=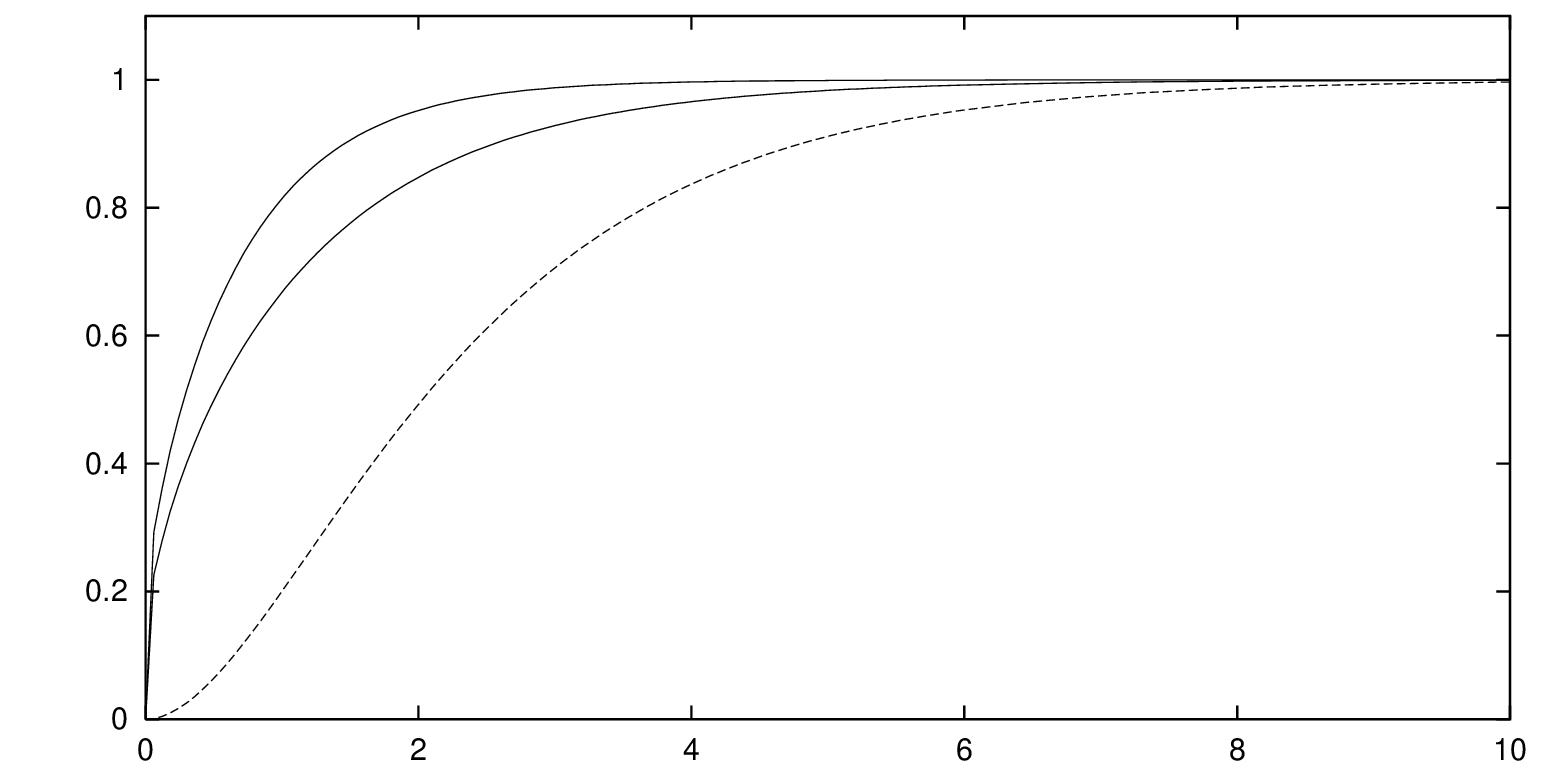,width=\myfigwidth}}
\vskip -2.2 in \hskip 1.1 in $h_-(r)$ 
\vskip 0.35 in \hskip 1.5 in $h_+(r)$
\vskip 0.2 in \hskip 1.9 in $b(r)$
\vskip 0.9 in \hskip 3.1 in $\etae r$
\caption{Plot of electroweak field profiles for $\WR{}{}$--string.}
\label{ewwstr fig}
\end{figure}
\renewcommand{\baselinestretch}{\myblstr}

The field equations (\ref{EWWphi},\ref{EWWgauge}) can be solved
numerically. The electroweak fields are shown in figure~\ref{ewwstr fig}.  
The parameters used were the same as those for figure~\ref{ewzstr fig}, 
thus $\eta_+ = \sqrt{2}\times 10^{-14.5}$, $\eta_- = 2\eta_+$,
and all coupling constants were set to 1.

\begin{figure}
\centerline{\psfig{file=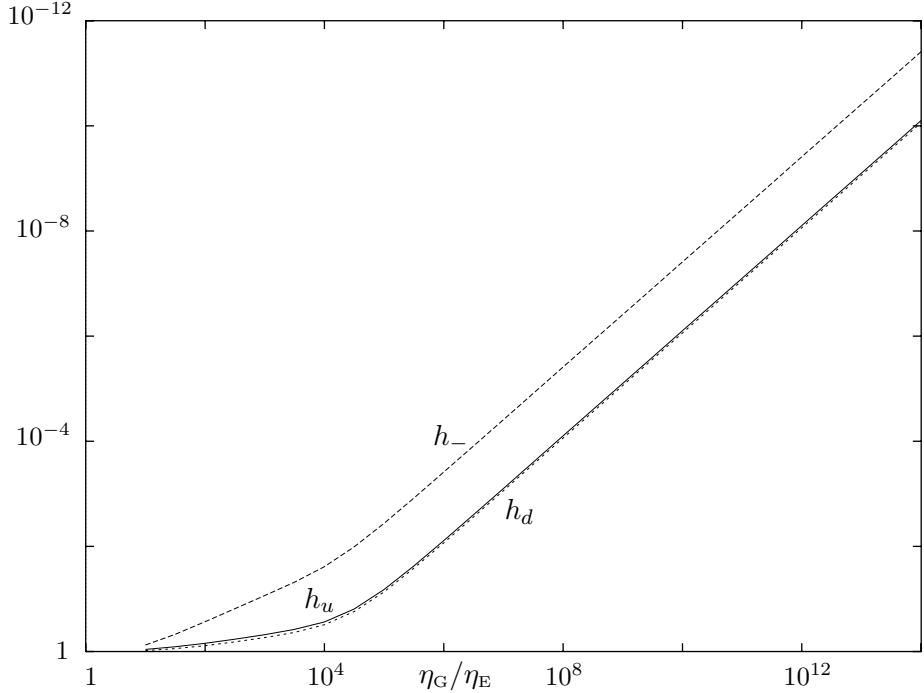 ,width=\myfigwidth}}
{\small
\vskip -3.6 in \hskip 0.5 in \ $10^{-12}$ 
\vskip 0.95 in \hskip 0.5 in \ \ $10^{-8}$ 
\vskip 0.9 in \hskip 0.5 in \ \ $10^{-4}$ } \hskip 1.8 in $h_-$ 
\vskip 0.2 in \hskip 3.15 in $h_\sbd$
\vskip 0.3 in \hskip 2.1 in $h_\sbu$
{\small
\vskip 0.1 in \hskip 0.8 in  $1$ 
\vskip -0.05 in \hskip 0.95 in $1$ \hskip 1.05 in $10^4$ 
	\hskip 1.0 in $10^8$ \hskip 1.0 in $10^{12}$ }
\vskip -0.2 in \hskip 2.7 in $\etag/\etae$
\caption{Variation of electroweak Higgs field at the centres of the
abelian and $\WR{}{}$ strings.}
\label{ewh0 fig}
\end{figure}

If $n=1$, $h_-(0)$ need not be zero. As with the $U(1)$ string we can
estimate its value by extending the trial solution \bref{SO10 trial}
to $r < \rs$. This gives $h_-(0) \approx \sqrt{\rs/\rew} \sim
10^{-7}$, so $\ph{10} \approx 0$ inside the string when $n=1$. Thus
electroweak symmetry is fully or almost fully restored there for all $n$. In
contrast, the symmetry breaking caused by the GUT Higgs fields is only
partially restored at the centre of the string, because $|\ph{126}|$
is about $\etag/\sqrt{2}$ there. Figure~\ref{ewh0 fig} shows the
numerically determined variation of $h_-(0)$ with respect to
$\etag/\etae$. As with the abelian string,
$h_- (0) \sim \etag/\etae$ when $\etag \gg \etae$, which is less
than the value suggested by the trial solution.

\renewcommand{\baselinestretch}{1}
\renewcommand{\baselinestretch}{\myblstr}

If $\etau=\etad$, \bref{EWWansatz} simplifies, and
\be 
\ph{10} = e^{i(n\ts + \tew)\th} \phvac{10} h_- (r)\ ,
\ee
so $\ph{10}$ has just one winding number, instead of the usual two. In
this respect it resembles the equivalent abelian case. The field
equations for $h_-$ and $b$ are then the same as
(\ref{EWWphi},\ref{EWWgauge}), but with $\eta_+ = 0$.

If $n$ is even an electroweak gauge field is not required. Putting
$\tew=0$ into \bref{EWWansatz} gives a suitable solution. In this case
$m=m'=n/2$. The resulting field equations are then \bref{EWWphi}
with $b=1$. In this case the region of electroweak symmetry
restoration is limited to the string core.

\subsection{$\Yg{}{}$ SU(2) Strings}
\label{SO10 EWYsect}

The fourth type of nonabelian string has $\ts$ proportional to an
appropriate linear combination of the generators of
the \Ygpm\ gauge fields. The different combinations are all equivalent
under $SU(3)_c \times U(1)_Q$, and so without loss of generality we
can take 
$\ts = \left(\gen{\Yg{1}{+}} + \gen{\Yg{1}{-}}\right)/(2\sqrt{2})$, 
and $n > 0$. As with the $\WR{}$--string 
$e^{2\pi in\ts}\phvac{10} = (-1)^n \phvac{10}$, so we will need a
non-zero $\tew$. Using
\bea
\ts (\Hu{0} \mp \Hd{1}) &=& \pm \frac{1}{2}(\Hu{0} \mp \Hd{1}) \ , \nonumber \\
\ts (\Hd{0} \pm \Hu{1}) &=& \pm \frac{1}{2}(\Hd{0} \pm \Hu{1}) \ , \nonumber \\
\ts \Hu{-} = \ts \Hd{+} &=& 0 \ ,
\label{SO10 tsYevec}
\eea
we can split $\evecud{\pm}$ into eigenstates of $\ts$. This can then
be used to determine the winding number of each component, and hence
conditions for $\ph{10}$ to be single valued (as we did with the other
strings). For a general $\tew$ with $w \neq 0$ the required conditions
are that
\be
\evalud{\pm} + \frac{n}{2} \ , \ \ \evalud{\pm}  \ , \ \  
\evalud{\pm} - \frac{n}{2} \ ,
\ee
are all integers. If $n$ is odd, no choice of $\tew$ satisfies
them. If however we take $\tew$ to be proportional to the generator
of the $Z$ boson, then $\evecud{} = \Hud{0}$ and there are no $\Hu{-}$
or $\Hd{+}$ components to worry about. This reduces the number of
constraints. Now only $\evalud{} + \frac{n}{2}$ and 
$\evalud{} - \frac{n}{2}$ need to be integers. These conditions are
satisfied by $|z| = \sqrt{10}/8$.

Unlike all the other cosmic strings we have considered, $[\ts,\tew] \neq 0$
for this string generator. This introduces several
complications. As with the Higgs fields we have to split $\tew$ up into
eigenvalues of $\ts^2$. If we define 
$T_1 = [\ts,[\ts,\tew]]$ and $T_0 = \tew - T_1$, then 
$[\ts,[\ts,T_j]] = T_j$, as required. The angular dependence of
$A_\th$ leads to a non-trivial $r$ component of the gauge field
equations \bref{SO10 gaugeequ}. In order for them to be satisfied a
non-zero $A_r$ field proportional to $[\ts,\tew]$ is needed. 

Additionally, we need to alter the angular dependence of
$\ph{10}$. Although introducing $\tew$ terms gives a single valued
$\ph{10}$ away from the string, it does not ensure that the
eigenstates of $\ts^2$ which make it up are single valued. If
$\tew = \sigma \sqrt{10} \gen{Z}/8$, with $\sigma = \pm 1$, a suitable
ansatz is
\bea
\ph{10} &=&  e^{in\ts\th} e^{i\tew\th} e^{i\tew'\th} 
		\sum_\pm \left( \psi_\sbu^\pm h_{\sbu\pm}(r) 
	+ \psi_\sbd^\pm h_{\sbd\pm}(r) \right) \nonumber \\
	&=& \psi_\sbu^+ h_{\sbu+}(r) e^{i\sigma m'\th} 
	+ \psi_\sbu^- h_{\sbu-}(r) e^{-i\sigma m\th}  
+ \psi_\sbd^+ h_{\sbd+}(r) e^{i\sigma m\th} 
	+ \psi_\sbd^- h_{\sbd-}(r) e^{-i\sigma m'\th}  \ , \nonumber \\
A_\th &=& \frac{2na(r)}{gr}\ts + \frac{2b_0(r)}{gr} T_0 
	+ \frac{2b_1(r)}{gr} e^{in\ts\th} T_1 e^{-in\ts\th}  \ , \nonumber \\
A_r &=& -i\frac{2k(r)}{g} \frac{8}{3} e^{in\ts\th} [\ts,\tew] e^{-in\ts\th}
\ , \ \ A_\mu = 0 \mbox{ \ otherwise} \ ,
\label{EWYansatz}
\eea
where $m=(n-1)/2$, $m'=(n+1)/2$, and
\be
\psi_\sbu^\pm = \frac{\etau}{2} \left( \Hu{0} \mp \sigma \Hd{1} \right)
\ , \ \ 
\psi_\sbd^\pm = \frac{\etad}{2} \left( \Hd{0} \pm \sigma \Hu{1} \right) \ .
\ee
$\tew'$ is chosen to give a single valued $\ph{10}$ inside the
string, without altering the solution \bref{SO10 multiAnsatz}
outside the string. Thus it must be a combination of gluon and photon
generators. 
$\tew' = \left(\gen{G_8}/\sqrt{3} + \gen{G_3} - \gen{A}/\sqrt{6}\right)/4$
is a suitable choice.

As with the other strings we expect the back-reaction of the electroweak
string on the GUT string to be small. For simplicity we will neglect
it. We will also ignore all the cross terms in the potential, and the
effects of $\ph{45}$ (which couples to $A_r$, but not the other
electroweak fields). The field equations for the electroweak Higgs
field and the additional gauge fields are
\bea
\lefteqn{h''_{\sbu\pm} + \frac{h'_{\sbu\pm}}{r} \mp \frac{1}{2}\left[
2k h'_{\sbu\mp} + \left(k'+\frac{k}{r}\right)h_{\sbu\mp} \right]}
\nonumber \\ && {}
- \frac{1}{4}\left(
k^2 + \frac{[n(1-a) \pm (1-\frac{5}{8}b_0)]^2 + (\frac{3}{8}b_1)^2}{r^2}
\right) h_{\sbu\pm} + \frac{3(1-\frac{5}{8}b_0)b_1}{16r^2} h_{\sbu\mp}
\nonumber \\ && {} =
\frac{\lame}{4}\left[ \etau^2(h_{\sbu+}^2 + h_{\sbu-}^2)
+ \etad^2(h_{\sbd+}^2 + h_{\sbd-}^2) - 2\etae^2\right] h_{\sbu\pm}
\nonumber \\ && \ \ \ {}
+\frac{\lame'}{2}\left[h_{\sbu+}h_{\sbd-} + h_{\sbd+}h_{\sbu-} - 2
\right] \etad^2 h_{\sbd\mp} \ ,
\eea
\bea
\lefteqn{h''_{\sbd\pm} + \frac{h'_{\sbd\pm}}{r} \pm \frac{1}{2}\left[
2k h'_{\sbd\mp} + \left(k'+\frac{k}{r}\right)h_{\sbd\mp} \right]}
\nonumber \\ && {}
- \frac{1}{4}\left(
k^2 + \frac{[n(1-a) \mp (1-\frac{5}{8}b_0)]^2 + (\frac{3}{8}b_1)^2}{r^2}
\right) h_{\sbd\pm} + \frac{3(1-\frac{5}{8}b_0)b_1}{16r^2} h_{\sbd\mp}
\nonumber \\ && {} =
\frac{\lame}{4}\left[ \etau^2(h_{\sbu+}^2 + h_{\sbu-}^2)
+ \etad^2(h_{\sbd+}^2 + h_{\sbd-}^2) - 2\etae^2\right] h_{\sbd\pm}
\nonumber \\ && \ \ \ {}
+\frac{\lame'}{2}\left[h_{\sbu+}h_{\sbd-} + h_{\sbd+}h_{\sbu-} - 2
\right] \etau^2 h_{\sbu\mp} \ ,
\eea
\bea
\lefteqn{b_1'' - \frac{b_1'}{r}  -  k^2 b_1 + 
\frac{8n}{3}\left[ 2a'k - (1-a)\left(k'-\frac{k}{r}\right) \right] =} 
\ \ \ \nonumber \\ & -g^2&\!\!\!\! 
\left\{ \etag^2 \left[ 2 f_0 f_1 b_0 - (f_0^2 + f_1^2)b_1 \right] 
+\left[\etau^2 h_{\sbu+} h_{\sbu-} 
+ \etad^2 h_{\sbd+} h_{\sbd-}\right]\frac{8-5b_0}{6}
\right. \nonumber \\ && \left.
{}- \frac{1}{4}\left[ \etau^2(h_{\sbu+}^2 + h_{\sbu-}^2)
+ \etad^2(h_{\sbd+}^2 + h_{\sbd-}^2)\right]b_1 
 \right\} \ , 
\eea
\setstrut{14.5 pt}
\bea
\lefteqn{b_0'' - \frac{b_0'}{r} = -\frac{g^2}{31} \left\{\mystrut 9\etag^2 
\left[ 2 f_0 f_1 b_1 - (f_1^2 + f_0^2)b_0 \right] 
\right.} \nonumber \\ &&\hspace{-.2in} \left. {}
+ 10 \left[ \etau^2(h_{\sbu+}^2 + h_{\sbu-}^2)
+ \etad^2(h_{\sbd+}^2 + h_{\sbd-}^2)\right]\left(1-\frac{5}{8}b_0\right)
\right. \nonumber \\ &&\hspace{-.2in} \left.
{}-\frac{15}{2}\left[\etau^2 h_{\sbu+} h_{\sbu-} 
+ \etad^2 h_{\sbd+} h_{\sbd-}\right] b_1
\right. \nonumber \\ &&\hspace{-.2in} \left.
+ 10 n\left[ \etau^2(h_{\sbu+}^2 - h_{\sbu-}^2)
- \etad^2(h_{\sbd+}^2 - h_{\sbd-}^2)\right](1-a) \mystrut \right\} \ , 
\eea
\setstrut{17 pt}
\bea
\lefteqn{\frac{1}{r^2}\left\{ 
\left[ n^2(1-a)^2 + \left(\frac{3}{8}b_1\right)^2  \right] k
-\frac{3n}{8}(b_1 a' + b'_1 (1-a))\right\} =}
\nonumber \\  & -g^2&\!\!\!\! \left\{ \mystrut \left(
\etag^2f_0^2 + \frac{1}{4}\left[ \etau^2(h_{\sbu+}^2 + h_{\sbu-}^2)
+ \etad^2(h_{\sbd+}^2 + h_{\sbd-}^2)\right] \right)k
\right. \nonumber \\ &&\left.{}
+ \frac{\etau^2}{2}(h_{\sbu+}h'_{\sbu-} - h_{\sbu-}h'_{\sbu+}) 
- \frac{\etad^2}{2}(h_{\sbd+}h'_{\sbd-} - h_{\sbd-}h'_{\sbd+}) \right\} \ .
\eea
The boundary conditions of the above functions are 
$b_i(0)=h_{\sbu+}(0)=h_{\sbd-}(0)=k(0)=k(\infty)=0$,
 $b_i(\infty)=h_{\sbud\pm} (\infty)=1$, and
if $n\neq 1$ then $h_{\sbu-}(0)=h_{\sbd+}(0)=0$ too. The potential
terms considered in deriving the above equations are not enough to give the
required boundary conditions at $r=\infty$. The effect of $\ph{45}$'s
coupling to $\ph{10}$ also needs to be considered, although for
simplicity we have neglected it in the derivation of the field
equations. 

We find that the above equations are independent of
$\sigma$, the sign of $\tew$. The energy of the string is also
independent, so the most favourable choice of electroweak gauge field
is degenerate. As with the $\WR{}$--string, if $n$ is even no extra
gauge fields are needed. The field equations are then the same as
above, but with $k = 0$, $b_i = 1$, $h_{\sbu+}=h_{\sbu-}$ and
$h_{\sbd+}=h_{\sbd-}$ everywhere. The resulting electroweak symmetry
restoration will be restricted to the string core.

\subsection{Summary}

Electroweak symmetry is restored and electroweak string gauge fields
are present around the abelian and the $\WR{}$ and $\Yg{}{}$
nonabelian GUT strings. This generally occurs in a region around the
string whose size is inversely proportional to the electroweak Higgs
VEV, and is much bigger than the string core. If $n$ is a multiple
of 5 for the abelian, or 2 for the nonabelian strings, the region is
approximately the same as the string core, and there are no extra
string gauge fields.

It is also possible that $\ph{10}$ will wind. For the abelian string its
winding number is the closest integer to $n/5$, and hence
zero for $n=1$. For the $\WR{}$ and $\Yg{}{}$ nonabelian strings it is
$n/2$ for even $n$. For odd $n$, different parts of $\ph{10}$ have different
winding numbers, a bit like the corresponding GUT string. They are
$(n-1)/2$ and $(n+1)/2$. The remaining two nonabelian strings
($\Xs{}{}$ and $\Xg{}{}$) have no effect on $\ph{10}$ at all.

With the $\Yg{}{}$--string the electroweak gauge field has angular
dependence, which leads to a non-zero radial gauge field
component. Also, the choice of electroweak gauge field varies with
distance from the string's centre. Both these additional effects are
restricted to the string core.

\section{Other Symmetry Restorations}
\label{SO10 OtherSymmRest}

\subsection{The Intermediate Symmetry Restoration}
\label{IntermedSymRest}

So far the effect of the string on the second Higgs field $\ph{45}$ has
been neglected, because it is far less significant. $\ph{45}$ is
in the adjoint representation, and so its covariant derivative takes
the form
$D_\mu\ph{45} = \partial_\mu\ph{45} 
	- \frac{1}{2}ig\left[A_\mu,\ph{45}\right]$.
The generator corresponding to the $Z'$ particle ($P$) commutes with
$\ph{45}$, so the gauge fields
of the abelian string will not stop $\ph{45}$ from taking its usual
vacuum expectation value everywhere. Thus it has no effect on the electroweak 
symmetry breaking, and gives no additional contribution to the energy.
The other strings will give non-vanishing covariant derivatives at
infinity. This is avoided by allowing $\ph{45}$ to wind like a string
\be 
\ph{45} = e^{in\th \ts} \pzero{45}(r)e^{-in\theta \ts} \ ,
\label{medansatz}
\ee
where $\pzero{45}(\infty)$ is equal to the usual vacuum expectation value of
$\ph{45}$. Conveniently $e^{2\pi n i\ts}$ (for all choices of $\ts$) and 
$\pzero{45}$ commute, so $\ph{45}$ will be single valued for
all $n$, and no extra gauge terms are needed. As with the $\ph{126}$ and
$\ph{10}$ fields it is necessary to split $\ph{45}$ up into eigenstates of
$\ts^2$. Thus, using the fact that $\phvac{45}$ and
$\ts^2$ commute, and that $\ts^3 = \frac{1}{4}\ts$
\be
\ph{45} = e^{in\th \ts} \psi_1 s_1(r) e^{-in\th \ts} + \psi_0 s_0(r)
\ee
with
\bea
\psi_1 &=& 2\ts^{2} \phvac{45} - 2\ts \phvac{45} \ts \ , \\
\psi_0 &=& \phvac{45}  - \psi_1 \ .
\eea
The field equations for $s_{0,1}(r)$ will be similar to \bref{nabequa}
and \bref{nabequb}, with the similar boundary conditions
$s_{0,1}(\infty)=1$ and $s_1(0)=0$. Since the gauge contribution
($1-a(r)$) vanishes for $r>\rs$, $\pzero{45}(r) = \phvac{45}$
is a solution outside the string. Thus the region of symmetry
restoration will be of radius $\rs$ (order $|\phvac{126}|^{-1}$).
This is in contrast to the other Higgs fields, which restore symmetry
in regions of order the reciprocal of their own values at $r=\infty$.

After the electroweak symmetry breaking the form of the solution for
the $\Yg{}{}$--string will be altered slightly due to the effect of the
non-zero $A_r$ field. As with $\ph{126}$, we expect the effect to be
tiny.

\subsection{The Minimal Energy Choice of $\ts$}

After the first symmetry breaking, there were only two gauge inequivalent
possible strings, and the nonabelian strings had the lowest
energy. After the second symmetry breaking there are 3 inequivalent
types of nonabelian string. The $\WR{}$--string has the highest energy,
the $\Xs{}{}$--string has less, while the $\Xg{}{}$ and $\Yg{}{}$
strings, which are gauge equivalent under $SU(2)_L$, have the lowest
energy. The final symmetry breaking gives an additional contribution to the
$\Yg{}{}$--string, so the most energetically favourable choice of string
generator is made up of \gen{\Xg{i}{\pm}}, and does not have any
effect on electroweak symmetry. Just because the other strings are not
energetically favourable does not mean that they will not form,
just that they are less likely to form (but see section \ref{SO10 OtherGUT}). 

\subsection{Non-gauge field symmetry restoration}
\label{NgaugeSymRest}

Even when a Higgs field is unaffected by a string's gauge fields, it
is still possible for symmetry restoration to occur via the potential
terms. This has previously been discussed for an abelian string in
\cite{Goodband}. For example, before electroweak symmetry restoration
occurs and in the absence of strings (so $\ph{10} = 0$), the
potential takes the form
\bea
V(\ph{126},\ph{45}) &\!\!=&\!\!
 \frac{\lamg}{4} \left(|\ph{126}|^2 - \etag^2\right)^2 
+ \frac{\lammed}{4} \left(|\ph{45}|^2 - \etamed^2\right)^2
\nonumber \\ &&
{}+ \lamx' \left(|\ph{126}|^2 - \etag^2\right)
	\left(|\ph{45}|^2 - \etamed^2\right) \ ,
\eea
where $\etamed$ is the usual VEV of $\ph{45}$, and
$|\ph{126}\cdot\ph{45}|^2$-like cross terms have been ignored. This is
minimised by setting $\ph{126}$ and $\ph{45}$ to their usual
VEVs. However, in the presence of an abelian string $|\ph{126}|$ is
proportional to $f(r)$, so writing $|\ph{45}| = s(r)\etamed$, the
potential becomes
\be
\frac{\lammed \etamed^4}{4} 
\left(s^2- 1 - \frac{2\lamx'\etag^2(1-f^2)}{\lammed\etamed^2}\right)^2 
+ \frac{\lamg\lammed - 4(\lamx')^2}{4\lammed} \etag^4 (f^2-1)^2 \ ,
\ee
which for small $r$ is no longer minimised by $s = 1$, if 
$\lamx' \neq 0$. Thus if the theory's parameters take appropriate values,
$|\ph{45}|$ will be lower than usual, or even zero at the string's
centre. Alternatively $|\ph{45}|$ could be higher than usual there. 

If symmetry restoration by this mechanism occurs at all, it will only
be in the region $r < \rs$, since $|\ph{126}|$ and hence
$V(\ph{126},\ph{45})$ take their usual values at larger $r$. Unlike
the corresponding symmetry restoration by gauge fields, $\ph{45}$ will
not wind. If there is a very strong $|\ph{126}\cdot\ph{45}|^2$ term
present, $\ph{45}$ may be forced to wind. Such terms cause $\ph{45}$
to be orthogonal to $\ph{126}$, so it may be more favourable for
$\ph{45}\cdot\ph{126}$ to remain zero than for $\ph{45}$ to have no
angular dependence. When $\ph{10}$ is not equal to its usual VEV, it could also
cause $|\ph{45}|$ to vary, although this effect will be very small for
most parameter ranges. In this case the symmetry restoration could
take place in the larger $r<\rew$ region.  

A similar situation can occur with $\ph{10}$ in the presence of an
$\Xg{}{}$ or $\Xs{}{}$ string. In this case both $|\ph{45}|$ and
$|\ph{126}|$ are lower than usual for $r < \rs$. For nonabelian
strings the potential is more complicated since it involves $f_1$
and $f_0$ terms, as well as the corresponding $\ph{45}$ terms. Even
inside the string, $f_{0}$ and $s_{0}$ are non-zero, so the variation
of the potential is likely to be less substantial than the abelian
case, and hence extra symmetry restoration is less likely to
occur. Even if it does, $\ph{10}$ will take its usual VEV outside
the string. Again we do not expect $\ph{10}$ to wind, unless there is
a very large $|\ph{10}\cdot\ph{45}|^2$ term, or something similar.

\section{Other Related Grand Unified Theories}
\label{SO10 OtherGUT}

Although only one particular $SO(10)$ GUT has been discussed, many of
the results apply to different symmetry breakings. Any theory of the form 
\be
SO(10) \cdots \pharrow{126} \cdots 
 SU(3)_c \times SU(2)_L \times U(1)_Y \times Z_2
 \pharrow{10} SU(3)_c \times U(1)_Q \times Z_2 
\label{gensymbreak}
\ee
could have string solutions of the form \bref{SO10 abansatz} or
\bref{SO10 nabansatz}, which would cause electroweak symmetry restoration
at low temperatures in the same way as \bref{symbreak}. The form of the 
other Higgs fields will not make much difference, as long as they are single
valued in the presence of a string (like $\ph{45}$). If they are, it
will not be necessary to add extra gauge fields, and so $\ph{10}$ will
have the same behaviour as in \bref{symbreak}.

The Higgs fields which gain their VEVs after $\ph{126}$ will
determine the most energetically favourable choice of string
generator, as $\ph{45}$ did in \bref{symbreak}. 
If a GUT of the form \bref{gensymbreak} has Higgs fields which take
non-zero VEVs before $\ph{126}$, the choice of $\ts$ will be more
restricted. If a generator has already been broken, the formation
of the corresponding string will not occur.

One theory of the form \bref{gensymbreak} is
\bea
SO(10) & \pharrow{A} & SU(5) \times U(1)_P 
\pharrow{45} SU(3)_c \times SU(2)_L \times U(1)_Y \times U(1)_P
\nonumber \\
& \pharrow{126} & SU(3)_c \times SU(2)_L \times U(1)_Y \times Z_2
\nonumber \\
& \pharrow{10} & SU(3)_c \times U(1)_Q \times Z_2 \ .
\label{SUxUpsymbreak}
\eea
The $\ph{A}$ Higgs field transforms under either the {\bf 45} or {\bf 210}
representation of $SO(10)$, and is an $SU(5)$ singlet. Unlike
\bref{symbreak}, only abelian strings can form in this theory, since
the only generator that $\ph{126}$ breaks is $P$. This means that
electroweak symmetry restoration will always occur in the presence of a string.
$\ph{A}$ and $\ph{45}$ will both take their usual VEVs, so the only
symmetries restored in
the string core will be $U(1)_P$, and the electroweak symmetry.
Another interesting feature of this theory is that strings can
form at energies close to the electroweak scale, and so
$\rs$ could be of similar size to $\rew$, although still smaller. This
string is a candidate for defect mediated electroweak 
baryogenesis~\cite{defectbaryogenesis}. Switching the second and third
symmetry breakings also gives a theory with similar solutions. 

A different unifying gauge group (instead of $SO(10)$), with similar
properties to \bref{SUxUpsymbreak} is $SU(5) \times U(1)_P$. It was
suggested in ref.~\cite{Witten}, and has two independent gauge coupling
constants. Unlike \bref{SUxUpsymbreak}, strings of all winding
numbers will be topologically stable, since $U(1)_P$ is broken to $Z$
instead of $Z_2$. They could still decay by splitting into several
strings with lower winding numbers. 
The field equations for the electroweak fields will
be the same as \bref{abEWphi} and \bref{abEWgauge}, but with
$1/5$ the ratio of the two couplings instead of just
$1/5$. If the ratio is $\alpha$, then $\ph{10}$'s winding
number will be the nearest integer to $\alpha n/5$ (with half
integers rounded towards zero), so if $|\alpha| > \frac{5}{2}$, $\ph{10}$
will always wind in the presence of a string.

A theory which is substantially different from \bref{symbreak} starts
with the symmetry breaking 
$SO(10) \pharrow{54} SU(4) \times SU(2)^2 \times Z^C_2$. The
$Z^C_2$ symmetry is not the $Z_2$ symmetry in
\bref{gensymbreak}. $\ph{10}$ is not invariant under it, so it must be
broken during or before the electroweak symmetry breaking. This will
lead to formation of domain walls, and so such a theory will have
substantially different properties to \bref{symbreak}, and is ruled
out cosmologically~\cite{domwall}.

Another type of theory closely related to \bref{gensymbreak} occurs
when $\ph{126}$ is replaced by $\ph{16}$, where the usual VEV of $\ph{16}$ is
proportional to $\Nr$. The gauge fields all gain masses in the
same way as the equivalent theory involving $\ph{126}$, but there
will be no discrete $Z_2$ symmetry, so there will be no topological
strings. However, solutions of the form
(\ref{SO10 abansatz},\ref{SO10 nabansatz}) can still form, although since 
$e^{2\pi i n \ts}$ will need to map $\Nr$ to $\Nr$ to give a
single valued $\ph{16}$, only solutions with even $n$ will occur. Of
course, if such strings are to be observed, they will need to be
stable, which will only happen for certain values of the theory's
parameters. Embedded defects similar to these have been discussed
previously~\cite{Vachaspati,Nathan}.
 
Yet another set of related theories can be obtained from
\bref{symbreak} by choosing a different VEV of $\ph{45}$. Adding a
multiple of $P$ to it will not affect any of the $SU(5)$ symmetry
breaking since all the $SU(5)$ fields commute with it, thus it will
not alter which gauge bosons become superheavy. It will alter the
sizes of the masses of the $SO(10)$ fields. The most energetically
favoured choice of nonabelian string will be the one with the lowest
energy contribution at the $\ph{45}$ symmetry breaking. This will be the
one whose string generator corresponds to gauge fields with the
lowest mass. So by choosing $\ph{45}$ appropriately, a different
nonabelian string could become most favourable. The $\Xg{}{}$ and $\Yg{}{}$
strings are gauge equivalent at this stage, but since the $\Xg{}{}$
string contributes nothing at the electroweak symmetry breaking, it
will always be more favourable than the $\Yg{}{}$ string. Thus any of the
$\Xg{}{}$, $\Xs{}{}$ or $\WR{}$--strings could be energetically
favourable. If it is the $\WR{}$--string, then it is most probable that
electroweak symmetry restoration will occur. The same sort of freedom
does not exist with $\ph{126}$ and $\ph{10}$, since any such change will
give different fermion mass terms, and radically alter the theory.

\section{Conclusions}
\label{SO10 Conc}

In this chapter we have uncovered a very rich microstructure for
$SO(10)$ cosmic strings. In particular, we have found four nonabelian
strings as well as one abelian string. We have examined the effect of
the strings on the subsequent symmetry breakings. Our results are
summarised in the table

\begin{center}
\begin{tabular}{@{}cc|ccc|c|c@{}}
Gauge & Type & \multicolumn{3}{|c|}{Symmetry restoration}
& EW & EW \\
field &  & $SO(10)$ & $SU(5)$ & EW & fields & windings\\
\hline
$Z'$ & $U(1)$ & yes & no & yes & $Z$ & $\pm(n/5-\{n/5\})$ \\
\Xspm & $SU(2)$ & partial & partial & no & ---  & 0 \\
\Xgpm & $SU(2)$ & partial & partial & no & ---  & 0 \\
\WRpm & $SU(2)$ & partial & partial & yes & \WLpm & $\pm(n/2\pm\{n/2\})$ \\
\Ygpm & $SU(2)$ & partial & partial & yes & 
		$Z$ (+ others) & $\pm(n/2\pm\{n/2\})$
\end{tabular} \vspace{.1in}
\end{center}

We have defined $\{x\}$ to be the fractional part of $x$, with
$|\{x\}| \le 1/2$. It seems that electroweak symmetry
restoration by GUT strings is quite likely. The exact results are
dependent on the details of the theory and the choice of string
generator. For the $SO(10)$ theory considered, electroweak symmetry is
restored for the abelian string, and half the possible nonabelian
strings, although the most energetically favourable of these
does not restore electroweak symmetry. However, other closely related
$SO(10)$ GUTs have different minimal energy string solutions,
which will restore the symmetry, such as \bref{SUxUpsymbreak}, or
\bref{symbreak} with a different choice of $\ph{45}$. Thus our
results generalise to a range of theories.

The size of the region of electroweak symmetry restoration for the
topologically stable ($n=1$) strings is determined by the electroweak
scale, and is much larger than the string core. For nonabelian strings, with
higher winding number, the region will be the same if they are
topologically equivalent to the $n=1$ string (i.e.\ odd $n$), and
restricted to the string core if they are topologically equivalent to
the vacuum (i.e.\ even $n$). There is no such distinction between
topological and non-topological abelian strings, which restore
symmetry in the larger region if the winding number is not a multiple of 5. 
Some of the $SU(5)$ symmetry is also restored by all of the nonabelian
strings, but not the abelian string. This is only within the string
core, irrespective of the string winding number, and since $\ph{45}$ is
not zero there, the restoration is only partial.   

For any choice of $\ts$, ignoring possible potential driven
symmetry restoration, the GUT will not be fully restored at the
string's centre. All the nonabelian strings have non-zero (although
smaller than usual) $\ph{126}$ and $\ph{45}$ fields at their
centre, so the $SO(10)$ (apart from the electroweak fields) symmetry
is only partially restored inside the string. For the abelian string
$\ph{126}$ is zero in the string core, but $\ph{45}$ takes its usual
value, so with the exception of $U(1)_P$, most of the $SO(10)$
symmetry is broken. The resulting gauge boson masses are smaller, but
still superheavy. However, for the abelian and the $\WR{}$ and $\Yg{}{}$
nonabelian strings, there is almost full restoration of electroweak symmetry 
in a larger region than the string core.

It is also possible for additional symmetry to be restored by the
potential terms of the theory. However this is far less significant
than that arising from the string gauge field, and will require some
tuning of parameters.

Although the profile of the electroweak Higgs field obeys the same
boundary conditions as a string, its exact form has a closer
resemblance to a string with non-integer winding number. For the abelian
string the actual winding number of $\ph{10}$ is less than that of the
GUT string (about $1/5$). The same is true for the nonabelian
string, which has the winding number (or numbers if $n$ is odd) of
$\ph{10}$ about $1/2$ that of the string itself. The winding of the
electroweak Higgs field has implications for the existence of massless
fermion currents, as we will discuss in chapter~\ref{Ch:Index}.

In our analysis we have only considered terms occuring in the tree
level Lagrangian. One-loop corrections are likely to induce couplings
between the nonabelian string field and the electroweak Higgs. This
may result in electroweak symmetry restoration around the $\Xg{}{}$
and $\Xs{}{}$ strings. However, the electroweak Higgs field is
unlikely to wind in this region.

\newcommand{\nfer}{{n_{\rm f}}}
\newcommand{\np}{{n_+}}
\newcommand{\nm}{{n_-}}
\newcommand{\nh}{{n}}
\newcommand{\nr}{N_R}
\newcommand{\nl}{N_L}
\newcommand{\TF}[1]{{\rm I}\left[{#1}\right]}
\newcommand{\pos}[1]{\left[{#1}\right]_+}
\newcommand{\Iall}{{\cal I}}
\newcommand{\Idiff}{{\cal I}_\Delta}

\newcommand{\qp}{q^+}
\newcommand{\qm}{q^-}
\newcommand{\qA}{q^A}
\newcommand{\qB}{q^B}

\chapter{A Fermion Zero Mode Index Theorem}
\label{Ch:Index}

\section{Introduction}

It has been realised in the past few years that many cosmologically significant
effects take place inside the core of strings~\cite{wbp&acd}. One
example is the formation of currents. These may provide a method of
detecting strings. If they form at high energy scales, and are charged,
the resulting electromagnetic field may be detectable~\cite{Witten}.
It is also possible that currents will allow vortons (stable string
loops) to exist. This may conflict with observations, in which case
the corresponding theory can be ruled out~\cite{vortbounds}. There are
several processes which can create currents on strings. These
include interaction with the plasma, and collisions between cosmic
strings. Charged currents can also be generated by magnetic or
electric fields.

In this chapter we investigate massless fermion currents. We do this by looking
for non-trivial zero energy fermion solutions, or zero
modes~\cite{Jackiw}. If they exist it is trivial to show that the
string has light-like fermion currents (see section~\ref{In FermZM}).

As has been shown in chapter~\ref{Ch:SO10}, subsequent phase transitions can
have a considerable effect on the microphysics of cosmic strings. Zero
modes on cosmic strings can be both created~\cite{acd&wbp} and
destroyed, thus creating or destroying the currents on the
strings. If the current is destroyed by subsequent microphysical
processes, then loops of cosmic string will no longer be stable, and vorton
bounds on the theory will be evaded. In this chapter we discuss the
fate of vortons and string superconductivity as the strings encounter
subsequent phase transitions in a systematic fashion.

In section~\ref{Ind theorem} we derive an index theorem giving the number of
zero modes for a general mass matrix. Whilst index theorems have been
derived before~\cite{Weinberg,Ganoulis} they have been more restrictive 
in their validity and have only been able to determine the difference
in right-moving and left-moving zero modes. Our index theorem has much
more general applicability and can give a bound on the number of zero modes. 

In section~\ref{Ind SO10ZeroModes} we discuss the existence of zero modes
on all the different strings formed at the breaking of the $SO(10)$
Grand Unified symmetry discussed in chapter~\ref{Ch:SO10}. We show
that at the electroweak phase transition zero modes can acquire a
small mass which leads to dissipation of the string current. This
allows vortons to decay and weakens the cosmological bounds on such
models~\cite{acd&wbp3}.

In section~\ref{Ind spec} we consider the
implications of spectral flow. An important feature
that allows zero modes to be removed is the presence of a
massless particle (possibly another zero mode) that mixes with the
zero mode after the transition. The implications of such couplings for
current build up before the transition are also considered.

We generalise some of the results of section~\ref{Ind SO10ZeroModes}
to other related theories in section~\ref{Ind OtherGUTs}. In other
models the zero modes survive subsequent transitions allowing the
associated vortons to persist. Such behaviour is displayed by a toy
model discussed in section~\ref{Ind persist}, where we explicitly
construct the zero mode solutions after the symmetry
breaking. Finally, we summarise our conclusions in 
section~\ref{Ind Conclusions}.

\section{Zero Mode Index}
\label{Ind theorem}
\newcommand{\half}{\mbox{$\frac{1}{2}$}}

Cosmic strings form in models with vacuum manifolds which are not
simply connected. For example in a $U(1)$ model, with potential
$(|\phi|^2 - \eta^2)^2$, stable solutions exist with $\phi = \eta e^{i\th}$
at $r=\infty$. In order for the total energy to be finite, a non-zero
gauge field is needed to give a vanishing covariant derivative at
$r=\infty$. In a more general theory, involving a larger group, $G$, string
solutions take the form
\bea
\phi(r,\th) = e^{i\ts\th} \phi(r) \ \ \ , \ \ \ 
A_\th = \frac{1}{er} T(r) \ ,
\eea
where $\ts$ is a generator of $G$ that is broken by $\phi$. The choice
of $\ts$ is restricted by the fact that $\phi$ must be single valued. $\phi(r)$
is equal to the usual VEV of $\phi$ at $r=\infty$, and must be regular
at $r=0$. $T(r)$ obeys $T(0)=0$, $T(\infty)=\ts$.

In a general theory, $\ts$ will affect different components of $\phi$
differently. This means that the various parts of $\phi$ can have a
wide range of winding numbers. In a theory with multiple phase
transitions, the additional Higgs fields will be affected in the same
way. It may also be necessary to alter $\ts$ at phase transitions to
make the new Higgs fields single valued.

In a theory with $\nfer$ two-component fermions, the fermionic part of
the Lagrangian is 
\be 
\Lagf = \bar{\psi}_\ga i\sig{\mu} D_\mu \psi_\ga 
- \frac{1}{2}i\bar{\psi}_\ga \Mab \psi^c_\gb + \hconj \ ,
\label{Ind Lag}
\ee
where $\psi^c_\gb = i\sig{2} \psi_\gb^\ast$. If $\Mab$ depends on
$\th$, as would be expected if $\Mab$ arose from the Higgs field of
the string, then it is possible that the field equations will have
non-trivial zero energy solutions. Solutions with only $r$ and $\th$
dependence can be split up into eigenstates of $\sig{3}$:
$\psi^L_\ga$, $\psi^R_\ga$. Such solutions have zero energy. If we
solve the equations of motion in the background of a cosmic string,
the field equations become
\be
e^{i\th}\left(\dr + \frac{i}{r}\dth + eA_\th \right) \psi^L_\ga 
+ M_{\ga \gb} \psi^{L\ast}_\gb = 0 \ ,
\label{Ind FEleft}
\ee
\be
e^{-i\th}\left(\dr - \frac{i}{r}\dth - eA_\th \right) \psi^R_\ga 
- M_{\ga \gb} \psi^{R\ast}_\gb = 0 \ ,
\label{Ind FEright}
\ee
where $A_\th$ is the string gauge field.
If $z$ and $t$ dependence is added to the solutions they will
correspond to currents flowing along the string. Their direction is
left for those corresponding to \bref{Ind FEleft}, and right for
\bref{Ind FEright}. In order to be physically relevant the solutions must
be normalisable. Let $\nl$ and $\nr$ be the number of such solutions
to \bref{Ind FEleft} and \bref{Ind FEright} respectively. We attempt to derive
an expression for them by generalising the analysis in
ref.~\cite{Jackiw}, which involves removing the $\th$ dependence of
the problem, and then considering solutions near $r=\infty$ and $r=0$.

Choose the $\psi_\ga$s to be eigenstates of the string gauge field,
with eigenvalues $q_\ga$. The $q_\ga$ will depend on the fermion
charges and the winding numbers of the various components of the Higgs
fields. Since the mass terms in \bref{Ind Lag} are gauge invariant, the
angular dependence of the mass matrix must be 
\be
M_{\ga \gb}(r,\th) = \Cab(r)e^{i(q_\ga + q_\gb)\th} \ 
\ \ \ (\mbox{no summation}) \ .
\label{MassNth}
\ee
The $\th$ dependence can also be factored out of the $\psi_\ga$s. 
\bea
\psi^L_\ga &=& e^{i(q_\ga - \frac{1}{2})\th}
		(U^L_\ga(r)e^{il\th} + V^{L\ast}_\ga(r)e^{-il\th}) \ ,\\
\psi^R_\ga &=& e^{i(q_\ga + \frac{1}{2})\th}
		(U^R_\ga(r)e^{il\th} + V^{R\ast}_\ga(r)e^{-il\th}) \ . 
\label{Ind PsiNth}
\eea
$l$ can take any value which gives a single valued $\psi$. Thus $l$ is
integer or half-integer when the $q_\ga$ are half-integer or integer
respectively. 

First consider left moving zero modes.
Putting (\ref{MassNth},\ref{Ind PsiNth}) into \bref{Ind FEleft} gives
equations for $U^L_\ga$ and $V^L_\ga$.
As $r \rightarrow \infty$, $\Cab = O(1)$ and 
$eA_\th \longrightarrow \ts/r$, so
\bea
\left(\dr + \frac{1}{2r} - \frac{l}{r} \right) U^L_\ga 
+ \Cab(\infty) V^L_\gb = 0 \ , \label{BigrFEU} \\
\left(\dr + \frac{1}{2r} + \frac{l}{r} \right) V^L_\ga
+ \Cab(\infty) U^L_\gb = 0 \ . \label{Ind BigrFEV}
\eea
Diagonalising $\Cab$, we find that the $2\nfer$ solutions to
these equations are modified Bessel functions. To leading order they are
proportional to $e^{\pm \la_i r}/\sqrt{r}$, where $\la_i$ are $\Cab$'s
eigenvalues. Half of these large $r$ solutions are
normalisable at $r=\infty$. If $\la_i = 0$ the corresponding solutions are
$r^{\pm l-1/2}$. If $l \neq 0$ (which is certain if the $q_\ga$ are not
half integer), half of these solutions are acceptable. If $l=0$ the
states corresponding to the solutions are physical, but are not
localised to the string, and so are not of interest to us. 

Thus if all $\la_i \neq 0$ or the $q_\ga$ are not half integer,
exactly $\nfer$ of the large $r$ solutions are normalisable at
$r=\infty$. We will assume that this is the case in the following
analysis. If these conditions do not hold, the index theorem derived
below may overestimate the number of allowed solutions, and so will only
provide an upper bound for the number of zero mode solutions.

The Higgs fields, and hence $\Mab$, are regular at the origin, so
as $r \rightarrow 0$, $\Cab = O(1)$, and $eA_\th = O(r)$. Thus
\bea
\left(\dr - \frac{q_\ga - \frac{1}{2} + l}{r}\right) U^L_\ga 
+ \Cab V^L_\gb = 0 \ , \label{SmallrFEU} \\
\left(\dr - \frac{q_\ga - \frac{1}{2} - l}{r}\right) V^L_\ga 
+ \Cab U^L_\gb = 0 \ . \label{SmallrFEV}
\eea 
To leading order, the small $r$ solutions are
\be \begin{array}{lll}
U^L_\ga &\sim& r^{q_\ga - \frac{1}{2} + l} \ , \\
V^L_\gb &\sim& O(1)r^{q_\ga + \frac{1}{2} + l} \ \ \forall \gb \ , \\
U^L_\gb &\sim& O(1)r^{q_\ga + \frac{3}{2} + l} \ \ \forall \gb \neq \ga \ ,
\end{array} \label{leftUsol}
\ee
where each choice of $\ga=1 \ldots \nfer$ gives one complex solution, and
\be \begin{array}{lll}
V^L_\ga &\sim& r^{q_\ga - \frac{1}{2} - l} \ , \\
U^L_\gb &\sim& O(1)r^{q_\ga + \frac{1}{2} - l} \ \ \forall \gb \ , \\
V^L_\gb &\sim& O(1)r^{q_\ga + \frac{3}{2} - l} \ \ \forall \gb \neq \ga \ .
\end{array} \label{leftVsol}
\ee
This gives a total of $2\nfer$ independent complex solutions. For
given $l$ and $\ga$, \bref{leftUsol} will be normalisable (for small $r$) if 
$l \geq -q_\ga + 1/2$. If $l \leq q_\ga - 1/2$ then
\bref{leftVsol} will be normalisable. Thus for a given $l$ the number
of well behaved small $r$ solutions is
\be
\nl^0 (l) = \sum^\nfer_{\ga=1} \TF{l \leq q_\ga - \half} + 
	\TF{l \geq -q_\ga + \half} 
\ee
where $\TF{X}$ equals 1 if $X$ is true, and 0 if $X$ is false.
What we are actually interested in is the number of solutions that are
normalisable for all $r$ ($\nl(l)$). Each such solution will be equal to
some combination of the $\nfer$ well behaved solutions to
(\ref{BigrFEU},\ref{Ind BigrFEV}) at large $r$, and a combination of the $\nl^0(l)$
suitable solutions to (\ref{SmallrFEU},\ref{SmallrFEV}) for small $r$.
If there are only $\nfer$, or less, suitable small $r$ solutions, then
in general any combination of the large $r$ solutions will not be well
behaved at $r=0$. If there are $\nfer + k$ suitable small $r$
solutions, then $k$ independent combinations of the large $r$
solutions will be well behaved everywhere. It may be possible to get
more solutions by fine tuning the theory, in which case the index
derived would be a lower bound.

The number of normalisable solutions for a given $l$ is
\be
\nl(l) = \pos{\nl^0 (l) - \nfer} \ ,
\ee
where $\pos{x}$ is defined to be equal to zero if $x<0$, and $x$ if $x \geq 0$.

This is not true if the equations obtained from \bref{Ind FEleft} can be
split into several independent sets. This will occur when 
$\Mab$ is a direct sum of mass matrices. In this case the
mass matrix can be split up into smaller matrices, which can be
analysed individually. Even when $\Mab$ is not a direct sum
of other matrices, it may still be possible to split the equations
into two independent sets. This case will be considered separately later. 

Since $U^L_\ga$ and $V^L_\ga$ are determined by real equations, each
complex solution gives two real solutions. This suggests that the total
number of left moving zero modes, $\nl$, is $2\sum_l \nl(l)$. However,
as can be seen from \bref{Ind PsiNth}, solutions for $l=k$ and $l=-k$ are
equal. For $l=0$ $U^L_\ga = \pm V^L_\ga$, so one of $\psi^L_\ga$'s
solutions is zero. Thus the total number of independent real solutions
is
\be
\nl = \sum_l \nl(l) = \sum_l \pos{\sum^\nfer_{\ga=1} \left(\TF{l \leq q_\ga 
	- \half} + \TF{l \geq -q_\ga + \half}\right) - \nfer} \ .
\label{nl1}
\ee
The summation is over all values of $l$ that give single valued
$\psi$. Since all the Higgs fields which make up $M_{\ga \gb}$ are
single valued, \bref{MassNth} implies all $q_\ga$ or all $q_\ga - 1/2$ are
integers (assuming $M_{\ga \gb}$ is not a direct sum of smaller
matrices), in which case respectively $l - 1/2$ or $l$ is an integer.

A similar analysis can be applied to right moving zero modes. For large
$r$ the behaviour is the same. For small $r$, solutions are well
behaved if $l \geq q_\ga + 1/2$ or $l \leq -q_\ga - 1/2$. This gives
\be
\nr = \sum_l \nr(l) = \sum_l \pos{\sum^\nfer_{\ga=1} \left(\TF{l \leq 
	-q_\ga - \half} + \TF{l \geq q_\ga + \half}\right) - \nfer} \ .
\label{nr1}
\ee

If $q_\ga$ is positive then one or two of the $\TF{\ldots}$ terms in
\bref{nl1} will be non-zero. If $q_\ga$ is negative, one or zero
of them will be non-zero. By splitting $q_\ga$ into positive and
negative eigenvalues, and ordering them, \bref{nl1} and \bref{nr1} can
be simplified. If there are $\np$ positive and $\nm$ negative $q_\ga$s
then, after reordering,
$q_\ga$ = ($\qp_1$, $\qp_2$ $\ldots$ $\qp_\np$, $\qm_1$, $\qm_2$
$\ldots$ $\qm_\nm$, 0 $\ldots$ 0), where $\qp_j>0$, $\qm_j<0$. Clearly if the
string gauge eigenvalues are not integer, there are no zeros. The
$\TF{\ldots}$ terms in \bref{nl1} and \bref{nr1} can be combined to give
\be
\nl = \sum_l \pos{
\sum^\np_{j=1} \TF{-\qp_j + \half \leq l \leq \qp_j - \half}
- \sum^\nm_{j=1} \TF{\qm_j - \half < l < -\qm_j + \half} } \! ,
\label{nl2} \\
\ee \be
\nr = \sum_l \pos{
\sum^\nm_{j=1} \TF{\qm_j + \half  \leq l \leq -\qm_j - \half}
- \sum^\np_{j=1} \TF{-\qp_j - \half < l < \qp_j + \half}  } \! .
\label{nr2}
\ee
These expressions can be further simplified if the eigenvalues are
ordered. If $\qp_1 \geq \qp_2 \geq \ldots \geq \qp_\np > 0$ and 
$\qm_1 \leq \qm_2 \leq \ldots \leq \qm_\nm < 0$, it is possible to evaluate
the $l$ summation by considering cancellation of the $\qp_j$ and
$\qm_j$ terms. This gives
\bea
\nl &=& \sum^{\min(\nm,\np)}_{j=1} 2\pos{\qp_j + \qm_j}
			+ \sum^\np_{j=\nm+1} 2 \qp_j \ , \label{Ind nl3} \\
\nr &=& \sum^{\min(\nm,\np)}_{j=1} 2\pos{-\qm_j - \qp_j} 
			- \sum^\nm_{j=\np+1} 2 \qm_j \ . \label{Ind nr3}
\eea
Taking the difference of these results gives
\bea
\Idiff &\!\!=\!\!& \nl - \nr \nonumber \\
&\!\!=\!\!& \sum^{\min(\nm,\np)}_{j=1} 2(\qp_j + \qm_j) +
\sum^\np_{j=\nm+1} 2\qp_j + \sum^\nm_{j=\np+1} 2\qm_j \nonumber \\
 &\!\!=\!\!& \sum^\nfer_{\ga=1} 2q_\ga 
 = \frac{1}{2 \pi i} \left[\ln \det M \right]^{2\pi}_{\th=0} \ .
\label{diffindex}
\eea
This is in agreement with other index theorems obtained elsewhere
\cite{Weinberg,Ganoulis}. The other index theorems were obtained by a
different method and only gave $\Idiff$, not $\nl$ and $\nr$.

The total number of zero modes is
\be
\Iall = \nl+\nr = \sum^{\min(\nm,\np)}_{j=1} 2\left|\qp_j + \qm_j\right| +
\sum^\np_{j=\nm+1} 2\qp_j - \sum^\nm_{j=\np+1} 2\qm_j \ ,
\label{allindex}
\ee
where only one of the last 2 terms contributes, depending on whether
$\np$ or $\nm$ is bigger. This is also true of \bref{diffindex} and
(\ref{Ind nl3},\ref{Ind nr3}).

If $\Iall$ is to be zero, then for every positive $q_\ga$ there
must be one negative $q_\gb$ with the same magnitude. If every fermion
field couples to a Higgs field with winding number zero, this will be
the case.

The above approach fails if $\Cab$ is of the form
\be
\bmat{cc} 0 & A_{\ga \gb} \\ B_{\ga \gb} & 0 \emat \ .
\label{Ind case2}
\ee
If $\Cab$ is assumed to have no zero eigenvalues, $A_{\ga \gb}$
and $B_{\ga \gb}$ are both $\nh \times \nh$ matrices, where $\nh = \nfer/2$.
In this case when (\ref{MassNth},\ref{Ind PsiNth}) are substituted into
\bref{Ind FEleft}, two independent sets of equations are
obtained. Expressions for $\nl$ and $\nr$ can found by considering
just one set of these solutions. Putting
\be
\psi^L_\ga = e^{i(q_\ga - \frac{1}{2})} 
\left\{ \begin{array}{ll}
	U^L_\ga(r)e^{il\th} & \ga = 1 \ldots \nh \\
	V^{L\ast}_\ga(r)e^{-il\th} & \ga = \nh +1 \ldots \nfer
\end{array} \right. 
\ee
and \bref{MassNth} into \bref{Ind FEleft} gives
(\ref{BigrFEU},\ref{Ind BigrFEV}) for large $r$ and
(\ref{SmallrFEU},\ref{SmallrFEV}) for small $r$, but with a more restricted
range on the indices. In this case the allowed values of $l$ need no
longer be integers or half-integers, but the difference between any 2
values is still an integer. For large $r$, $\nh$ of the $\nfer$ complex
solutions are normalisable. For a given $l$ at small $r$, there is one
normalisable solution for each $q_\ga$ satisfying 
$l \geq -q_\ga + 1/2$ ($\ga = 1 \ldots \nh$) or 
$l \leq q_\ga - 1/2$ ($\ga = \nh +1 \ldots \nfer$).

Matching the solutions for large and small $r$ gives
\be
\nl(l) = \pos{\sum^\nh_{\ga=1} \left( \TF{l \leq q_{\ga+\nh} - \half} 
		+ \TF{l \geq -q_\ga + \half}\right) - \nh} \ . 
\label{nlspec}
\ee
The solutions for different $l$ are independent (unlike the previously
considered cases), so the total number of real solutions ($\nl$) is just twice
the number of complex solutions, thus $\nl = 2\sum_l \nl(l)$. 
A similar expression can be obtained for $\nr$.

The $\TF{\ldots}$ terms in \bref{nlspec} can be combined. Defining 
$\qB_j = -q_j$ and $\qA_j = q_{j+\nh}$ for $j = 1 \ldots \nh$,
\bref{nlspec} and the corresponding expression for $\nr$, become 

\be
\nl = 2\sum_l \pos{\sum^\nh_{j=1} \left( 
\TF{-\qB_j + \half \leq l \leq \qA_j - \half} 
- \TF{\qA_j - \half < l < -\qB_j + \half} \right)} \! , 
\label{Ind nl2b}
\ee
\be
\nr = 2\sum_l \pos{\sum^\nh_{j=1} \left( 
\TF{\qA_j + \half \leq l \leq -\qB_j - \half} 
- \TF{-\qB_j - \half < l < \qA_j + \half} \right)} \! .
\label{Ind nr2b}
\ee
If the string gauge eigenvalues are reordered, so that $\qA_1 \geq
\ldots \geq \qA_{\nh}$ and $\qB_1 \leq \ldots \leq \qB_{\nh}$, the
expressions reduce to 
\bea
\nl &=& \sum^n_{j=1} 2\pos{\qA_j + \qB_j}  \ , \label{Ind nlAB} \\
\nr &=& \sum^n_{j=1} 2\pos{-\qB_j - \qA_j}  \ , \label{Ind nrAB}
\eea
which are similar to \bref{Ind nl3} and \bref{Ind nr3}, with
$\np=\nm=\nh$. This is not identical to the previous result, since the
$q_\ga$ are divided up differently.

When there are just two fermion fields involved, all the results reduce to 
\bea
\Idiff &=& 2(q_1 + q_2) \ , \\
\Iall &=& |\Idiff| \ .
\label{Ind n2fer}
\eea

\section{Fermion Zero Modes on SO(10) Cosmic Strings}
\label{Ind SO10ZeroModes}

One example of a phenomenologically credible grand unified theory (GUT)
has the symmetry breaking
\bea
SO(10) & \pharrow{126} & SU(5) \times Z_2 \\ 
& \pharrow{45} & SU(3)_c \times SU(2)_L \times U(1)_Y \times Z_2 \\
& \pharrow{10} & SU(3)_c \times U(1)_Q \times Z_2 \ .
\eea
This theory was discussed in greater detail in chapter~\ref{Ch:SO10}.
Appendix~\ref{Ch:app} contains details of the fermion fields and their
mass terms. A significant feature of its fermions are the massive
neutrinos. It has been conjectured that rather than being massless,
the left-handed neutrino actually has a very small mass. It would then
provide a dark matter candidate. Recent experimental results suggest
that neutrinos are massive, giving further credibility to the above
model and those like it~\cite{numass}. We can use these results to
obtain an estimate of the right-handed neutrino mass and the $SO(10)$
breaking scale. We will consider the top quark/tauon family, since
renormalisation effects (which we will neglect) are likely to be less
significant. The left-handed neutrino mass is approximately equal to
$m_t^2/m_{\nu(\tau R)}$. The results of ref.~\cite{numass} indicate
that $\nu_\mu$ mixes with another type of neutrino which has a different
mass, thus at least one of these neutrinos must be massive. Assuming
the detected mass difference is the result of 
$\nu_\tau \leftrightarrow \nu_\mu$ mixing, 
$\Delta m^2 =  m^2_{\nu(\tau L)} - m^2_{\nu(\mu L)} = 
(5\times 10^{-4} - 6\times 10^{-3})\mbox{eV}^2$. 
We will assume that $m_{\nu(\mu)} \ll m_{\nu(\tau)}$, and so
$m^2_{\nu(\tau R)} \approx m_t^4/\Delta m^2$. Taking 
$m_t = (180\pm12)$GeV~\cite{quarkmass}, we obtain 
$m_{\nu(\tau R)} \approx (4  - 16)\times 10^{14}$GeV, suggesting that the
ratio of the grand unification and electroweak scales is approximately
$(2 - 9)\times 10^{12}$.

We will now apply the results of the previous section to the various
cosmic strings of this model. In several cases the string Higgs field
($\ph{126}$) causes the electroweak Higgs field ($\ph{10}$) to take a
string-like solution. It is then possible that $\ph{10}$, like
$\ph{126}$ will give rise to fermion zero
modes~\cite{Witten,acd&wbp,Stern}. The fermionic part of this theory's
Lagrangian is
\be 
\Lagf = \flb i\sig{\mu} D_\mu \fl 
- \frac{1}{2} i\yuke \flb \ph{10} \flc \\
- \frac{1}{2} i\yukg \flb \ph{126} \flc + \hconj \ ,
\label{fermLag}
\ee
where $\sig{\mu} = (-I,\sig{i})$ since $\fl$ is a two component
spinor. For simplicity we will only consider one family of
fermions. Varying $\flb$ in \bref{fermLag} gives the field equations
\be
 i\sig{\mu} D_\mu \fl 
- i\yuke \ph{10} i\sig{2} \fl^\ast
- i\yukg \ph{126} i\sig{2} \fl^\ast = 0 \ .
\label{fermEL}
\ee 
In section~\ref{Ind theorem}, an expression for the number of zero mode
solutions to this type of equation was derived. The numbers and
types of fermion zero modes can easily be found once \bref{fermLag} is
put into the same form as \bref{Ind Lag}. This involves calculating
the mass matrix, $M$, for each type of string, and splitting it up
into irreducible parts. Then the fermion fields coupling to each part, $\psi$,
need to be expressed in terms of eigenstates of $\ts$. The eigenvalues
of these states are then divided into two sets and inserted into the
appropriate expressions (\ref{Ind nl3},\ref{Ind nr3},\ref{allindex}) and 
(\ref{Ind nlAB},\ref{Ind nrAB}). The way the eigenvalues are divided
up depends on the mass matrix. Usually they will be split into
positive and negative eigenvalues (see remarks before
\bref{Ind nl3}), but if the mass matrix has the same type of
degeneracy as \bref{Ind case2}, they are split according to which
part of the matrix they couple to (see remarks before and after
(\ref{Ind nl2b},\ref{Ind nr2b})).

\subsection{Zero Modes for the Abelian String}
\label{Ind abstrZM}

\subsubsection{High Temperature Neutrino Zero Modes}

At high temperatures $\ph{10}$ is zero, and so with the exception of
$\nu^c$, none of the fermion fields are affected by Higgs
fields. Thus the relevant part of the theory is just a two-component
spinor coupling to an abelian string. This case has been discussed in
section~\ref{In FermZM}, and in greater detail in ref.~\cite{Jackiw}.

There are $|n|$ normalisable zero energy solutions. If $z$ and $t$
dependence are added, they all move to the left along the string if
$n>0$. If $n<0$ they move to the right. Thus conjugate neutrino zero modes
always exist at high temperatures in the presence of an abelian
string. For $r > \rs$ (assuming, for simplicity, that all coupling
constants are approximately 1) the solutions decrease exponentially,
so the zero modes are confined to the string core.
 
\subsubsection{High Temperature Non-Neutrino Zero Modes}

Although there is no Higgs field acting on the other fermion fields, it is
possible for zero modes to be generated by the string gauge fields, as
discussed by Stern and Yajnik~\cite{Stern}. The index theorem
discussed in section~\ref{Ind theorem} does not apply in this case, since it
assumes that all the fermions considered couple to Higgs fields.

Labelling the upper and lower components of the fermion fields $\la^L$
and $\la^R$ respectively (as in (\ref{Ind FEleft},\ref{Ind FEright})),
where $\la = u_i, d_i^c, \mbox{etc.}$ (not $\nu^c$), the fermion field
equations reduce to
\be
\left( \dr + \sigma^{L,R}\left[\frac{i}{r}\dth + 
p_{\la}n\frac{a(r)}{r} \right] \right) \la^{L,R} = 0 \ ,
\label{Sternequ}
\ee
where $\sigma^{L,R} = \pm 1$, according to which component is being used.
$p_\la$ is the eigenvalue of the field with respect to $P/10$:  
$p_{e^-} = p_{\nu} = p_{d^c_i} = -3/10$,
$p_{u_i^c} = p_{u_i} = p_{d_i} = p_{e^+} = 1/10$. There are
normalisable solutions if $|np_\la| > 1$, all of which can be
found analytically. The number of solutions is equal to the highest integer
that is less than $|np_{\la}|$. Thus $|n|$ must be at least
4 for any zero modes of this type to exist. If the only stable
strings have winding number 1, then only conjugate neutrino zero modes will
be present at high temperatures around an abelian string. 

\subsubsection{Low Temperature Non-Neutrino Zero Modes}

At lower temperatures $\ph{10}$ is non-zero and couples to all the
fermion fields. Now none of the particles are massless, and all zero
modes can be found using the index theorem. The electroweak phase
transition changes the string generator used in 
section~\ref{Ind theorem} to $\ts + \tew$, with $\tew$ defined in
section~\ref{SO10 EWZsect}. Before applying the theorem, the
mass matrix needs to be split into irreducible parts. In this case
there are 8 of them, one for each particle type. With the exception of
the neutrino fields, there is just one Higgs field coupling to
them. The Lagrangians for the non-neutrino fields are then \bref{Ind Lag}, 
with
\be
M = \bmat{cc} 	0 & m_\la h_\la(r) e^{\pm im\th} \\ 
		m_\la h_\la(r) e^{\pm im\th} & 0 \emat \ , \ 
\psi = \bmat{c} \la^c \\ \la \emat \ ,
\label{abLowZM}
\ee
where $\la$ can be $d_i$, $u_i$, or $e^-$. We have defined
$m_u = \yuke \etau/4$, $m_\sbe = m_\sbd = \yuke \etad / 4$, and 
$h_\sbe (r) = h_\sbd(r)$. The upper sign is taken
for the $u_i$ and $u_i^c$ fields, which couple to the $\Hu{0}$
component of $\ph{10}$. The lower sign applies for the $d_i$, $d_i^c$
and $e^\pm$ fields, since they couple to the $\Hd{0}$ component. 
Applying \bref{Ind n2fer} reveals that there are $2|m|$
solutions per particle type ($14|m|$ total, not counting
neutrinos). The field equations have previously been discussed
in ref.~\cite{Stern}. 

If $n$ is not a multiple of 5, the solutions decay exponentially
outside $r=\rew$. When $n$ is a multiple of 5 (in which case 
$m = n/5$), they decay outside $r=\rs$. Thus the zero modes are
confined to the region of symmetry restoration. 
The difference in sign between up and down quarks in \bref{abLowZM}
has physical significance when $z$ and $t$ dependence are added to the
solutions. The up quark currents flow in the opposite direction to the
down quark and electron currents.

\subsubsection{Low Temperature Neutrino Zero Modes}

The situation is more complex for the neutrino fields since they are
affected by two Higgs fields at the same time. In this case
\bref{abLowZM} is replaced by
\be
M = \bmat{cc}   m_\sbG f(r) e^{in\th} & m_\sbu h_\sbu(r) e^{im\th} \\ 
		m_\sbu h_\sbu(r) e^{im\th} & 0 \emat \ , \ 
\psi = \bmat{c} \nu^c \\ \nu \emat \ ,
\ee
where $m_\sbG = \yukg \etag$. This case is more complex
than \bref{abLowZM}, although \bref{Ind n2fer} still applies. It
implies there are $2|m|$ zero modes. Surprisingly this does not depend
directly on $n$. As with the other particle zero modes, they will be
confined to the region of symmetry restoration. They move left if
$m>0$ (which only happens if  $n \ge 3$), or right if $m<0$ ($n \le -3$).

Considering all fermions, there are a total of $16|m|$ zero modes,
half of which are left moving, half are right moving. For a
topologically stable string $m=0$, so there are no zero electroweak modes.
In the case of the neutrinos this is slightly surprising, since at
higher temperatures when $\ph{10}$ is zero, the abelian string does have
neutrino zero modes. Intuitively, since 
$\ph{10} \sim (\etae/\etag) \ph{126} \sim 10^{-14}\ph{126}$,
the situation could be expected to be the same for lower temperatures. 

Since $|2m| < |n|$, some of the neutrino zero modes will be destroyed
by the electroweak phase transition. For a
stable $n=1$ string all zero modes are destroyed. Thus, since higher $n$
strings usually decay, there are zero modes before, but not
after the electroweak phase transition. It is expected that the
neutral current in the string will disperse~\cite{HillWidrow} in which
case any vortons formed will dissipate after about
$10^{-10}$sec~\cite{acd&wbp3}. Before the electroweak phase transition
from about $10^{10}$GeV -- $10^2$GeV the universe would undergo a
period of matter domination. Once the vortons dissipate there would be
some reheating of the universe. However the electroweak interactions
and physics below the phase transition would be unaffected.

\subsection{Zero Modes of the SU(2) Strings}

There are two additional complications with nonabelian strings.
Firstly the particle states are not eigenstates of the string
generator, although this is easily solved by re-expressing the problem
in terms of gauge eigenstates. Secondly, there are effectively twice
as many Higgs fields, since each Higgs field has parts with two
different winding numbers and different profiles.

\subsubsection{High Temperature Neutrino Zero Modes}

At high temperatures the gauge fields are proportional to $\ts$. Since
$\nu^c$ is not an eigenstate of $\ts$, the problem must be expressed in
terms of $\chi^{c(\pm)}= (\nu^c \pm 2\ts \nu^c)/\sqrt{2}$, 
which are eigenvectors of $\ts$. Their eigenvalues are $\pm 1/2$.
The relevant part of the Lagrangian is \bref{Ind Lag}, with
\be
M = \frac{m_\sbG}{2} \bmat{cc} 	f_1(r) e^{in\th} & f_0(r) \\ 
	               		f_0(r) & f_1(r) e^{-in\th} \emat \ , \ 
\psi = \bmat{c} \chi^{c(+)} \\ \chi^{c(-)} \emat \ .
\label{nabHighNuZM}
\ee
Since $f_1(\infty)=f_0(\infty)=1$ one of the fermion fields is
massless at large $r$. This means that \bref{Ind n2fer} does not apply
in this case, as its derivation assumed that either the mass matrix
has no zero eigenvalues at large $r$, or that $\ts$ does not have half
integer eigenvalues. However, \bref{Ind n2fer} can still be used to give an
upper bound on the number of zero modes. Applying it to
\bref{nabHighNuZM} reveals the upper bound to be zero. Hence there are
no neutrino zero modes on the $SU(2)$ strings for any values of $n$.

\subsubsection{High Temperature Non-Neutrino Zero Modes}

For the fermion fields that do not couple to $\ph{126}$ it is possible
for zero modes to exist by the same mechanism as
(\ref{Sternequ}). However, unlike the abelian case, some fermion
fields are annihilated by $\ts$, so $p_\la$ is effectively zero,
and they cannot have zero energy solutions for any value of $n$. For
instance, the $u_i$, $d_i$, $\nu$ and $e^-$ fields are all zero
eigenstates of the string generator for the high temperature
$\WR{}$--string. Thus, in the presence of this type of string, solutions
can only occur for the conjugate fields. Defining
$\chi^{(\pm)} = (\la \pm 2\ts \la)/\sqrt{2}$, where $\ts\la$ is not
proportional to $\nu^c$ or $\ts\nu^c$, or equal to zero, the
nonabelian equivalent of \bref{Sternequ} is
\be
\left( \dr + \sigma^{L,R} \left[\frac{i}{r}\dth \pm 
\frac{1}{2}n\frac{a(r)}{r} \right] \right) \chi^{(\pm)L,R} = 0 \ .
\ee
It has the solutions
\be
\chi^{(\pm)L,R} = r^l \exp \left( \sigma^{L,R}\left\{il\theta 
\mp \frac{n}{2} \int_0^r ds\frac{a(s)}{s} \right\} \right) \ .
\ee
The solutions are normalisable if 
$0 \leq l < \pm n\sigma^{L,R} / 2 - 1$. Thus the total number
of solutions, per number of particles (6 in this case), will be the
largest integer below $n/2$. In order for any such solutions
to exist $n$ must be at least 3, so they do not occur for
topologically stable strings.

\subsubsection{Low Temperature $\Xs{}{}$ and $\Xg{}{}$ String Zero Modes}

At low temperatures $\ph{10}$ is non-zero and couples to all the fermion
fields. When $\ts$ is made up of generators of the $\Xs{}{}$ or
$\Xg{}{}$ fields, $\ph{10}$ just takes its usual vacuum expectation
value. For the fermion fields that are not affected by $\ph{126}$ there
is effectively no string and so no zero modes.
For the fields affected by $\ph{126}$ the solutions of the field
equations will be at least as divergent as those of
\bref{nabHighNuZM}, so there will no normalisable solutions.

\subsubsection{Low Temperature $\WR{}$--String Non-Neutrino Zero Modes}
\label{LTWstrNnZM}

The neutrino and electron fields all couple to $\ph{126}$ in the
presence of a $\WR{}$--string, while the quark fields are only affected by
$\ph{10}$. The string generator is now of $\ts + \tew$. Its
non-neutrino fermion eigenstates are $\chi_i^{(\pm)} = (u_i \pm d_i)/\sqrt{2}$ 
and $\chi_i^{c(\pm)} = (u_i^c \mp d_i^c)/\sqrt{2}$, where $i=1\ldots 3$. 
They are eigenstates of both $\ts$ and $\tew$, with $\tew$ defined in
section~\ref{SO10 EWWsect}. The fermion mass matrix is obtained from
\bref{EWWansatz}. It is reducible into 4 parts. One couples the
neutrino and electron fields, and the others couple the
$\chi_i^{(\pm)}$ to the corresponding $\chi_i^{c(\pm)}$. The
appropriate expressions to insert into \bref{Ind Lag} are then
\bea
M &=& \bmat{cccc} 0 & 0 & m_+ h_+ e^{im'\th} & m_- h_- e^{im\th} \\
		0 & 0 & m_- h_- e^{-im\th} & m_+ h_+ e^{-im'\th} \\
		m_+ h_+ e^{im'\th} & m_- h_- e^{-im\th} & 0 & 0 \\
		m_- h_- e^{im\th} & m_+ h_+ e^{-im'\th} & 0 & 0 \emat
\ , \\ \nonumber && \\ \nonumber \psi^T &=& 
\bmat{cccc} 	\chi^{c(+)}_i & \chi^{c(-)}_i & 
		\chi^{(+)}_i & \chi^{(-)}_i \emat^T \ .
\label{nabWLowZM}
\eea
where $m_\pm = (m_\sbu \mp m_\sbd)/2$. When $n$ is odd the winding
numbers of the Higgs components are $m = (n-1)/2$ and $m' = (n+1)/2$

In this case the second version of the index theorem is needed since
\bref{nabWLowZM} is in the same form as \bref{Ind case2}. If
$m_\sbu \neq m_\sbd$, (\ref{Ind nlAB},\ref{Ind nrAB}) reveal that there
are $2m = n-1$ left moving and $2m$ right moving zero modes. If
$m_\sbu = m_\sbd$, the $e^{\pm im'\th}$ terms are not present in
\bref{nabWLowZM}, since $m_+ =0$. The mass matrix is then
reducible. Applying \bref{Ind n2fer} to the 2 parts shows there are
$2m$ left and right moving modes in this case as well.

Since there are 3 choices of $i$ there are a total of $12m$ different
zero modes for the $\WR{}$--string after electroweak symmetry breaking
(assuming no neutrino zero modes). The solutions are contained in
the $r<\rew$ region. Since $m=0$ for the energetically stable $n=1$
string, it has no fermion zero modes.

When $n$ is even (so the string is actually topologically equivalent
to the vacuum), $M$ takes the same form as above, but with $m=m'=n/2$.
Thus the results for the odd $n$ strings can be applied to even $n$
strings, and there are $12(n/2)$ normalisable solutions.

\subsubsection{Low Temperature $\Yg{}{}$--String Non-Neutrino Zero Modes}

The $\Yg{}{}$--string can be approached in a similar way to the
$\WR{}$--string, although there are additional complications due to the
form of the gauge fields. To put the problem in the same form as
\bref{Ind Lag}, it needs to be expressed in terms of the fermion
eigenstates of the string generator, which in this case is
$\ts+\tew+\tew'$. $\tew$ and $\tew'$ are defined in
section~\ref{SO10 EWYsect}. Defining $v_i = (u_1,u_2^c,u_3^c,\nu)$ and
$w_i = (e^+,-d_3,d_2,-d_1^c) \sigma$ (where $\sigma$ is the sign of
$\tew+\tew'$), the eigenstates are  $v_i$, $w_i$ and 
$\chi^{(\pm)}_i = (v_i^c \pm w_i^c)/\sqrt{2}$. Obtaining the fermion
mass matrix from \bref{EWYansatz} we find, for $i=1\ldots 3$, that
\bea
M &=& 
\bmat{@{}cc@{}cc@{}} 0 & 0 & 
	\frac{m_\sbu}{\sqrt{2}} h_{\sbu+} e^{i\sigma m'\th} & 
	\frac{m_\sbd}{\sqrt{2}} h_{\sbu-}e^{-i\sigma m\th} \\
	0 & 0 & \frac{m_\sbu}{\sqrt{2}} h_{\sbd+} e^{i\sigma m\th} & 
	-\frac{m_\sbd}{\sqrt{2}} h_{\sbd-} e^{-i\sigma m'\th} \\
	\frac{m_\sbu}{\sqrt{2}} h_{\sbu+}e^{i\sigma m'\th} &
	\frac{m_\sbd}{\sqrt{2}} h_{\sbd+}e^{i\sigma m\th} & 0 & 0 \\
	\frac{m_\sbu}{\sqrt{2}} h_{\sbu-}e^{-i\sigma m\th} &
	-\frac{m_\sbd}{\sqrt{2}} h_{\sbd-}e^{-i\sigma m'\th} & 0 & 0 \emat
\ , \\ \nonumber && \\ \nonumber
\psi &=& \bmat{cccc} v_i & w_i & \chi^{(+)}_i & \chi^{(-)}_i \emat^T \ .
\eea
This is similar to the mass matrix for the $\WR{}$--string. However
in this case the electroweak gauge fields couple $\chi^{(+)}_i$
to $\chi^{(-)}_i$. Although their effect is zero away from the string,
they still mix the fermion solutions. Thus, unlike \bref{nabWLowZM},
the first form of the index theorem \bref{allindex} is
used. We find there are no normalisable zero mode solutions. If $n$ is
even there are no electroweak gauge fields, and the situation is the
same as the $\WR{}$--string. There are then $6n$ zero modes.

\subsubsection{Low Temperature $\WR{}$ and $\Yg{}{}$ String Neutrino
Zero Modes}

For the fields affected by both $\ph{126}$ and $\ph{10}$ in the presence
of a $\WR{}$--string, the mass matrix is \bref{nabWLowZM}, with the
$M$ of \bref{nabHighNuZM} replacing the top left $2\times 2$
submatrix. The eigenstates are now $\chi_i^{(\pm)} = (\nu \pm e^-)/\sqrt{2}$ 
and $\chi_i^{c(\pm)} = (\nu^c \mp e^+)/\sqrt{2}$. This time however,
the first version of the index theorem \bref{allindex} is used. It
implies that there are no normalisable zero mode solutions for any
choice of the parameters.

For the $\Yg{}{}$--string we take the fermion eigenstates to be $v_4$,
$w_4$ and $\chi^{(+)}_4$ (defined above). The situation is
similar to the $\WR{}$--string, and so none of the $SU(2)$ cosmic strings
have low temperature zero modes involving the conjugate neutrino field.

\subsection{Summary}
The only fermion zero modes that form at high temperatures are $\nu^c$
zero modes around abelian strings (in which case there are $|n|$ of
them), or those that involve fermion fields that just couple to the
string gauge fields, and not $\ph{126}$. This latter type of solution
will only occur for higher $n$ strings ($n \geq 3$ for $SU(2)$
strings, $|n| \geq 4$ for abelian strings).

At low temperatures there are a total of $16|m|$ different
zero modes on an abelian string ($|m|$ for each particle type), where
$m$ is the winding number of $\ph{10}$. $m=0$ when $|n|<3$, so there are
no zero modes around topologically stable abelian strings, and hence
they can only be superconducting at low temperatures in the presence
of an unusual Higgs potential~\cite{Witten}.

If $m \neq 0$, and $z$ and $t$ dependence are added to the solutions,
they will correspond to massless fermion currents. The electron
and down quark currents will then flow in the opposite direction to
the neutrino and up quark currents. 

In the presence of a $\Xg{}{}$ or $\Xs{}{}$ nonabelian string there
are no zero modes at any temperature. After the electroweak phase
transition the $\WR{}$--string has $12m$
zero modes ($m$ for each particle type not coupling to $\ph{126}$),
where $m$ is the winding number of the part of $\ph{10}$
which winds least. $m = n/2$ for even $n$, and $m = (n-1)/2$ for
odd $n$. The $\Yg{}{}$--string has same number of zero modes as the
$\WR{}$--string if $n$ is even, and zero if $n$ is odd. Thus for a
minimal energy, topologically stable string there are no fermion zero
modes, although there is still the possibility of superconductivity
due to gauge boson zero modes. Thus even the $\Xg{}{}$ and
$\Xs{}{}$--strings may be superconducting~\cite{ABC}. Indeed, it has
been shown that such strings do become current carrying by gauge boson
condensation~\cite{Yates}.

The currents corresponding to any fermion zero modes present on
the $SU(2)$ strings do not consist of single particle types, as those
around an abelian string do. Instead they are made up of eigenstates
of the string generator. Also, unlike the abelian case, currents containing
each particle type flow in both directions along the string.

\section{Index Theorems and Spectral Flow}
\label{Ind spec}
\renewcommand{\baselinestretch}{1}
\begin{figure}
\vskip 0.5 in
\centerline{\psfig{file=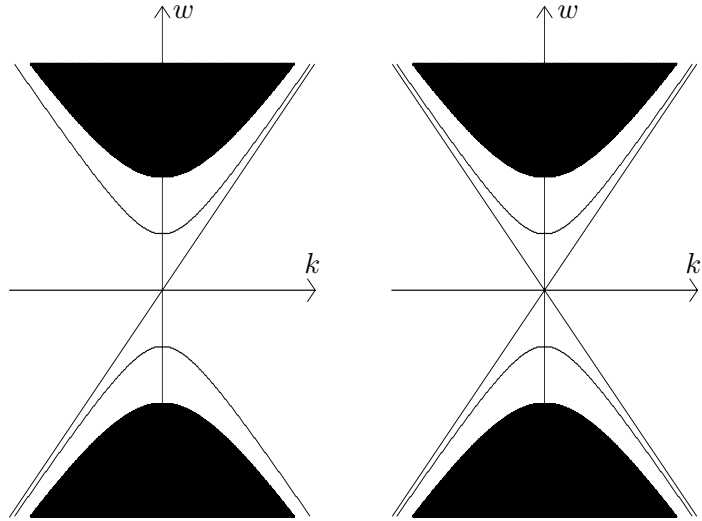,width=0.7\myfigwidth}\ \hspace{0.8in} \ }
\vskip -3.2 in 
\hskip 1.92 in $w$
\hskip 1.85 in $w$
\vskip 1.15 in 
\hskip 2.61 in $k$
\hskip 1.86 in $k$
\vskip 1.2 in
\caption[The Dirac spectrum of cosmic strings with a zero mode and a
very low lying bound state.]{The Dirac spectrum of cosmic strings with
a zero mode (left) and a very low lying bound state (right). Both
spectra also have a bound state and continuum.} 
\label{specfig}
\end{figure}
\renewcommand{\baselinestretch}{\myblstr}

We have shown that zero modes can acquire masses at subsequent phase
transitions. No matter how small this mass is, the spectrum of the Dirac
operator changes significantly. If we compare the Dirac spectrum 
with a zero mode and a low lying bound state with infinitesimal mass
(fig.~\ref{specfig}), we see that an arbitrarily
small perturbation to the zero mode introduces an entire new branch to
the spectrum. Any massive state gives a spectrum that is symmetric
about both the $w$ and $k$ axes, there is always a reference frame in
which the particle is at rest and others where it is moving up or
down the string. Conversely the zero mode, which is massless, can only
move in one direction along the string and its spectrum is asymmetric.
The transition from zero mode to low lying bound state causes drastic
changes in the spectrum and can be brought about by infinitesimal
changes in the value of one Higgs field. If we consider the species
with the zero mode alone, this infinite susceptibility to the
background fields appears unphysical. However, when we include the
massless neutrino in the $SO(10)$ model the spectral changes are less
worrying. For a small coupling between the two neutrinos, both the
before and after spectra have a continuum of massless or nearly
massless states. These states can be used to build the extra branch of
the perturbed zero mode spectrum, allowing small changes in the
overall spectrum for small changes in the background fields.

This observation leads us to conjecture that zero modes can be removed
only if they become mixed with other massless states. 

The coupling between the left and right handed neutrinos and the
electroweak Higgs field need not be artificially small, the small mass
of the light neutrino can be generated by the seesaw mechanism
\cite{seesaw}. This coupling is present prior to the electroweak phase
transition and allows transitions of the form $\nu_L+\bar\nu_R \to f \bar f$,
where $f$ is any light fermion from the standard model and the
intermediate state is an electroweak Higgs. Such interactions allow
zero modes on the string to scatter from massless neutrinos in the
surrounding plasma and provide a current damping mechanism that
affects current build up prior to the electroweak transition.

\section{Fermion Zero Modes in Other Theories}
\label{Ind OtherGUTs}

As discussed in section~\ref{SO10 OtherGUT}, various symmetry breaking
schemes give rise to the type of strings considered in
section~\ref{Ind SO10ZeroModes}. The resulting strings will have the
same kind of zero modes as theory \bref{symbreak},
provided none of the other Higgs fields couple to the fermions. The
only Higgs fields that can couple to fermions are those which
transform under a representation contained in the ${\bf 16} \times
{\bf 16}$, since fermion mass terms transform as a product of
{\bf 16}s. The only such representations are {\bf 126}, {\bf 10}, and
{\bf 120} (which is antisymmetric), so the results of this section apply to
a wide range of $SO(10)$ theories.

One alternative symmetry breaking is to use $\ph{16}$ instead of
$\ph{126}$. Since $\ph{16}$ does not couple to the fermions, $\yukg$ will be
zero in \bref{fermLag}, and so the neutrinos will have the same kind
of zero modes as all the other particles (As would be the case if
$\ph{126}$ were present, but $\yukg$ were zero). However such a theory
has left-handed neutrinos with significant masses, and an observable
right-handed neutrino, so it is not compatible with the standard model
(unless some other mechanism is introduced to alter the neutrino masses).

If the size of the coupling of the string gauge field to the electroweak
Higgs field were different (as in the $SU(5) \times U(1)_P$ theory
mentioned in section~\ref{SO10 OtherGUT}), it is possible for $|m|$ to
be non-zero on a topological string. In this case fermion zero modes can
be present. This also means that neutrino zero modes may survive the
electroweak phase transition. In fact if $|2m/n| > 1$, extra neutrino
zero modes will be created at the phase transition.

Although we have concentrated on $SO(10)$, many of the results apply
to a very wide range of theories. For example, we expect the
destruction of zero modes at phase transitions will frequently
occur. In a theory with Majorana mass terms, such as the right handed
neutrino term in section~\ref{Ind abstrZM}, the first version of the
index theorem can be used \bref{allindex}. After a phase transition,
if every fermion couples to a non-winding Higgs field, such as
$\ph{10}$, then all previously formed zero modes will be destroyed
(see remarks after \bref{allindex}). 

Another type of GUT with cosmic string solutions which couple to fermions are
those with the symmetry group $E_6$~\cite{Witten,E6GUT}. Such theories
have a larger fermion sector (27 fields rather than 16), more phase
transitions, and can have far more Yukawa couplings. $E_6$ also has
massive neutrinos, a total of 5 compared to $SO(10)$'s 2. The details will
be more complicated, but we expect currents to be formed and destroyed as
in the $SO(10)$ theories. We also expect the $SU(2)$ strings of the
theory to have zero, or very few, massless fermion currents.

\section{A Model with Persistent Zero Modes}
\label{Ind persist}

In this section we consider a model in which a zero mode survives a
subsequent symmetry breaking, despite coupling to a Higgs field with a
non-winding component. Consider a Majorana fermion similar to that in
\bref{In Lagferm}, but with a different Higgs field.
In this case the Higgs field has two parts, a winding part from the string
and a constant part from a second symmetry breaking,
\be
\phi = \eta\left(f(r) e^{i\th} + p \right) \ .
\ee
Such a Higgs field could only occur as the result of a global
symmetry being broken. Note that because of this the above $\phi$ cannot be
written in the form used in \bref{MassNth}, so the index theorem
developed in section~\ref{Ind theorem} does not apply in this case.
We take only the upper component of $\psi$ to be non-zero, in which
case its field equations reduce to
\be
e^{i\th}\left(\dr + \frac{i}{r}\dth\right) \psi 
+ \mass{f} \left(f(r) e^{i\th} + p \right) \psi^\ast = 0 \ .
\ee
Changing variables to $X$,
\be
X = \int_0^r f(\rho) d\rho +pr\cos\th + c \ ,
\ee
we have
\be
\left(\frac{\partial X}{\partial r} + 
		\frac{i}{r}\frac{\partial X}{\partial \th}\right) 
	= f(r) + p\left[\cos\th + \frac{i}{r}(-r\sin\th)\right] \ .
\ee
Thus
\be
\left(f(r) e^{i\th} + p \right)
\left(\partial_X \psi + \mass{f}\psi^\ast \right) = 0 \ .
\ee
This is solved by
\be
\psi = e^{-\mass{f}X} \ .
\ee
We have explicitly constructed a zero mode after the second phase transition.
The $X$ coordinate is similar to the usual radial coordinate, but
`centres' on the effective zero of the resultant Higgs field, rather
than the core of the string.

\section{Conclusions}
\label{Ind Conclusions}

In this chapter we have seen that the microphysics of cosmic strings
can be influenced by subsequent phase transitions. Fermion zero modes,
and consequently superconductivity, of the strings can be created
or destroyed by such phase transitions. In determining whether or
not a cosmic string is superconducting it is not enough to just consider
this at formation, but to follow the microphysics through the multiple 
phase transitions that the system undergoes. 

The existence of fermion zero modes at high temperatures enables the
string to carry a current, and thus leads to the formation of vortons
\cite{vorton}. Normally, vortons formed at such high temperatures
result in the theory being ruled out cosmologically
\cite{vorton,vortbounds}. We have shown that it is possible
for currents formed at high energy to dissipate after a subsequent phase
transition if the relevant fermion zero mode does not survive the
phase transition. The vortons then cease to be stable, and cannot be
used to rule out the theory~\cite{acd&wbp3}. This is the
case with $\nu_R$ zero modes on an $SO(10)$ string. Prior to
dissipation there could be a period of vorton domination. After the
phase transition the universe would reheat and then evolve as normal.

To enable a systematic analysis of this effect we have derived a
generalised index theorem. Our index theorem is especially applicable 
to theories where the fermions acquire mass from more than one Higgs field.
We applied the index theorem and also considered spectral flow. As a
result we conjecture that zero modes are destroyed when they mix with
other fermions that acquire mass at a subsequent phase transition from
a non-winding Higgs field.

We applied the index theorem to a realistic GUT, with an $SO(10)$
symmetry group. The theory has 5 distinct types of string, all of
which we considered.

For the abelian string it is the winding number of $\ph{10}$ that
determines the existence of fermion zero modes after electroweak
symmetry breaking. The number of zero modes is 16 times its winding number, so
there will be none for topologically stable strings, which
have $|n|=1$ and $m=0$. Neutrino zero modes can always exist at
high temperatures, but they do not survive the electroweak phase
transition (for $|n|=1$). This has interesting implications as
mentioned above.

In the presence of a nonabelian string, different parts of $\ph{10}$ can
have different winding numbers. In the cases considered it is
generally the part with the lowest winding number which determines the
number of zero modes. Its winding number is equal to $n/2$ rounded down
to the nearest whole integer for $\WR{}$--strings, and for the
$\Yg{}{}$--strings with even $n$. The other strings have no zero modes.
When present there are a total of 12 times this number of
possible fermion zero modes. The fields coupling to $\ph{126}$ (part of
which has winding number 0) do not have such solutions at any
temperature. As with the abelian string there are no zero energy
fermion solutions for topologically stable strings, and so fermion
zero modes on strings are not as common as would be expected.

At high temperatures, it is also possible (for higher $n$) for fermion
fields which couple to the string gauge fields (but not the Higgs
fields) to have zero modes. The number of such
zero modes is not determined by the index theorem as it only
considers the effect of the Higgs field. However, such zero modes are
easy to spot from the field equations. This can occur
for the nonabelian strings, and for the non-conjugate neutrino fields
around abelian strings. This effect is always overridden if
the fermion field couples to a non-zero Higgs field.

We expect qualitatively similar results in most other grand unified theories.

\newcommand{\al}{\alpha}
\newcommand{\ald}{{\dot{\alpha}}}
\newcommand{\sm}{{\sigma^\mu}}
\newcommand{\sigb}[1]{\bar{\sigma}^{{#1}}}
\newcommand{\sgth}{\sigma^\vp}
\newcommand{\sgr}{\sigma^r}
\newcommand{\sgz}{\sigma^z}
\newcommand{\lab}{\bar{\lambda}}
\newcommand{\psb}{\bar{\psi}}
\newcommand{\phb}{\bar{\phi}}
\newcommand{\chb}{\bar{\chi}}
\newcommand{\dmu}{{\partial_\mu}}
\newcommand{\dvp}{{\partial_\vp}}
\newcommand{\evp}[1]{e^{{#1}\vp}}

\chapter{$N=1$ Supersymmetric Abelian Cosmic Strings}
\label{Ch:SUSY}

\section{Introduction}

In this chapter, we investigate the microphysics of cosmic string solutions 
admitted by supersymmetric (SUSY) field theories~\cite{susybook}. This
is important for at least two reasons. First, SUSY field theories
include many popular candidate theories of physics above the
electroweak scale. Second, the recent successes of duality in SUSY
Yang-Mills theories may mean that the physics of nonperturbative
solutions such as topological solitons may be easier to understand
than in non-supersymmetric theories. As in early studies  of non-SUSY
defects~\cite{Kibble}, we work in the context of the simplest models
and in particular with versions of the abelian Higgs model obeying the
supersymmetry algebra with one SUSY generator ($N=1$). We demonstrate
that the particle content and interactions dictated by SUSY naturally
give rise to cosmic string superconductivity in these models. Furthermore,
by using SUSY transformations, we are able to find solutions for the
fermion zero modes responsible for superconductivity in terms of the
background string fields. A special case of the solutions discussed in
this chapter has been obtained in a similar model by other
authors using different techniques~\cite{Garriga}.

The effect of supersymmetry on the $U(1)\times U(1)$ bosonic superconducting
Witten model has been discussed by Morris~\cite{Morris}. The conclusion
there is that simple bosonic superconducting strings do not survive in the
transition to the supersymmetric theory. Instead, in order to implement 
bosonic superconductivity in these theories, it is necessary to construct quite
complicated models with a minimum of five chiral superfields.
In the present chapter we study a different aspect of supersymmetric
cosmic strings, namely the fermionic sector of the theories. In
contrast to the results of ref.~\cite{Morris}, we find that, even in
the simplest SUSY abelian Higgs models, fermionic superconductivity is an 
inevitable result of the couplings and particle content required by the SUSY
algebra. This powerful result leads us to the strong conclusion that
all supersymmetric abelian cosmic strings are superconducting.

The structure of the chapter is as follows. In section~\ref{SS defs} we
present the $N=1$ SUSY abelian Higgs models. Such simple SUSY models
are well-known in particle physics (for example see ref.~\cite{Fayet I}). 
However, we believe the cosmological relevance of the solutions
we explore here to be new. In order to make contact with both the
supersymmetry and cosmology literature, we employ both the superfield
and component formalisms and repeat a number of well-established facts and
conventions for the sake of clarity. Spontaneous symmetry breaking (SSB)
in these models can be implemented in two distinct ways, leading to different
theories with different particle content. We call these distinct
models theory F and theory D respectively to refer to the origin of the
SSB term in the Higgs potential. In section~\ref{SS ZeroModesI} we
focus on theory F. We demonstrate how the cosmic string solution can
be constructed in the bosonic sector, and determine the number of
fermion zero modes using the index theorem in section~\ref{Ind
theorem}. We then employ SUSY transformations to solve these equations
in terms of the background string fields. In section~\ref{SS
ZeroModesII} we repeat the analysis for theory D. The type of symmetry
breaking in theory D is peculiar to theories with an abelian gauge
group and we therefore expect theory F to be more representative of
models with nonabelian gauge groups, such as grand unified
theories. In section~\ref{SS Bogo} we check our results for the
special case discussed in ref.~\cite{Garriga}. In fact, for theory D,
the solutions are already of this special form. Finally, in
section~\ref{SS Conc}, we comment on the possible implications of our
findings.

\section{Supersymmetric Abelian Higgs Models and 
Spontaneous Symmetry Breaking}
\label{SS defs}
Let us begin by defining our conventions. Throughout this chapter we use the
Minkowski metric with signature $-2$, the antisymmetric 2-tensor
$\eps_{21} = \eps^{12} = 1$, 
$\eps_{12} = \eps^{21} = -1$ and the Dirac gamma matrices in the
representation
\be
\gamma^\mu = \bmat{cc} 0 & \sm \\ \sigb{\mu} & 0 \emat \ ,
\ee
with $\sm = (-I,\sig{i})$ and $\sigb{\mu} = (-I,-\sig{i})$, where
$\sigma^i$ are the Pauli matrices.

We consider supersymmetric versions of the spontaneously broken gauged
$U(1)$ abelian Higgs model. These models are related to or are simple
extensions of those found in ref.~\cite{Fayet I}. In superfield notation, 
such a theory consists of a vector superfield $V$ and $m$ chiral
superfields $\Phi_i$, ($i=1\ldots m$), with $U(1)$ charges $q_i$. In
the Wess-Zumino gauge these may be expressed in component notation as
\bea
V(x,\th,\thb) & = & -(\th\sm\thb)A_\mu(x) + i\th^2\thb\lab(x) 
		- i\thb^2\th\la(x) + \frac{1}{2}\th^2\thb^2 D(x) \ ,
\nonumber \\
\Phi_i(x,\th,\thb) & = & \phi_i(y) + \sqrt{2}\th\psi_i(y) + \th^2 F_i(y) \ ,
\label{superfldDef}
\eea
where $y^\mu = x^\mu + i\th\sm\thb$~\cite{susybook}.
Here, $\phi_i$ are complex scalar fields and $A_\mu$ is a vector field. These
correspond to the familiar bosonic fields of the abelian Higgs model.
The fermions $\psi_{i \alpha}$, $\lab_{\alpha}$ and
$\la_{\alpha}$ are Weyl spinors and the complex bosonic fields, $F_i$, and 
real bosonic field, $D$, are auxiliary fields. Finally, $\th$ and $\thb$ are
anticommuting superspace coordinates. In the component formulation of the 
theory one eliminates $F_i$ and $D$ via their equations of motion and
performs a Grassmann integration over $\th$ and $\thb$.
Now define
\bea 
D_\al & = & \frac{\partial}{\partial\th^\al} +
i\sigma^\mu_{\al \ald} \thb^\ald \dmu \ , \nonumber  \\
\bar{D}_\ald & = & -\frac{\partial}{\partial\thb^\ald} -
i\th^\al \sigma^\mu_{\al \ald} \dmu \ , \nonumber \\
W_\al & = & -\frac{1}{4}\bar{D}^2 D_\al V \ , 
\eea
where $D_\al$ and $\bar{D}_\ald$ are the supersymmetric covariant derivatives
and $W_\al$ is the field strength chiral superfield.
The superspace Lagrangian density for the theory is then given by
\be
{\tilde \Lag} = \frac{1}{4}
\left( W^\al W_\al|_{\th^2} + \bar{W}_\ald \bar{W}^\ald|_{\thb^2} \right) 
+ \bar{\Phi}_i e^{g q_i V} \Phi_i |_{\th^2\thb^2}
+  W(\Phi_i)|_{\th^2} +\bar{W}(\bar{\Phi}_i)|_{\thb^2} + \xi D \ .
\label{susyLag}
\ee
In this expression $W$ is the superpotential, a holomorphic function of the 
chiral superfields (i.e.\ a function of $\Phi_i$ only and not $\bar{\Phi}_i$) 
and $W|_{\th^2}$ indicates the $\th^2$ component of $W$.
The term linear in $D$ is known as the Fayet-Iliopoulos
term~\cite{Fayet II}. Such a term can only be present
in a $U(1)$ theory, since it is not invariant under more general gauge
transformations. 

For a renormalizable theory, the most general superpotential is
\be
W(\Phi_i) = a_i \Phi_i + \frac{1}{2}b_{ij} \Phi_i\Phi_j 
			+ \frac{1}{3}c_{ijk} \Phi_i\Phi_j\Phi_k \ ,
\ee
with the constants $b_{ij}$, $c_{ijk}$ symmetric in  their indices.
This can be written in component form as
\be
W(\phi_i, \psi_j, F_k)|_{\th^2}
  = a_i F_i + b_{ij}\left(F_i\phi_j - \frac{1}{2}\psi_i\psi_j\right)
	+ c_{ijk} \left(F_i\phi_j\phi_k - \psi_i\psi_j\phi_k \right) 
\ee
and the Lagrangian \bref{susyLag} can then be expanded in Wess-Zumino gauge
in terms of its component fields using \bref{superfldDef}.
The equations of motion for the auxiliary fields are
\be
F^\ast_i + a_i + b_{ij}\phi_j + c_{ijk}\phi_j\phi_k = 0 \ ,
\ee
\be
D + \xi + \frac{g}{2} q_i \phb_i \phi_i = 0 \ .
\label{deqn}
\ee
Using these to eliminate $F_i$ and $D$ we obtain the Lagrangian density 
in component form as
\be
\Lag = \Lag_B + \Lag_F + \Lag_Y - U \ ,
\label{nsusyLag}
\ee
with
\bea
\Lag_B & = & (D^{i\ast}_\mu \phb_i) (D^{i\mu} \phi_i)
		- \frac{1}{4} F^{\mu\nu}F_{\mu\nu} \ , \\
\Lag_F & = & -i\psi_i \sm D^{i\ast}_\mu \psb_i - i\la \sm \dmu \lab \ , \\
\Lag_Y & = & \frac{ig}{\sqrt{2}} q_i \phb_i \psi_i \la 
 	  - \left(\frac{1}{2}b_{ij} + c_{ijk}\phi_k \right) \psi_i \psi_j 
	  + \cconj \ , \\
   U   & = & |F_i|^2 + \frac{1}{2}D^2
 = |a_i + b_{ij}\phi_j + c_{ijk}\phi_j\phi_k|^2 
     + \frac{1}{2}\left(\xi + \frac{g}{2} q_i \phb_i \phi_i \right)^2 \ ,
\label{Ueqn}
\eea
where $D^i_\mu = \dmu + \frac{1}{2}ig q_i A_\mu$ and 
$F_{\mu\nu} = \dmu A_\nu - \partial_\nu A_\mu$. 

Now consider spontaneous symmetry breaking in these theories.
Each term in the superpotential must be gauge invariant.
This implies that $a_i \neq 0$
only if $q_i =0$, $b_{ij} \neq 0$ only if $q_i + q_j =0$, and 
$c_{ijk} \neq 0$ only if $q_i + q_j + q_k=0$. 
The situation is a little more complicated than in non-SUSY theories, since
anomaly cancellation in SUSY theories implies the existence of more than one 
chiral superfield (and hence Higgs field). In order to break the gauge
symmetry, one may either
induce SSB through an appropriate choice 
of superpotential, or, in the case of the $U(1)$ gauge
group, one may rely on a non-zero Fayet-Iliopoulos term.

We shall refer to the theory with superpotential SSB (and, for simplicity, 
zero Fayet-Iliopoulos term) as theory F and
the theory with SSB due to a non-zero Fayet-Iliopoulos term as theory D. 
Since the
implementation of SSB in theory F can be repeated for more general gauge 
groups, we
expect that this theory will be more representative of general defect-forming
theories than theory D for which the mechanism of SSB is specific to the 
$U(1)$ gauge group.

\section{Theory F: Vanishing Fayet-Iliopoulos Term}
\label{SS ZeroModesI}
The simplest model with vanishing Fayet-Iliopoulos term 
($\xi=0$) and spontaneously broken gauge symmetry contains three
chiral superfields. It is not possible to construct such a model with
fewer superfields which does not either leave the gauge symmetry
unbroken or possess a gauge anomaly. The fields are two charged fields
$\Phi_\pm$, with respective $U(1)$ charges $q_\pm = \pm 1$, and a
neutral field, $\Phi_0$. A suitable superpotential is then
\be
W(\Phi_i) = c \Phi_0 (\Phi_+ \Phi_- - \eta^2) \label{susyW} \ ,
\ee
with $\eta$ and $c$ real.
The potential $U$ is minimised when $F_i=0$ and $D=0$. This occurs when
$\phi_0=0$, $\phi_+ \phi_- = \eta^2$, and $|\phi_+|^2 = |\phi_-|^2$.
Thus we may write $\phi_\pm = \eta e^{\pm i\al}$, where $\alpha$ is some 
function. It is interesting to note that $U$ also has a local minimum
at $\phi_+ = \phi_- = 0$. $\phi_0$ is undetermined. This minimum has
non-zero vacuum energy, and will cause the universe to expand
exponentially. This process is called inflation~\cite{inflation}, and
is a potential solution to many cosmological problems. It will stop as
$\phi_0$ approaches 0.

We shall now seek the Nielsen-Olesen~\cite{Nielsen} solution 
corresponding to an infinite straight cosmic string.  
We proceed in the same manner as for non-supersymmetric
theories. Consider only the bosonic fields (i.e.\ set the fermions to
zero) and in cylindrical polar coordinates $(r,\vp, z)$ write
\bea
\phi_0 & = & 0 \ , \\
\phi_+ \  = \  \phi_-^\ast & = & \eta e^{in\vp}f(r) \ , \\
A_\mu & = & -\frac{2}{g} n \frac{a(r)}{r}\delta_\mu^\vp \ , \\
F_\pm \  = \  D & = & 0 \ , \\
F_0 & = & c \eta^2 (1 - f(r)^2) \ ,
\label{SS StringSol}
\eea
so that the $z$-axis is the axis of symmetry of the defect. The profile 
functions, $f(r)$ and $a(r)$, obey 
\be
f''+\frac{f'}{r} - n^2\frac{(1-a)^2}{r^2} = c^2 \eta^2 (f^2 -1)f \ ,
\label{fEqn}
\ee
\be
a''-\frac{a'}{r} = -g^2 \eta^2(1-a)f^2 \ ,
\label{aEqn}
\ee
with boundary conditions $f(0)=a(0)=0$ and $f(\infty)=a(\infty)=1$.
Note here, in passing, an interesting aspect of topological defects in
SUSY theories. The ground state of the theory is supersymmetric but 
spontaneously breaks the gauge symmetry, while in the core of the defect the
gauge symmetry is restored but, since $|F_i|^2 \neq 0$ in the core, 
SUSY is spontaneously broken there. 

We have constructed a cosmic string solution in the bosonic sector of the 
theory. Now consider the fermionic sector.
With the choice of superpotential \bref{susyW} the component form of the
Yukawa couplings becomes
\be
\Lag_Y = i\frac{g}{\sqrt{2}}
	 \left(\phb_+ \psi_+ - \phb_- \psi_-\right) \la
		 - c \left(\phi_0 \psi_+ \psi_- + \phi_+ \psi_0 \psi_- 
		+ \phi_- \psi_0 \psi_+ \right) + \cconj
\label{SS YukI}
\ee
As with a non-supersymmetric theory, non-trivial zero energy fermion
solutions can exist around the string. The number of solutions can be
determined with the index theorem derived in section~\ref{Ind theorem}.
$z$ and $t$ dependence can easily be added to the solutions. They then
correspond to massless currents flowing along the string. The index
theorem also determines their direction of travel. When the
string solution \bref{SS StringSol} is substituted into 
\bref{SS YukI}, the non-zero terms can be written in the same form as
\bref{Ind Lag}, with
\be
M = \bmat{cccc} 0 & 0 & -c \evp{-i} & i\frac{g}{\sqrt{2}}\evp{-i} \\
		0 & 0 & -c \evp{i} & -i\frac{g}{\sqrt{2}}\evp{i} \\
		-c \evp{-i} & -c \evp{i} & 0 & 0 \\
		i\frac{g}{\sqrt{2}}\evp{-i} & -i\frac{g}{\sqrt{2}}\evp{i}
		& 0 & 0 \emat \eta f(r) \ , \ 
\psi = \bmat{c} \psi_+ \\ \psi_- \\ \psi_0 \\ \la \emat \ .
\ee
This mass matrix is in the same form as \bref{Ind case2}, so the
second form of the theorem (\ref{Ind nlAB},\ref{Ind nrAB}) is used. It
shows that there are $2|n|$ complex (or $4|n|$ real) solutions. Half
their corresponding currents move left along the string, and the other
half move right.

In general, in non-supersymmetric theories, it is difficult to find 
solutions for fermion zero modes in string backgrounds. However, in
the supersymmetric case, SUSY transformations relate the fermionic 
components of the superfields to the bosonic ones and we may use this to
obtain the fermion solutions in terms of the background string fields.
A SUSY transformation is implemented by the operator 
$G=e^{\eps Q + \epb \bar{Q}}$, where $\eps_{\alpha}$ are Grassmann
parameters and $Q_{\alpha}$ are the generators of the SUSY algebra
which we may represent by
\bea
Q_\al & = & \frac{\partial}{\partial\th^\al} 
- i\sigma^\mu_{\al \ald} \thb^\ald \dmu \ , \\
\bar{Q}^\ald & = & \frac{\partial}{\partial\thb_\ald}
		- i\bar{\sigma}^{\mu \ald \al} \th_\al \dmu \ .
\label{SusyTransform}
\eea
In general such a transformation will induce a change of gauge. It is then 
necessary to perform an additional gauge transformation to return to the
Wess-Zumino gauge in order to easily interpret the solutions. For an abelian 
theory, supersymmetric gauge transformations are of the form
\bea
\Phi_i & \longrightarrow & e^{-i\Lambda q_i}\Phi_i \ , \\
\bar{\Phi}_i & \longrightarrow & e^{i\bar{\Lambda} q_i}\bar{\Phi}_i \ , \\
V & \longrightarrow & V + \frac{i}{g}\left(\Lambda - \bar{\Lambda}\right) \ ,
\eea
where $\Lambda$ is some chiral superfield. 

Consider performing an infinitesimal SUSY transformation on 
\bref{SS StringSol}, using $\dmu A^\mu = 0$. The appropriate $\Lambda$ to
return to Wess-Zumino gauge is
\be
\Lambda = ig\epb\sigb{\mu}\th A_\mu (y) 
\ee
The component fields then transform in the following way
\bea
\phi_\pm(y) & \longrightarrow & \phi_\pm(y)  
+ 2i\th\sm\epb D_\mu \phi_\pm(y) \ , \\ 
\th^2 F_0(y) & \longrightarrow & \th^2 F_0(y) + 
2\th\eps F_0(y) \ , \\
-\th\sm\thb A_\mu(x) & \longrightarrow & -\th\sm\thb A_\mu(x) 
+ \frac{1}{2}i\th^2 \thb \sigb{\mu}\sigma^\nu \epb F_{\mu \nu}(x) 
- \frac{1}{2}i\thb^2 \th \sm\sigb{\nu} \eps F_{\mu \nu}(x)
\ .
\eea
Writing everything in terms of the background string fields,
only the fermion fields are affected to first order by the transformation.
They are given by
\bea
\la_\al &\longrightarrow& \frac{2na'}{gr}i(\sgz)^\beta_\al \eps_\beta \ , \\
(\psi_\pm)_\al &\longrightarrow& \sqrt{2} \left(if'\sgr \mp
\frac{n}{r}(1-a)f \sgth\right)_{\al \ald} \epb^\ald \eta \evp{\pm in} \ , \\
(\psi_0)_\al &\longrightarrow& \sqrt{2}c\eta^2(1-f^2)\eps_\al \ ,
\eea
where we have defined
\bea
\sgth & = & \bmat{cc} 0 & -i\evp{-i} \\ i\evp{i} & 0 \emat \ , \\
&& \nonumber \\
\sgr & = & \bmat{cc} 0 & \evp{-i} \\ \evp{i} & 0 \emat \ .
\eea
Let us choose $\eps_{\alpha}$ so that only one component is non-zero.
Taking $\eps_2 = 0$ and $\eps_1 = -i\delta/(\sqrt{2}\eta)$, where $\delta$
is a complex constant, the fermions become
\bea
\la_1 & = & \delta\frac{n\sqrt{2}}{g\eta}\frac{a'}{r} \ , \\ 
(\psi_+)_1 & = & \delta^\ast \left[f'+\frac{n}{r}(1-a)f\right]\evp{i(n-1)} 
\ , \\
(\psi_0)_1 & = & -i\delta c\eta(1-f^2) \ , \\
(\psi_-)_1 & = & \delta^\ast \left[f'-\frac{n}{r}(1-a)f\right]\evp{-i(n+1)} \ .
\label{SS ZMsolF}
\eea
It is these fermion solutions which are responsible for the string
superconductivity.
Similar expressions can be found when $\eps_1 = 0$. It is clear from
these results that the string is not invariant under supersymmetry,
and therefore breaks it. However, since $f'(r), a'(r), 1-a(r)$ and $1-f^2(r)$ 
are all approximately zero outside of the string core, the SUSY breaking and
the zero modes are confined to the string. We note that this method gives us 
two zero mode solutions. Thus, for a winding number one string, we obtain the
full spectrum, whereas for strings of higher winding number, only a partial
spectrum is obtained.

\section{Theory D: Non-vanishing Fayet-Iliopoulos Term}
\label{SS ZeroModesII}

Now consider theory D in which there is just one primary charged chiral
superfield involved in the symmetry breaking and a non-zero Fayet-Iliopoulos
term. In order to avoid gauge anomalies, the model must contain other 
charged superfields. These are coupled to the primary superfield
through terms in the superpotential such that the expectation values
of the secondary chiral superfields are dynamically zero. 
One simple way to do this is add 8 charge $1/2$ chiral fields, $X_a$,
and a superpotential term:
\be
\sum_a g_a \Phi X_a X_a
\ee
The extra fermions cancel the anomaly, and the scalar potential
makes the bosonic parts of $X_a$ zero~\cite{D&S}.
The secondary superfields have no effect on SSB and are invariant under
SUSY transformations. Therefore for the rest 
of this section we shall concentrate on the primary chiral 
superfield which mediates the gauge symmetry breaking. 

Defining $\xi = -\frac{1}{2}g\eta^2$, the theory is spontaneously
broken and there exists a string solution obtained from the ansatz
\bea 
\phi & = & \eta e^{in\vp}f(r) \ , \\
A_\mu & = & -\frac{2}{g} n \frac{a(r)}{r}\delta_\mu^\vp \ , \\
D & = & \frac{1}{2}g\eta^2 (1-f^2) \ , \\
F & = & 0 \ .
\eea
Taking $n>0$, the profile functions $f(r)$ and $a(r)$ obey the first
order equations
\be
f' = n\frac{(1-a)}{r}f \ ,
\label{fEqnII}
\ee
\be
n\frac{a'}{r} = \frac{1}{4}g^2 \eta^2 (1-f^2) \ .
\label{aEqnII}
\ee
Now consider the fermionic sector of the theory, and apply the index
theorem \bref{Ind n2fer}. For $n>0$ there are $2n$ zero modes, which
move left. Performing a SUSY transformation, again using $\Lambda$ as
the gauge function to return to Wess-Zumino gauge, gives to first order
\bea
\la_\al &\longrightarrow& 
	\frac{1}{2}g\eta^2(1-f^2)i(I+\sgz)_\al^\beta\eps_\beta \ ,\\
\psi_\al &\longrightarrow& 
\sqrt{2}\frac{n}{r}(1-a)f(i\sgr-\sgth)_{\al \ald} \epb^\ald \eta \evp{in} \ . 
\eea
If $\eps_1=0$ both these expressions are zero. The same is true of all
higher order terms, and so the string is invariant under the
corresponding transformation. For other $\eps$, taking 
$\eps_1=-i\delta/\eta$ gives
\bea 
\la_1  &=& \delta g \eta(1-f^2) \ ,\\
\psi_1 &=& 2\sqrt{2} \delta^\ast \frac{n}{r}(1-a)f \evp{i(n-1)} \ . 
\eea
Similar results are obtained when $n<0$. In this case the zero modes
move right, and the non-zero fields are $\la_2$ and $\psi_2$.

In this theory supersymmetry is only half broken inside the
string. This is in contrast to theory F which fully breaks
supersymmetry in the string core. The theories also differ in that
theory D's zero modes will only travel in one direction, while the
zero modes of theory F  (which has twice as many) travel in both
directions. In both theories the zero modes and SUSY breaking are
confined to the string core.

\section{The Super-Bogomolnyi Limit}
\label{SS Bogo}
In non-supersymmetric theories it is usually difficult to find
solutions for fermion zero modes on cosmic string backgrounds. In such
theories one can, however, often obtain solutions in the 
{\it Bogomolnyi limit} which, in our theory, corresponds to choosing

\be
2c^2 = g^2 \ .
\label{BL}
\ee
In this limit, the energy of the vortex saturates a topological bound, there
are no static forces between vortices and the 
equations of motion for the string fields reduce to a pair of coupled first
order differential equations.
It is a useful check of the solutions obtained in the previous sections to
confirm that they reduce to those already known in the Bogomolnyi limit.

Imposing~(\ref{BL}) equations~(\ref{fEqn},\ref{aEqn}), together with
the requirement of finite energy, become
\bea
f' & = & n \frac{f}{r}(1-a) \ , \\
n\frac{a'}{r} & = & c^2\eta^2(1-f^2) \ .
\label{Bogomolnyi}
\eea
Note that these are similar to~(\ref{fEqnII},\ref{aEqnII}). We see
that all solutions to theory D are automatically Bogomolnyi solutions.
Imposing \bref{BL} on \bref{SS ZMsolF} gives the following solutions. 
\bea
\la_1 & = & \delta c\eta(1-f^2) \ , \\
(\psi_+)_1 & = & 2\delta^\ast n\frac{f}{r}(1-a)\evp{i(n-1)} \ , \\
(\psi_0)_1 & = & -i\delta c\eta(1-f^2) \ , \\
(\psi_-)_1 & = & 0 \ .
\eea
This limit, with $n=1$, was considered for a similar theory by Garriga
and Vachaspati~\cite{Garriga} and the above results are in agreement
with theirs. This is a useful check of the techniques we use.

\section{Concluding Remarks}
\label{SS Conc}
We have investigated the structure of cosmic string solutions to
supersymmetric abelian Higgs models. For completeness we have analysed two 
models, differing by their method of spontaneous symmetry
breaking. However, we expect theory F to be more representative of general
defect forming theories, since the SSB employed there is not specific to 
abelian gauge groups. 

We have shown that although SUSY remains unbroken outside the
string, it is broken in the string core (in contrast to the gauge
symmetry which is restored there). In theory F supersymmetry is broken 
completely in the string core by a non-zero $F$-term, while in theory D
supersymmetry is partially broken by a non-zero $D$-term. We have
demonstrated that, due to the particle content and couplings dictated
by SUSY, the cosmic string solutions to both theories are
superconducting in the Witten sense. We believe this to be quite a
powerful result, that all supersymmetric abelian cosmic strings are
superconducting due to fermion zero modes. Note that this result is in
contrast to those obtained in earlier analyses of purely bosonic
superconductivity in Witten-type SUSY models~\cite{Morris}. An
immediate and important application of the results of the present
chapter is that SUSY GUTs which break to the standard model and yield
abelian cosmic strings (such as some breaking schemes of $SO(10)$)
may face strong constraints from cosmology~\cite{vortbounds}. The
existence of zero modes around SUSY monopoles has been investigated
previously~\cite{mono}. However, in this chapter we have considered
cosmic strings since, unlike monopoles, supercurrents on strings are
cosmologically significant.

Although explicitly solving for such zero modes in the case of
non-supersymmetric theories is difficult, in the models we study it is
possible to use SUSY transformations to relate the functional form of
the fermionic solutions to those of the background string fields, which are
well-studied. For theory D the solutions all obey the Bogomolnyi equations 
exactly, and for theory F we have also checked that the solutions we find 
reduce to those already known in the special case of the Bogomolnyi limit.

While we have performed this first analysis for the toy model of an abelian 
string, we expect the techniques to be quite general and in fact to be more
useful in nonabelian theories for which the equations for the fermion zero
modes are significantly more complicated. The question of superconductivity
in nonabelian SUSY cosmic strings is considered in the next
chapter. We also examine the effect of soft SUSY breaking in all of
the models.

\newcommand{\Ch}{\Phi}
\newcommand{\Chc}{\tilde{\Phi}}
\newcommand{\bfCh}{{\bf \Phi}}
\newcommand{\bfChc}{{\bf \tilde{\Phi}}}
\newcommand{\Sca}{S}
\newcommand{\Scac}{{\tilde{S}}}
\newcommand{\Scao}{{S_0}}

\newcommand{\phc}{\tilde{\phi}}
\newcommand{\bfphi}{\mbox{\boldmath $\phi$}}
\newcommand{\bfphc}{\mbox{\boldmath $\phc$}}
\newcommand{\psc}{\tilde{\psi}}
\newcommand{\bfpsi}{\mbox{\boldmath $\psi$}}
\newcommand{\bfpsc}{\mbox{\boldmath $\psc$}}
\newcommand{\Fc}{\tilde{F}}

\newcommand{\sca}{s}
\newcommand{\scac}{\tilde{s}}
\newcommand{\scao}{s_0}
\newcommand{\scf}{\omega}
\newcommand{\scfx}[1]{\omega_{({#1})}}
\newcommand{\scfc}{\tilde{\omega}}
\newcommand{\scfo}{\omega_0}

\newcommand{\e}[1]{{\bf e}_{{#1}}}
\newcommand{\ept}{e^{i\vp}}
\newcommand{\emt}{e^{-i\vp}}

\newcommand{\sbrkA}{{\cal A}}
\newcommand{\mtil}{\tilde{m}}
\newcommand{\xtil}{\tilde{\xi}}

\chapter{Cosmic Strings and Supersymmetry Breaking}
\label{Ch:SUSY2}

\section{Introduction}

In recent years, supersymmetry (SUSY) has become increasingly favoured 
as the theoretical structure underlying fundamental particle interactions.
In this light it is natural to investigate possible cosmological implications
of SUSY. 

In chapter~\ref{Ch:SUSY} we discussed the effect of SUSY on
the microphysics of simple cosmic string solutions of abelian field theories.
In particular we developed and applied the technique of SUSY transformations 
to investigate the form of fermionic zero modes, required by SUSY, which
lead to cosmic string superconductivity. In the present work we extend our
original ideas to a more general class of field theories, namely those with
a nonabelian gauge group. Since nonabelian gauge theories underlie modern
particle physics and, in particular, unified field theories, this class
of theories is a realistic toy model for grand unified theories (GUTs). The 
particular example we examine, 
$SU(2)\times U(1)\rightarrow U(1)\times Z_2$, admits two types
of string solution, one abelian and the other nonabelian. This model has a
similar structure to $SO(10)$ and should provide insight into cosmic
strings in SUSY GUTs. Most of the features exhibited by this theory will
also appear in larger nonabelian theories. We apply the
technique of SUSY transformations to the nonabelian case and extract the 
behaviour of the zero modes as functions of the background string fields. We
then compare the results to those obtained in chapter~\ref{Ch:SUSY} for 
abelian strings.

Since SUSY is clearly broken in the universe today, it is important to know
how the SUSY zero modes behave when soft-SUSY breaking occurs.
We investigate this for both the abelian and nonabelian strings by explicitly 
introducing soft-SUSY breaking terms into the theory. The result is that
all the zero modes are destroyed in almost all the theories, the
exception being when a Fayet-Iliopoulos term is used to break the gauge
symmetry in an abelian model. We briefly comment on the physical reasons
for this and show how the effect may be seen through the breakdown of an
appropriate index theorem.

These results have a cosmological significance since fermion zero modes on the
string can be excited, causing a current to flow along the string. The
string then behaves as a perfect conductor. The existence of charge carriers
changes the cosmology of cosmic strings. In particular, they can stabilise
string loops, resulting in the production of vortons~\cite{vorton}.
Such vortons can dominate the energy density of the Universe, and have been
used to constrain GUT models with current-carrying strings~\cite{vortbounds}.
However, if the zero modes are destroyed at the SUSY breaking energy scale, 
then the current, and hence vortons, can dissipate. Thus, the underlying 
theory may be cosmologically viable.

The plan of this paper is as follows. In section~\ref{SS2 SU2xU1} we
construct a simple supersymmetric model based on the group
$SU(2)\times U(1)$ and display the abelian and nonabelian string
solutions. In section~\ref{SS2 ZeroModes}, we use an index theorem to
find its fermion zero modes. We then use SUSY transformations to
obtain the zero modes in terms of the background string fields. Soft
SUSY breaking terms are introduced in section~\ref{SS2 SUSYbreak}, and their 
effect on the string solution is considered in section~\ref{SS2 U1SSbreak}. In
section~\ref{SS2 breakzeromode} the effect of SUSY-breaking on the
zero modes is analysed using the index theorem. In this
section we consider both string solutions for the $SU(2)\times U(1)$
model and also for the $U(1)$ theories discussed in
chapter~\ref{Ch:SUSY}. For the nonabelian string the SUSY breaking
terms destroy the zero modes, while for the other string solutions the
situation is more complicated.

\section{An SU(2) $\times$ U(1) Model}
\label{SS2 SU2xU1}
There exist many nonabelian theories with breaking schemes giving rise to
cosmic strings. In general both abelian and nonabelian strings can be 
produced in such a process, depending on which part of the vacuum manifold
is involved in the winding. 

In this section we consider a simple example in which the gauge group 
$SU(2) \times U(1)$ is spontaneously broken down to the group 
$U(1) \times Z_2$ via the superpotential
\be 
W = c_1 \Scao (\bfCh \cdot \bfChc - \eta^2) 
    + c_2 (\Sca \bfCh^T \Lambda \bfCh + \Scac \bfChc^T \Lambda \bfChc) \ .
\ee
The chiral superfields $\Ch_i(\phi_i,\psi_i,F_i)$ and 
$\Chc_i(\phc_i,\psc_i,\Fc_i)$ are $SU(2)$ triplets with $U(1)$ 
charges $\pm1$ respectively. The other chiral superfields,
$\Scao(\scao, \scfo, F_\Scao)$, $\Sca(\sca, \scf, F_\Sca)$ and 
$\Scac(\scac, \scfc, F_\Scac)$, are $SU(2)$ scalars with $U(1)$ charges 0, 
$-2$ and $+2$ respectively. Finally, defining $T^4 = \sqrt{2/3} I$, 
the vector supermultiplets are $V^a(A^a_\mu,\la^a,D^a)$, $a=1, \ldots, 4$.
Since the constant matrix $\Lambda$ satisfies 
$\Lambda T^i = -(T^i)^\ast \Lambda$ 
($i=1, \ldots, 3$), and the $SU(2)$ gauge transformations are 
$\delta \bfCh = i T^a n^a \bfCh$ and 
$\delta \bfChc = -i T^{a\ast} n^a \bfChc$, the superpotential is gauge 
invariant.
The scalar potential, derived in the standard manner~\cite{susybook}, is then
\bea
U &=& c_1^2 | \bfphi \cdot \bfphc - \eta^2 |^2
    + c_2^2 | 2 \phi_1 \phi_3  - \phi_2^2 |^2
    + c_2^2 | 2 \phc_1 \phc_3  - \phc_2^2 |^2
\nonumber \\ &&
    {}+ | c_1 \scao \bfphc + 2 c_2 \sca \Lambda \bfphi |^2 
    + | c_1 \scao \bfphi + 2 c_2 \scac \Lambda \bfphc |^2
 + \frac{e^2}{3} (|\bfphi|^2 - |\bfphc|^2 - 2|\sca|^2 + 2|\scac|^2 )^2
\nonumber \\ &&
{}+ e^2 |(\phi_1 + \phi_3)\phi_2^\ast - (\phc_1 + \phc_3)\phc_2^\ast|^2
 + \frac{e^2}{2} (|\phi_1|^2 - |\phi_3|^2 - |\phc_1|^2 +|\phc_3|^2 )^2 
\ .  
\eea
This is minimised when all fields are zero except $\phi_1 = \phc_1 =
\eta$ or at any (broken) gauge transformation of this. We note also
that the theory has a local minimum with $\bfphi = \bfphc = 0$ and
that this structure can give rise to hybrid
inflation~\cite{inflation}. This is true even for the abelian theory
described in chapter~\ref{Ch:SUSY2}. In both cases inflation ends with defect
formation.

As we mentioned above, there are abelian and nonabelian string
solutions to this theory. The abelian solution is obtained from the ansatz
\bea
\phi_1 \ = \ \phc^\ast_1 &=& \eta f(r) \ept \ ,\\
A_\vp   &=& \frac{a(r)}{er} \sqrt{\frac{3}{5}} T^G \ ,\\
F_\Scao &=& c_1 \eta^2 (1 - f(r)^2)  \ ,
\eea
where $T^G = \sqrt{\frac{3}{5}}T^3 + \sqrt{\frac{2}{5}}T^4$. All other
fields are zero and the profile functions $a$ and $f$ obey the
boundary conditions $f(0)=a(0)=0$ and $f(\infty)=a(\infty)=1$.

The nonabelian solution is obtained from the ansatz
\bea
\bfphi \ = \ \bfphc^\ast &=& 
		\eta \left\{ \frac{1}{2} (\ept \e{+} + \emt \e{-}) f(r) 
			+ \frac{1}{\sqrt{2}} \e{0} g(r) \right\}\ , \\
A_\vp   &=& \frac{a(r)}{er} T^1 \ ,  \\
F_\Scao &=& \frac{1}{2} c_1 \eta^2 (2 - f(r)^2 - g(r)^2) \ , \\
F_\Sca \ = \ F_\Scac &=& \frac{1}{2} c_2 \eta^2 (f(r)^2 - g(r)^2) \ ,
\eea  
where $\e{k}$ are unit vectors obeying $T^1 \e{k} = k\e{k}$. In this
case $g(0)$ is finite, $g(\infty)=1$ and $f$ and $a$ 
obey the same boundary conditions as in the abelian case. 

Note that $f$, $g$ and $a$ are solutions to simple coupled second
order ordinary differential equations. Their forms can be obtained
numerically and are well known~\cite{Ma}.

\section{SUSY Transformations and Zero Modes}
\label{SS2 ZeroModes}

The string solutions obtained above have all the fermion fields set to zero.
In this section we investigate what happens when these fields are excited in
the background of the cosmic string. We can find the number of zero
modes with the index theorem in chapter~\ref{Ch:Index}.

We already know that there must exist fermion zero modes on the string. Rather
than attempting to solve the difficult fermion equations of motion to obtain 
them, we shall use the technique described in
section~\ref{SS ZeroModesI} which exploits the power of SUSY to obtain
some of the solutions.

\subsection{Abelian string}
The relevant part of the Lagrangian is the Yukawa sector which is entirely
determined by supersymmetry. In the abelian case, the non-zero Yukawa 
couplings are
\setstrut{17 pt}
\bea 
\Lag_Y &=& -\left[
	 c_1 (\ept \psc_1 + \emt \psi_1) \scfo
 	 +\sqrt{\frac{5}{6}}ie (\emt \psi_1 - \ept \psc_1) \la^G 
       \right. \nonumber \\ && \left. \hspace{0.3 in}
	{}+ 2 c_2 (\ept \psi_3 \scf + \emt \psc_3 \scfc) 
       \right. \nonumber \\ && \left. \hspace{0.3 in}
	 {}+ \frac{ie}{\sqrt{2}} (\emt \la_+ \psi_2 + \ept \la_- \psc_2)
       \mystrut \right] \eta f(r) + \cconj \ , 
\label{abYuk}
\eea
where $\la_\pm = (\la^1 \mp i\la^2)/\sqrt{2}$. With respect to the
string generator, the only fields with non-zero eigenvalues are
$\psi_1$ and $\la_+$ (eigenvalue 1) and $\psc_1$ and $\la_-$
(eigenvalue $-1$). The Yukawa Lagrangian can be split up into 5 independent
parts. Applying the index theorem (\ref{Ind nlAB},\ref{Ind nrAB}) to
these reveals that there are a total of three left moving and three
right moving complex zero modes.

Now, following the techniques in section~\ref{SS ZeroModesI}, we
perform an infinitesimal SUSY transformation (with Grassmann parameter
$\eps_{\alpha}$) followed by a gauge transformation to return to the
Wess-Zumino gauge. The string fields all transform quadratically and
so are unchanged to first order. However, the fermions transform
linearly and, in terms of the background string fields, it is possible
to find two complex (or 4 real) fermion zero mode solutions given by
\bea
\scfo &=& \sqrt{2} c_1 \eta^2 (1 - f^2) \eps \ , \\
\la^G &=& -i\sqrt{\frac{3}{5}} \frac{a}{er} \sig{z} \eps \ , \\
\psi_1 &=& i\sqrt{2} \eta \ept
	\left( f' \sig{r} + i\frac{1-a}{r}f \sig{\th} \right) \epb \ , \\
\psc_1 &=& i\sqrt{2} \eta \emt
	\left( f' \sig{r} - i\frac{1-a}{r}f \sig{\th} \right) \epb \ .
\label{abzeromodes}
\eea
Setting either component of $\eps$ to zero gives one of the zero modes. 
One is left moving, the other moves right.

\subsection{Nonabelian string}
In the nonabelian case it is convenient to split $\bfpsi$, $\bfpsc$
and $\la^i$ into eigenvectors of $T^1$, and label them by their
eigenvalues.  Defining $\chi_i = (\psi_i + \psc_i)/\sqrt{2}$, 
$\zeta_i = (\psi_i - \psc_i)/\sqrt{2}$ and 
$\scfx{\pm} = (\scf \pm \scfc)/\sqrt{2}$, the non-zero Yukawa couplings are
\bea 
\Lag_Y &\!\!=\!\!& -c_1\eta\left[ \chi_0 g(r) + 
	 \frac{1}{\sqrt{2}}(\chi_+\emt + \chi_-\ept)f(r)\right] \scfo
	 - \frac{ie\eta}{2} (\chi_+\emt - \chi_-\ept)f(r) \la^0
       \nonumber \\ &&
	 {}- c_2 \eta \left[-\sqrt{2}\chi_0 g(r)
	 + (\chi_+\emt + \chi_-\ept)f(r)\right]\scfx{+}
       \nonumber \\ &&
	 {}- c_2\eta \left[-\sqrt{2}\zeta_0 g(r)
	 + (\zeta_+\emt + \zeta_-\ept)f(r)\right]\scfx{-}
       \nonumber \\ &&
	 {}+ \frac{e\eta}{2}\left[\sqrt{2} (\zeta_-\la^+ - \zeta_+\la^-) g(r)
	 + \zeta_0(\la^+ \emt - \la^- \ept)f(r) \right]
       \nonumber \\ &&
	 {}- \frac{ie\eta}{\sqrt{6}}\left[ \sqrt{2} \zeta_0 g(r)
	 + (\zeta_+ \emt + \zeta_- \ept)f(r)\right] \la^4 + \cconj \ .
\label{nabLagY}
\eea
In this case $\chi_\pm$, $\zeta_\pm$ and $\la^\pm$ have eigenvalues
$\pm 1$. Applying (\ref{Ind nlAB},\ref{Ind nrAB}) to the two
irreducible parts of \bref{nabLagY} shows that there are just two complex
zero modes, moving in opposite directions.

Once again performing an infinitesimal SUSY transformation and a (nonabelian)
gauge transformation we obtain two complex zero modes
\bea
\scfo &=& \frac{1}{\sqrt{2}} c_1 \eta^2 (2 - g^2 - f^2) \eps \ , \\
\scfx{+} &=& c_2 \eta^2 (g^2 - f^2) \eps \ , \\
\la^0 &=& -i \frac{a}{er} \sig{z} \eps \ , \\
\chi_+ &=& i \eta \ept
	\left( f' \sig{r} + i\frac{1-a}{r}f \sig{\th} \right) \epb \ , \\
\chi_- &=& i \eta\emt
	\left( f' \sig{r} - i\frac{1-a}{r}f \sig{\th} \right) \epb \ , \\
\chi_0 &=& i \sqrt{2}\eta g' \sig{r} \epb \ . 
\label{nabzeromodes}
\eea
Thus in this case there are no zero modes beyond those implied by
SUSY, in contrast to the abelian case. This is related to the fact
that there are components of the Higgs fields that do not wind in the
nonabelian case. 

\section{Soft SUSY Breaking}
\label{SS2 SUSYbreak}

Perhaps the most attractive feature of supersymmetry arises from the
non-renormalisation theorems, which provide a solution to the
hierarchy problem. The mass of the electroweak Higgs field receives
quadratically divergent radiative corrections from the particles which
it couples to. In general this leads to conflict with
experiment. In a supersymmetric theory the contributions from the
fermion and boson fields cancel each other exactly. This ensures that
quadratic divergences are absent, and so any tree-level hierarchy of
scales is  protected from receiving quantum corrections.

Supersymmetry is not observed at everyday temperatures, and so must
have been broken as the universe cooled. It is crucial
that the quadratic divergences remain absent from the theory, so that
the hierarchy problem is still avoided. This is achieved by adding
only `soft' SUSY breaking terms. These are all either mass terms, or
couplings with positive mass dimension~\cite{softSSbreak}. Obviously
they must be non-invariant under SUSY too. All the allowed terms are
equivalent to one of three types. They lead to the following changes
to the Lagrangian

\begin{enumerate}
\item Arbitrary mass terms for scalar particles are added to the
scalar potential.

\item Bilinear and trilinear scalar terms in the superpotential, plus
their Hermitian conjugates, are added to the scalar potential with
arbitrary coupling.

\item Mass terms for the gauginos are added to the Lagrangian density.
\end{enumerate}

For a general theory the allowed terms are
\be
-\Lag_{\rm soft} = m^2_{ij} \phi^\ast_i \phi_j + \left(b'_{ij} \Phi_i \Phi_j 
	+ c'_{ijk} \Phi_i \Phi_j \Phi_k + \cconj \right) 
	+ m_\la \la^a \la^a + \cconj \ .
\label{SSbreak}
\ee
Of course, only those soft SUSY breaking terms which respect the
gauge and other symmetries of the theory are allowed. Because of this
the bilinear and trilinear terms resemble those of the superpotential, and are
conventionally written as multiples of them. Experiment suggests that
the SUSY breaking scale is around 1~TeV.

\section{SUSY breaking and an Abelian String Model}
\label{SS2 U1SSbreak}

In section~\ref{SS ZeroModesI} we referred to an abelian theory in
which the gauge symmetry is broken via an $F$ term as `theory
$F$'. The corresponding superpotential was
\be
W = c \Phi_0 (\Phi_+ \Phi_- - \eta^2) \ .
\ee
The allowed soft SUSY breaking contributions to the potential are
\be
m^2_i |\phi_i|^2 + c \sbrkA \phi_0 \phi_+ \phi_- + \cconj \ .
\ee

Defining $m^2 = (m^2_+ + m^2_-)/2$ 
and $\xi = (m^2_+ - m^2_-)/g$, the scalar potential becomes
\bea
  V &=& c^2 |\phi_+ \phi_- - \eta^2|^2 
+ |\phi_0|^2(c^2(|\phi_-|^2 + |\phi_+|^2) + m^2_0)
+ m^2 (|\phi_+|^2 + |\phi_-|^2)  \nonumber \\ &&
{}+ \frac{g^2}{8} \left(|\phi_+|^2 - |\phi_-|^2 + 
	 \frac{2}{g}\xi \right)^2 
+ c \sbrkA \phi_0 \phi_+ \phi_- + c \sbrkA (\phi_0 \phi_+ \phi_-)^\ast
- \frac{\xi^2}{2} \ .
\label{SS2 brkpotential}
\eea
We see that $\xi$ acts like a Fayet-Iliopoulos term. Any stationary
points of the potential will have
\be
\phi_0 = -\frac{c\sbrkA \phi_+^\ast \phi_-^\ast}{m^2_0 
	+ c^2 (|\phi_+|^2+ |\phi_-|^2)} \ .
\ee
Thus $\phi_0$ will acquire a non-zero expectation value if the
trilinear term is present. 

It is not generally possible to find the minimum of
\bref{SS2 brkpotential} analytically. Instead we will consider a couple of
special cases. It is convienient to define $\mtil = m/(c\eta)$, 
$\xtil = \xi/(g\eta^2)$ and $\beta = 2c^2/g^2$.

If $\sbrkA = 0$ the potential has stationary points at $\phi_i = 0$ and
\bea
\phi_\pm &=& \eta e^{\pm ia} 
	\sqrt{1 - \frac{\mtil^2}{\cosh \chi}} \, e^{\pm \chi/2} \ , \nonumber \\ 
\phi_0 &=& 0 \ ,
\eea
where $a$ is real and $\chi$ satisfies
\be
\sinh \chi + \mtil^2 (\beta - 1)\tanh \chi + \xtil = 0 
\label{chiEq1}
\ee
and
\be
\cosh \chi - \mtil^2 > 0 \ .
\label{chiEq2}
\ee
If $\mtil^2 \geq 1$ and $|\xtil/\beta|^2 \leq (\mtil^4 -1)$ then
\bref{chiEq1} has no real solutions which also satisfy
\bref{chiEq2}. The only minimum of $V$ is then $\phi_i = 0$, and so
the SUSY-breaking terms restore the broken $U(1)$ gauge symmetry. 

If $\mtil^2 < 0$, $\mtil^2(\beta-1) < -1$ and 
$|\xtil|^{2/3} < [-\mtil^2(\beta-1)]^{2/3} - 1$ then \bref{chiEq1} has
3 solutions, two of which are minima. Since the potential has
disconnected minima, it is possible for domain walls to form. Domain
wall formation occurs when a discrete vacuum symmetry is broken. In
this case it is the $\phi_+ \leftrightarrow \phi_-^\ast$
symmetry. If $\xi=0$ the minima have the same energy. The domain walls are then
stable, and will come to dominate the energy density of the universe
as it evolves, conflicting with observations~\cite{domwall}. If $\xi \neq 0$ 
the minima will have different energies, and the walls can decay by
quantum tunnelling. If this happened rapidly enough, conflict with
observations could be avoided. 

Other values of the parameters give a unique global minimum with
$\phi_\pm \neq 0$. Since we require the phase transition not to be
reversed, and that no domain walls form, the range of allowed
parameters is restricted. However, we expect $\mtil, \xtil \ll 1$
in general, so these restrictions are not very significant. If the
phase transition and supersymmetry breaking occur close together then
$\mtil, \xtil \sim 1$, and the allowed parameter range may be
significantly reduced.

If $\xi = m^2_0 = 0$, the vacuum is given by
\bea
\phi_\pm &=& \eta e^{\pm ia}
	\sqrt{1 + \frac{\sbrkA^2 - 4 m^2}{(2c\eta)^2}} \ , \nonumber \\ 
\phi_0 &=& -\frac{\sbrkA}{2c} \ .
\eea

If $\sbrkA = 0$ and $m^2_0 < 0$ the potential \bref{SS2 brkpotential}
resembles that of the bosonic superconductivity model in
ref.~\cite{Witten}. Although $\phi_0 = 0$ outside of a cosmic string,
it may be energetically favourable for it to be non-zero
inside. Excitations of this condensate could then form currents
flowing along the string, although these are likely to be small.

\section{Fermion Zero Modes after SUSY breaking}
\label{SS2 breakzeromode}

We will now consider the effect of soft SUSY breaking on the fermion
zero modes.
Since the techniques we have used for finding the zero modes are
strictly valid only when SUSY is exact, it is necessary to investigate
the effect of these soft terms on the zero modes we have identified.
As we have already commented, the existence of the zero modes can be seen as 
a consequence of the index theorem in section~\ref{Ind theorem}. The index is
insensitive to the size and exact form of the Yukawa couplings, as
long as they are regular for small $r$, and tend to a constant at
large $r$. In fact, the existence of zero modes relies only on
the existence of the appropriate Yukawa couplings and that they have the 
correct $\vp$ dependence. Thus there can only be a change in the number of zero
modes if the soft breaking terms induce specific new Yukawa couplings
in the theory and it is this that we must check for. Further, it was
conjectured in section~\ref{Ind spec} that the destruction of a zero
mode occurs only  when the relevant fermion mixes with another
massless fermion.

\subsection{U(1) Abelian models}

\subsubsection{Theory F}

We saw in section~\ref{SS2 U1SSbreak} that although the scalar mass
terms will alter the values of $\phi_+$ and $\phi_-$, they do not
generally produce any new Yukawa terms. Thus these soft SUSY-breaking
terms have no effect on the existence of the zero modes. The possible exception
was a negative $m_0^2$ term. This may lead to a $\phi_0$ condensate
inside the string which will produce a $\psi_+ \psi_-$ term.

The presence of the trilinear term gives $\phi_0$ a non-zero 
expectation value everywhere, and hence produces a Yukawa term
coupling the $\psi_+$ and $\psi_-$ fields. This destroys all the zero
modes in the theory since the left and right moving zero modes mix. 

For completeness note that a gaugino mass term also mixes the left and right 
zero modes, aiding in their destruction.
  
In terms of the index theorem, the change in the number of zero modes
arises because (\ref{Ind nl3},\ref{Ind nr3}) applies after the SUSY
breaking, while (\ref{Ind nlAB},\ref{Ind nrAB}) applied before
it. Although the fermion eigenvalues do not change, the expression
relating them to the zero modes does.
  
\subsubsection{Theory D}

The $U(1)$ theory with gauge symmetry broken via a Fayet-Iliopoulos term and 
no superpotential is simpler to analyse. New Higgs mass terms 
have no effect on the number of zero modes, as in the above case, and
there are no trilinear terms. Further, although the gaugino mass terms
also affect the form of the zero mode solutions, they do not affect
their existence, and so in theory $D$ the zero modes remain even
after SUSY breaking.

\subsection{SU(2) $\times$ U(1) model}

\subsubsection{Abelian strings}

The effect of soft SUSY breaking terms on the zero modes which were
found analytically in \bref{abzeromodes} is identical to the
equivalent $U(1)$ theory. Thus, all SUSY zero modes are destroyed in
this case. In this larger theory there are also other, non-SUSY, zero modes.
Not all of these are destroyed by a gaugino mass term, as some
do not involve the gaugino fields. However, if the trilinear terms (or
possibly a negative $m^2_0$ term) give $\scao$ a non-zero expectation
value, all the zero modes are destroyed. The extra Yukawa terms mean
there are fewer irreducible parts to the fermion mass matrices. This
results in more terms cancelling in (\ref{Ind nlAB},\ref{Ind nrAB}),
reducing the number of zero modes. As in the other cases, the physical
reason behind the destruction of the zero modes is that left and right
movers mix.

\subsubsection{Nonabelian strings}

As in the other cases above, non-zero gaugino mass or trilinear terms
destroy the zero modes that were found with SUSY transformations 
\bref{nabzeromodes}. Similarly (\ref{Ind nl3},\ref{Ind nr3}) are
required instead of (\ref{Ind nlAB},\ref{Ind nrAB}), implying that the
left and right moving modes mix. For  nonabelian strings these are the
only zero modes and so none remain after SUSY breaking.

\section{Comments and Conclusions}

We have examined the microphysics of abelian and nonabelian cosmic string 
solutions to the $SO(10)$ inspired supersymmetric $SU(2)\times U(1)$ model. 
By performing infinitesimal SUSY transformations on the background string
fields we have obtained the form of the fermionic zero modes responsible for
cosmic string conductivity. These solutions may be compared to those
found in chapter~\ref{Ch:SUSY}. 

Our results mean that fermion zero modes are always present around cosmic 
strings in SUSY. We conjecture that in theories with $F$-term gauge symmetry 
breaking, the zero modes given by SUSY always occur in pairs, one left and one
right moving. It also seems likely that such theories always have
hybrid inflation.

Furthermore, in the abelian case there were additional zero
modes that were not a consequence of supersymmetry. We expect that
similar extra zero modes will be present in a larger theory, even in
the nonabelian case. 

We have also analysed the effect of soft SUSY breaking on the existence 
of fermionic zero modes. The $SU(2)\times U(1)$ model and two simple abelian
models were examined. In all cases Higgs mass terms did not affect the
existence of the zero modes. In the theories with $F$-term symmetry
breaking, gaugino mass terms destroyed all zero modes which involved
gauginos and trilinear terms created extra Yukawa couplings which
destroyed all the zero modes present. 

All the theories with $F$-term symmetry breaking feature a chargeless
scalar field. If the SUSY-breaking contribution to its mass term is
negative, a chargeless condensate may form inside the string. This
would allow bosonic currents to flow along the
string. Additionally the condensate would destroy the
fermion zero modes, in a similar way to any trilinear terms.

In the abelian theory with $D$-term symmetry breaking, the zero modes
were unaffected by the SUSY breaking terms. It was conjectured
in section~\ref{Ind spec} that zero modes would only disappear when they mixed
with another massless fermion field and this is consistent with the
results obtained in this chapter. If the remaining zero modes survive
subsequent  phase transitions, then stable vortons could result. Such
vortons would dominate the energy density of the universe, rendering
the underlying GUT cosmologically problematic.

Therefore, although SUSY breaking may alleviate the cosmological disasters 
faced by superconducting cosmic strings~\cite{vortbounds}, there are classes of
string solution for which zero modes remain even after SUSY breaking. It
remains to analyse all the phase transitions undergone by specific SUSY
GUT models to see whether or not fermion zero modes survive down to the
present time. If the zero modes do not survive SUSY breaking, the
universe could experience a period of vorton domination beforehand,
and then reheat and evolve as normal afterwards.

If the zero modes do occur in pairs (one left and one right moving) in
$F$-term gauge symmetry breaking, it is possible that they could 
scatter off each other~\cite{BarrMatheson}. This would cause the current to 
decay, and could stop vorton domination.

There is the possibility that even if zero modes
are destroyed they become low-lying bound states. Such bound states may
still be able to carry a persistent current. If this is the case, even 
such theories may not be safe cosmologically. Work on this is under
investigation (and see chapter~\ref{Ch:Bound}).

It may also be possible to extend our analysis of the effect of SUSY breaking 
on the bosonic fields. The full potential is very complex, even in the
abelian case. It may be worth using some sort of approximation or a
numerical solution. We have already seen that domain walls can form in
some special cases, allowing the corresponding parameter range to be
ruled out. Other parameter ranges could be ruled out in the same way.

\newcommand{\Mg}{m_\sbG}
\newcommand{\Me}{m_\sbu}
\newcommand{\calM}{{\cal M}}
\newcommand{\calMt}{\widetilde{\cal M}}
\newcommand{\bfA}{\mbox{\boldmath $A$}}
\newcommand{\bfAt}{\widetilde{\mbox{\boldmath $A$}}}
\newcommand{\calD}{{\cal D}}
\newcommand{\calDt}{\widetilde{\cal D}}

\chapter{Massive Fermion Bound States}
\label{Ch:Bound}

\section{Introduction}

In section~\ref{In FermZM} and chapter~\ref{Ch:Index} we investigated fermion
zero modes and massless currents on cosmic strings. It is reasonable to
expect that cosmic strings may also have massive bound states and
currents. It is possible that such currents occur in models without
zero modes, in which case a greater range of models will have
conducting strings. We expect the maximum size of these currents
to be smaller than the massless case, since they will require less
momentum to escape from the string. 

If the theory contains particles which are massless off the string,
and these interact with the current carriers, we expect them to scatter into
those states and thus the current will dissipate. This is most likely
to happen in a GUT at high temperatures, since all the Standard Model
fields are massless then. This is not the case after the electroweak phase
transition, suggesting that massive currents will be most
significant at low temperatures. 

In chapter~\ref{Ch:SUSY} we showed that fermion zero modes occur in
all supersymmetric cosmic string theories. It seems that they will gain a
mass when supersymmetry is broken (see chapter~\ref{Ch:SUSY2}),
suggesting that massive fermion currents will occur in models with
broken supersymmetry.
 
The existence of bound states will also affect scattering off the
string. A stable bound state may help to catalyse certain
interactions. This could provide a way to probe a string's
internal structure, and determine the properties of the GUT in which
it formed.

We found that each of massless fermion currents considered in
chapter~\ref{Ch:Index} only moved in one direction along a cosmic
string. If a massive fermion current exists on a string, there will be
equal numbers of left and right moving states. If lower energy states
are always filled before higher energy ones, there will be equal
amounts of left and right movers. The total current is then zero.
The fermion states will still have non-zero angular momentum, and so
they may still stabilise string loops to form vortons. A net current
will develop if the states are not filled evenly, due to difficulties
in dissipating momentum.

In section~\ref{BS fieldeqn} we will derive the field equations for
fermion bound states in a general theory, and use them to show that
such states are always time or light like. We solve these equations
numerically for an abelian model in section~\ref{BS spectrum}. In
section~\ref{BS other} we speculate about the existence and type of
bound states in some other theories. The results are summarised in
section~\ref{BS sum}.

\section{Field Equations and a Positivity Condition}
\label{BS fieldeqn}

Consider a general fermion Lagrangian,
\be 
\Lagf = \bar{\psi}_\ga i\sig{\mu} D_\mu \psi_\ga 
- \frac{1}{2}i\bar{\psi}_\ga \Mab \psi^c_\gb + \hconj \ .
\label{BS Lag}
\ee
The field equations for $\psi_\ga$ can be found by varying \bref{BS Lag}
with respect to $\bar{\psi}_\ga$. We can then separate variables with the
aid of similar expressions to those used in chapter~\ref{Ch:Index}
\be
\psi_\ga = \bmat{c} e^{i(q_\ga-\frac{1}{2})\th} \left[
A^\ga_1(r) e^{il\th+i(wt+kz)} + A^{\ga\ast}_2 (r) e^{-il\th-i(wt+kz)}\right] \\
-ie^{i(q_\ga+\frac{1}{2})\th} \left[A^\ga_3(r) e^{il\th+i(wt+kz)} 
	- A^{\ga\ast}_4 (r) e^{-il\th-i(wt+kz)}\right] \emat \ .
\label{BS Aans}
\ee
$q_\ga$ are the charges of $\psi_\ga$ with respect to the string
generator. $l$ can take any value which gives a single valued $\psi_\ga$,
so it must be an integer if $2q_\ga$ are odd and a half integer if $2q_\ga$ are
even. Using the gauge invariance of \bref{BS Lag}, we deduce that the
angular dependence of the mass term is 
\be
M_{\ga \gb}(r,\th) = \Cab(r) e^{i(q_\ga + q_\gb)\th} \ 
\ \ \ (\mbox{no summation}) \ .
\ee
We will define the following operators
\begin{displaymath}
\calD_T^{(\ga)} = \dr + \frac{1}{2r}
+ \mbox{diag}\left(-\frac{l+q_\ga}{r} + eA_\th, 
\frac{l-q_\ga}{r} +eA_\th, \frac{l+q_\ga}{r} -eA_\th, 
-\frac{l- q_\ga}{r}-eA_\th \right) \ ,
\end{displaymath}
\be
\calD = \bmat{cc} 0 & (w+k)I \\ (w-k)I & 0 \emat \ , \ \ 
\calM_{\ga \gb} = \bmat{cc} \sig{1} & 0 \\ 0 & -\sig{1} \emat C_{(\ga \gb)}
\ .
\ee
Using \bref{BS Aans} the field equations reduce to 
\be 
\left(-\calD + \calD_T + \calM \right)\bfA = 0 \ ,
\label{BS Aeqn}
\ee
where $(\calD_T \bfA)^\ga = \calD_T^{(\ga)}\bfA^\ga$.
These have equal numbers of real and imaginary solutions.
If $l=0$ (which is only possible if $2q_\ga$ are odd) these equations are
reducible. We will consider real and imaginary solutions
separately, and set $A^\ga_2 = \pm A^\ga_1$ and 
$A^\ga_4 = \pm A^\ga_3$. Defining $\bfAt^\ga = (A^\ga_1, A^\ga_3)^T$,
\bref{BS Aeqn} becomes
\be
\left(-\calDt + \calDt_T \pm \calMt \right) \bfAt = 0 \ ,
\label{BS spAeqn}
\ee
with
\begin{displaymath}
\calDt^{(\ga)}_T = \dr + \frac{1}{2r} + \frac{1}{r}\mbox{diag}\left(
-\frac{q_\ga}{r} + eA_\th, \frac{q_\ga}{r} - eA_\th \right) \ ,
\end{displaymath}
\be
\calDt = \bmat{cc} 0 & w+k \\ w-k & 0 \emat \ , \ \
\calMt_{\ga \gb} = \bmat{cc} 1 & 0 \\ 0 & -1 \emat C_{(\ga \gb)} \ .
\ee
The upper and lower choices of sign correspond to real and imaginary
solutions respectively.

Now consider $\bfA^T \calD^2 \bfA$. Using \bref{BS Aeqn} and the
definition of $\calD$, we find
\be
\bfA^T \calD^2 \bfA = (w^2-k^2) |\bfA|^2 =
 \bfA^T (\calD_T + \calM)^2 \bfA \ .
\label{BS pos1}
\ee
We will now define the constant orthogonal matrix
\be
K = \bmat{cc} 0 & -I \\ I & 0 \emat \ .
\ee
Any physical solutions must be normalisable. We will scale $\bfA$ so
that $\int |\bfA|^2 r dr = 1$. Now \bref{BS pos1} implies
\be
w^2 - k^2 = \int \bfA^T (\calD_T + \calM)K^T K(\calD_T + \calM)\bfA r dr
= \int |K(\calD_T + \calM)\bfA|^2 r dr \ .
\ee
We have integrated by parts, and used the fact that $\calM$ and $\calD_T$
are symmetric. We see that $w^2 \ge k^2$, with equality occuring only when 
$(\calD_T + \calM)\bfA = 0$. Similar arguments apply to $\calDt$. Thus
there are no space-like fermion currents, and the only light-like ones
are generalisations of zero modes. This is in contrast to the case of
bosonic currents, where space-like currents exist and are the most
favoured~\cite{patrick}.

\section{Bound States in the Abelian String Model}
\label{BS spectrum}

For simplicity we will consider currents in the $U(1)$ model described in
chapter~\ref{Ch:Intro}. The fermionic part of its Lagrangian is given
by \bref{In Lagferm}, which leads to the field equations \bref{In U1Ferm}.
In this case there is only one fermion field, so the index $\ga$ in
\bref{BS Aans} can be dropped. For a string with winding number $n$,
$q = n/2$ and $M=\mass{f}f(r)e^{in\th}$.

Since there are no space-like currents we can set $k=0$ without loss
of generality. By changing the signs of $A_3$ and $A_4$, $w>0$
solutions can be changed into $w<0$ solutions, thus only positive $w$ need be
considered. We will take the string's winding number ($n$) to be 1, so
$q=1/2$ in \bref{BS Aans}. By interchanging $A_{1,3}$ and $A_{2,4}$
negative $l$ solutions can be obtained from the positive $l$
solutions. Thus it is only necessary to look for $l \ge 0$ solutions.

We will use a variation of the shooting method to determine the values
of $w$ which have normalisable fermion solutions on the string. At
large $r$ the solutions of \bref{BS Aeqn} have exponential
behaviour. Two of them decay and so are acceptable. In the case of the
small $r$ solutions, only two of them give a normalisable state.

Each of these 4 solutions can be numerically extended to some
intermediate value of $r$ (of order the string width). We can then see
if any non-trivial combinations of the large and small $r$ solutions
match up there. Since the equations are linear, this is sufficient to determine
if there are normalisable solutions for a given value of $w$. Thus we
need only consider variations of $w$. The special $l=0$
case can be treated similarly. There is then just one large $r$
and one small $r$ acceptable solution.

\begin{figure}
\centerline{\psfig{file=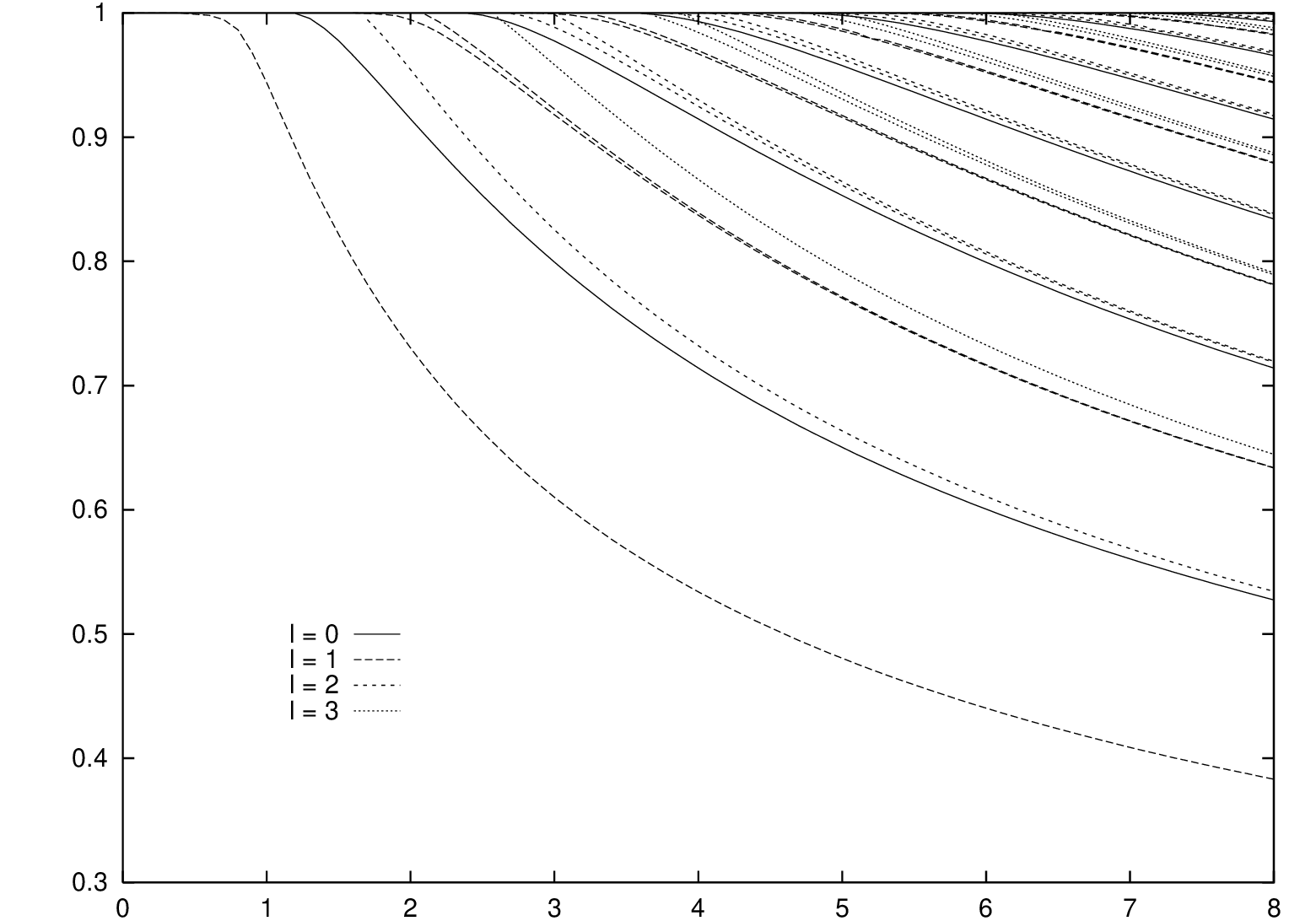,width=\myfigwidth}}
\vskip -1.9 in \hskip 0.5 in $w/\mass{f}$
\vskip 1.6 in \hskip 3.1 in $\mass{f}/\mass{s}$
\caption{Spectrum of $l=0\dots3$ fermion bound states.} 
\label{bsspec1 fig}
\end{figure}

\begin{figure}
\centerline{\psfig{file=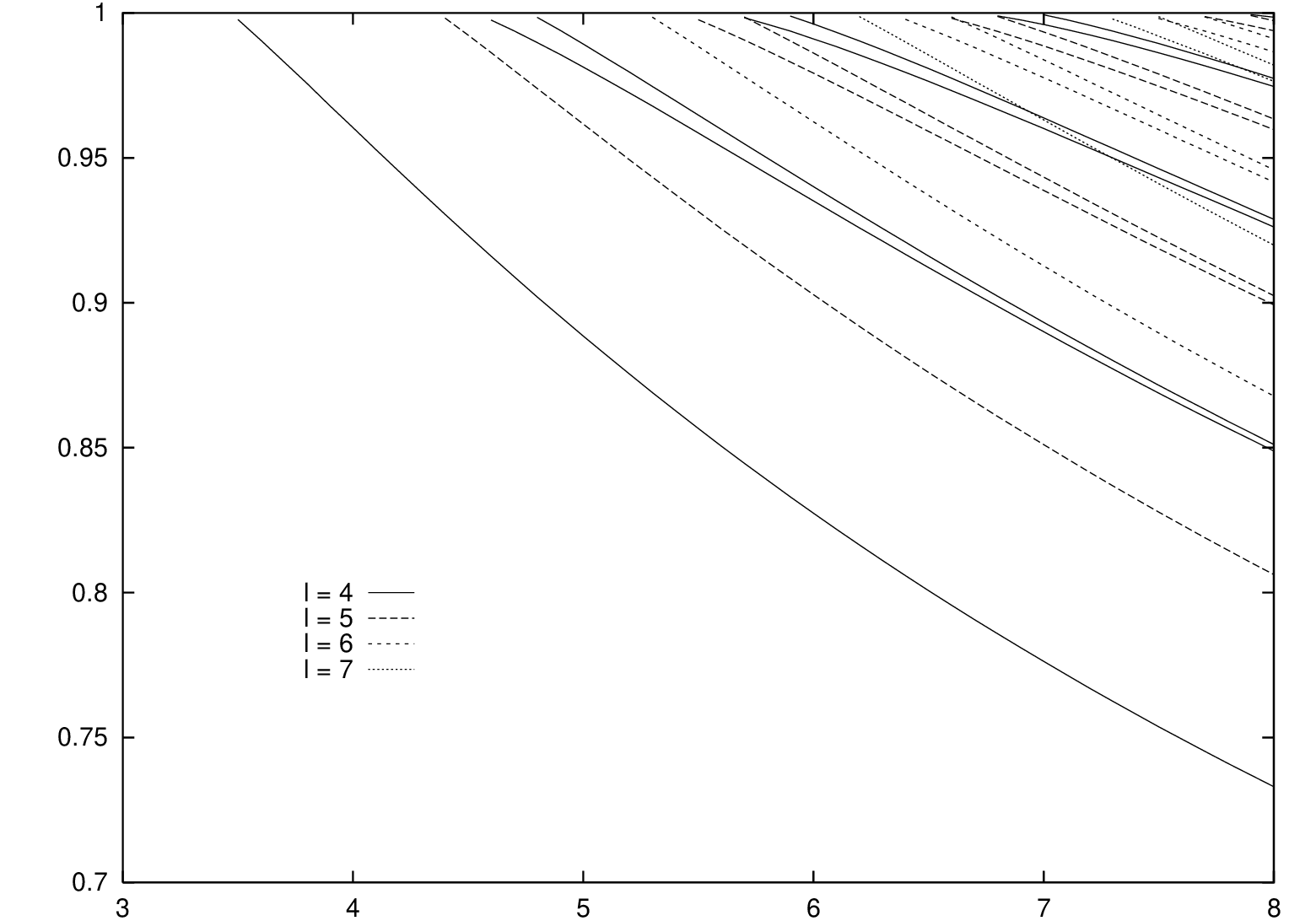,width=\myfigwidth}}
\vskip -2.1 in \hskip 0.5 in $w/\mass{f}$
\vskip 1.8 in \hskip 3.0 in $\mass{f}/\mass{s}$
\caption{Spectrum of $l=4\dots7$ fermion bound states.} 
\label{bsspec2 fig}
\end{figure}

Figures \ref{bsspec1 fig} and \ref{bsspec2 fig} show the variation of
the number of bound states with respect to the Yukawa coupling
($\mass{f}/\mass{s}$). Each line corresponds to one real and one
imaginary solution. For simplicity we have taken the Higgs and
gauge field masses ($\mass{s}$ and $\mass{v}$) to be equal. The string
field profiles are shown in figure~\ref{fa fig}. 

\begin{figure}
\centerline{\psfig{file=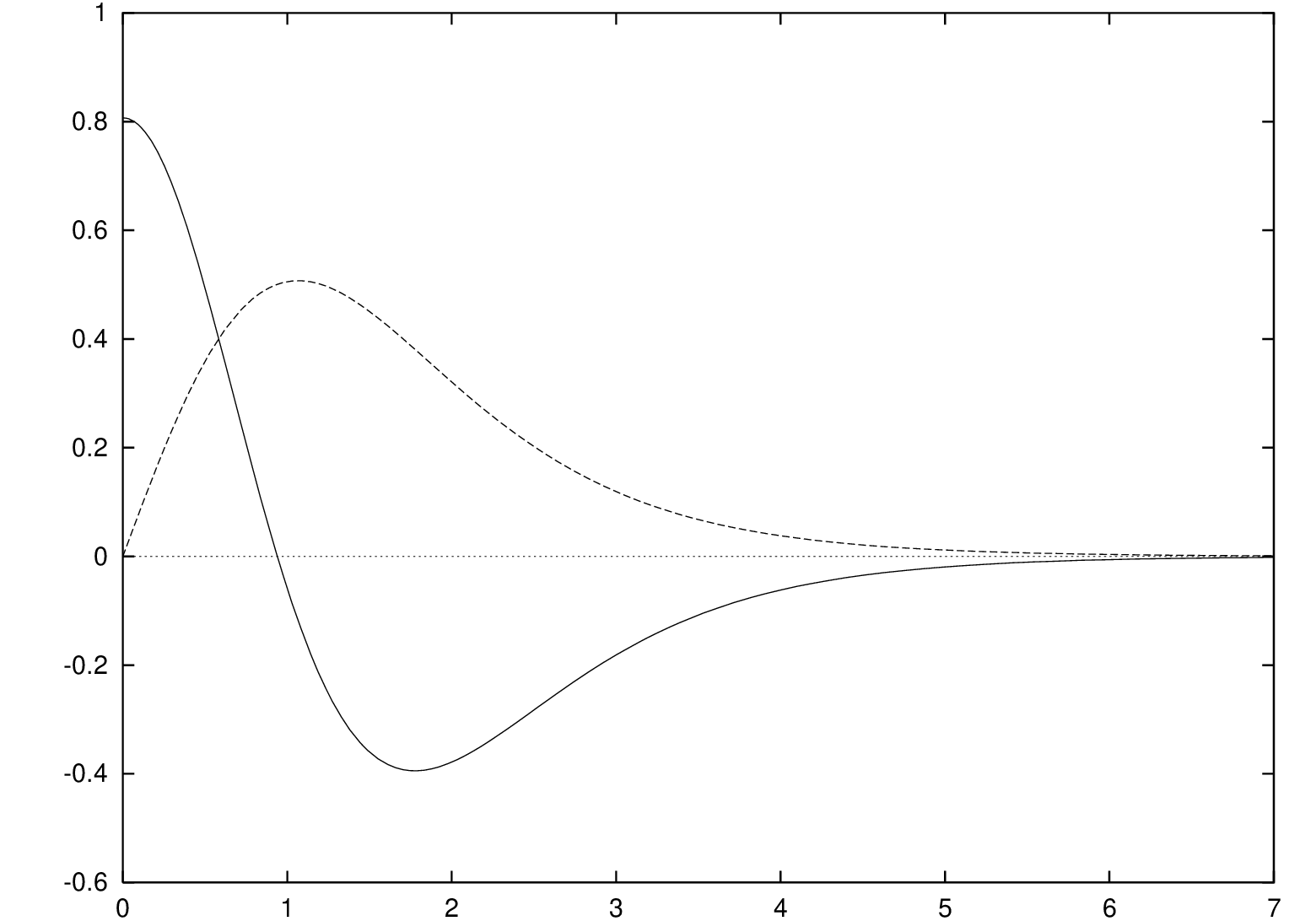,width=0.6\myfigwidth}
\psfig{file=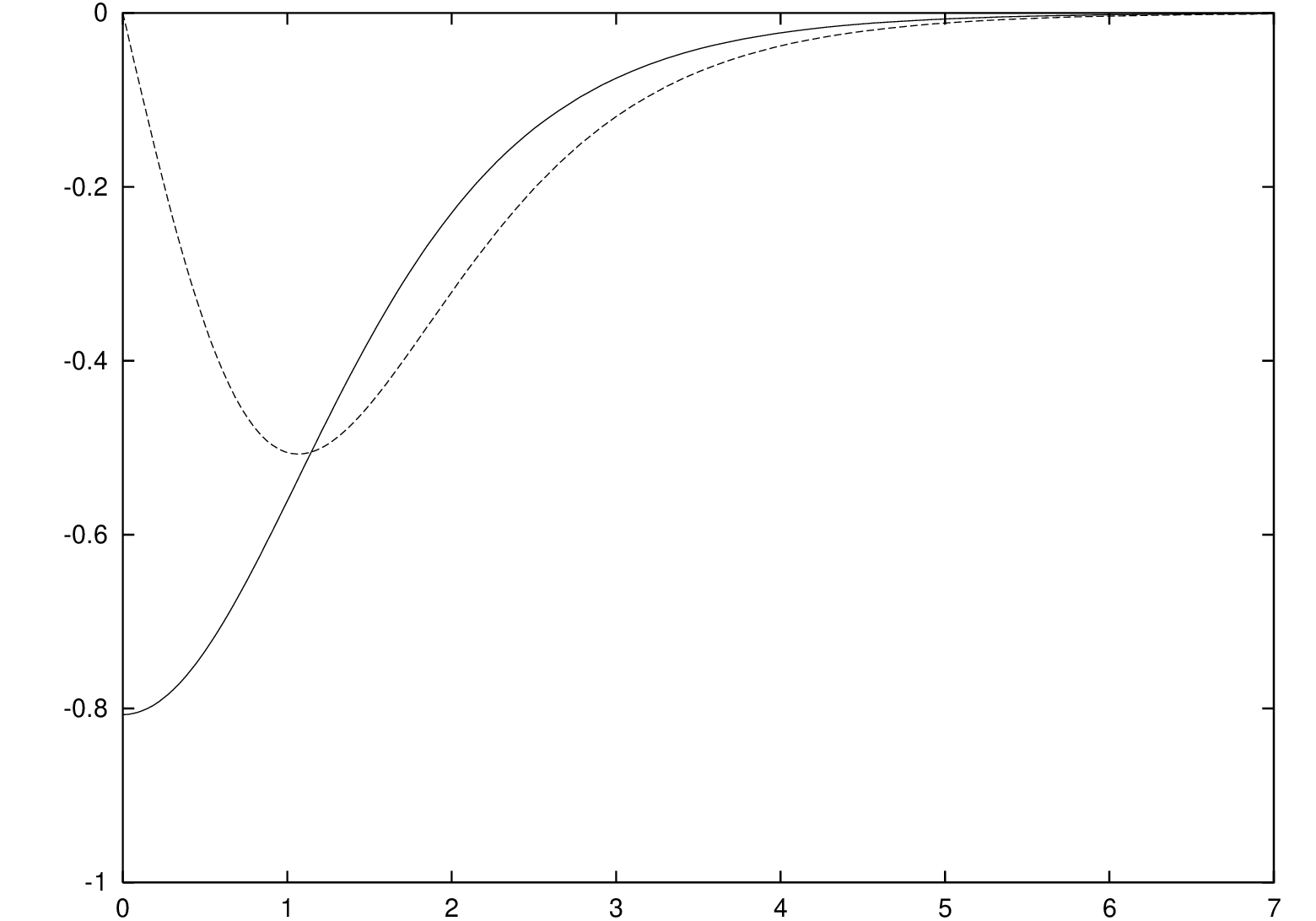,width=0.6\myfigwidth} \ }
\vskip 0.1 in
\centerline{\psfig{file=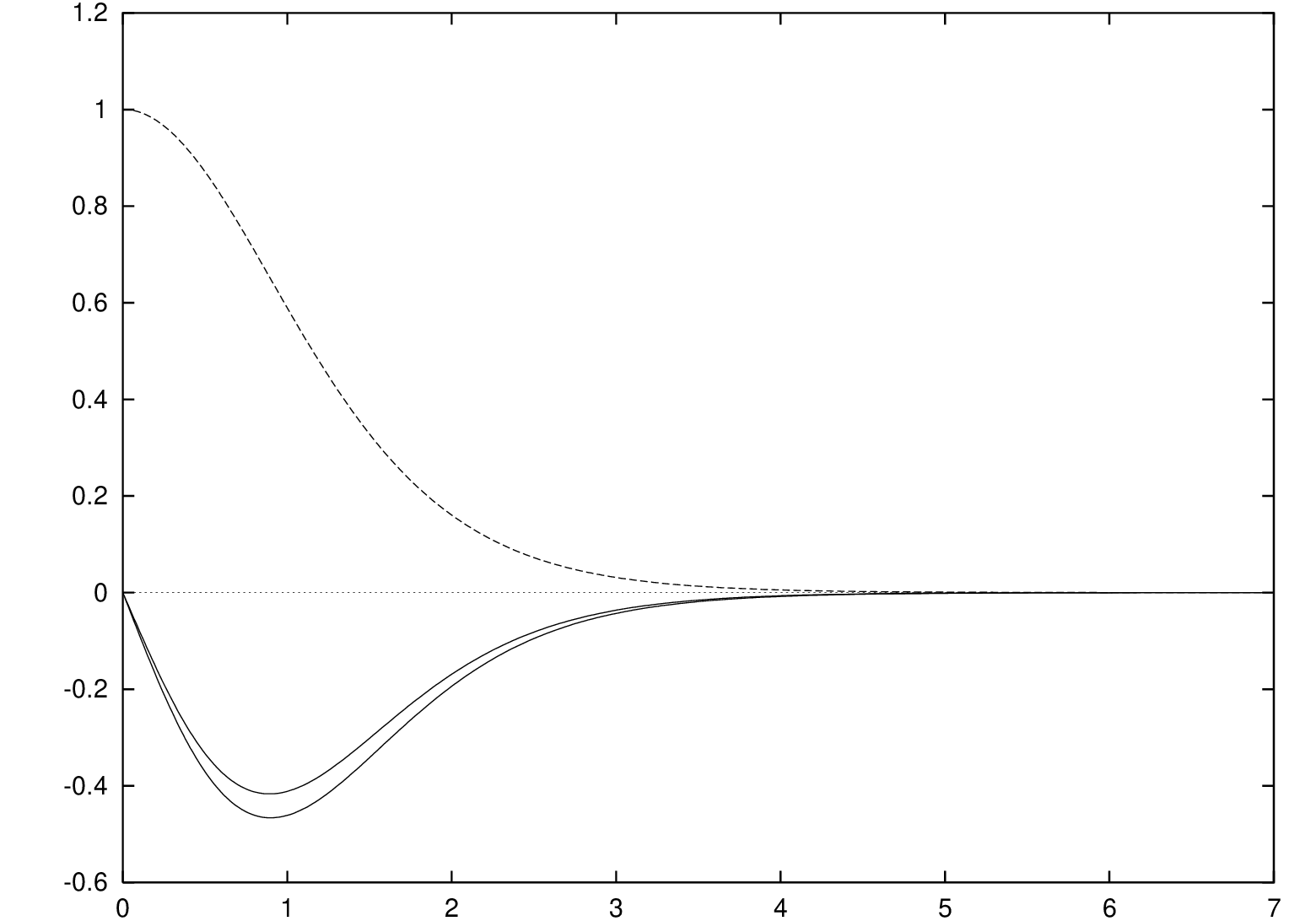,width=0.6\myfigwidth}
\psfig{file=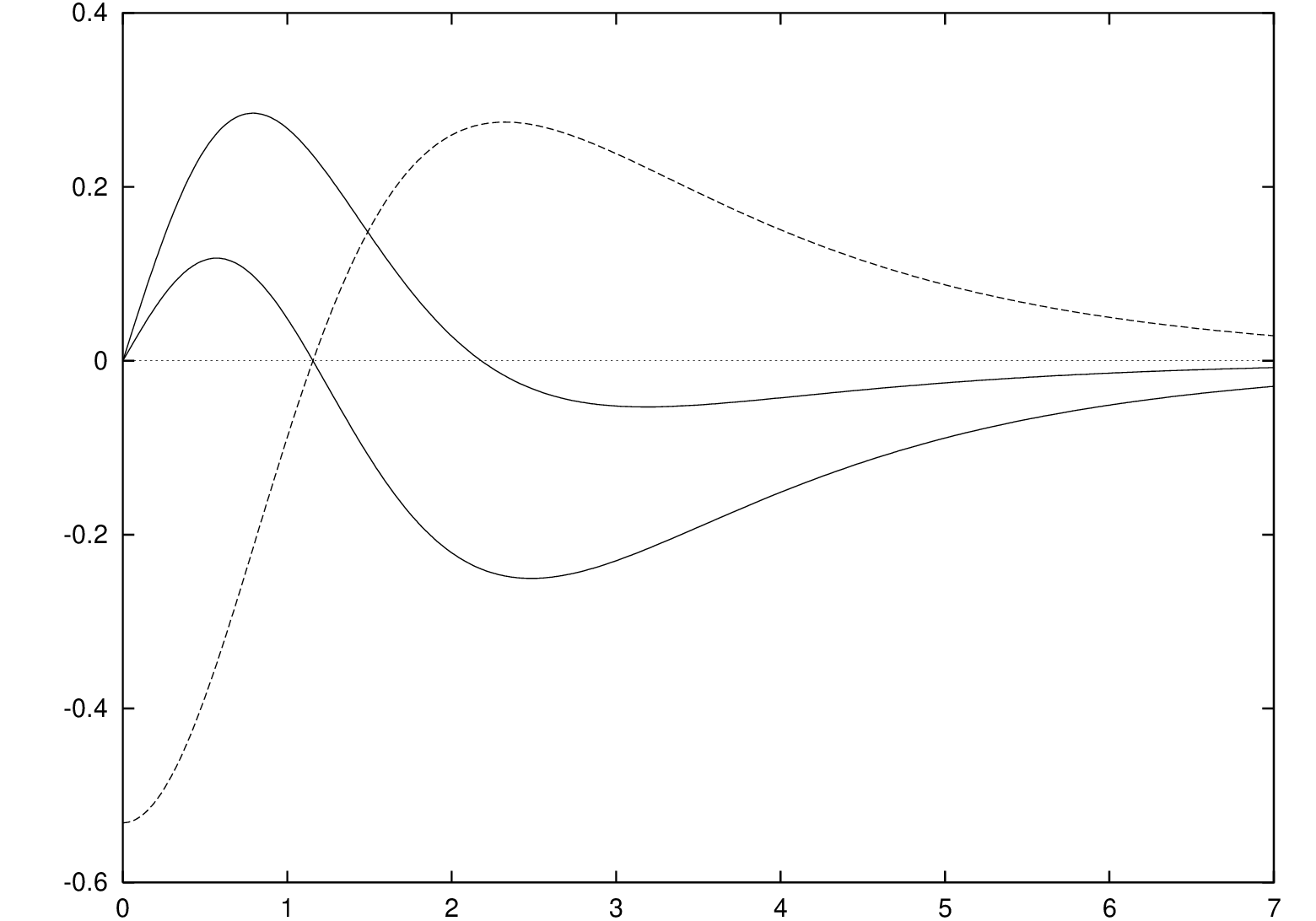,width=0.6\myfigwidth} \ }
\vskip 0.1 in
\centerline{\psfig{file=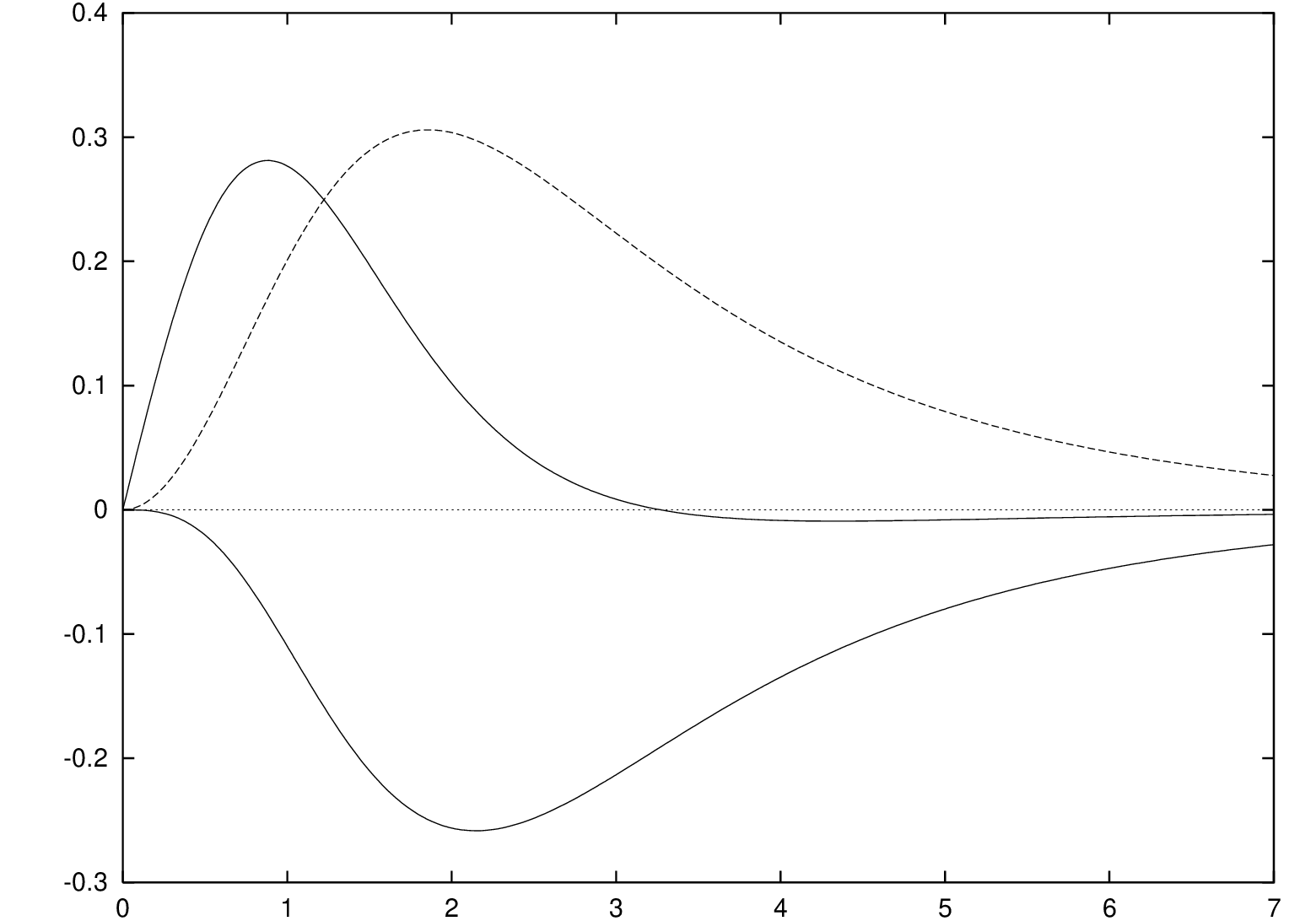,width=0.6\myfigwidth}
\psfig{file=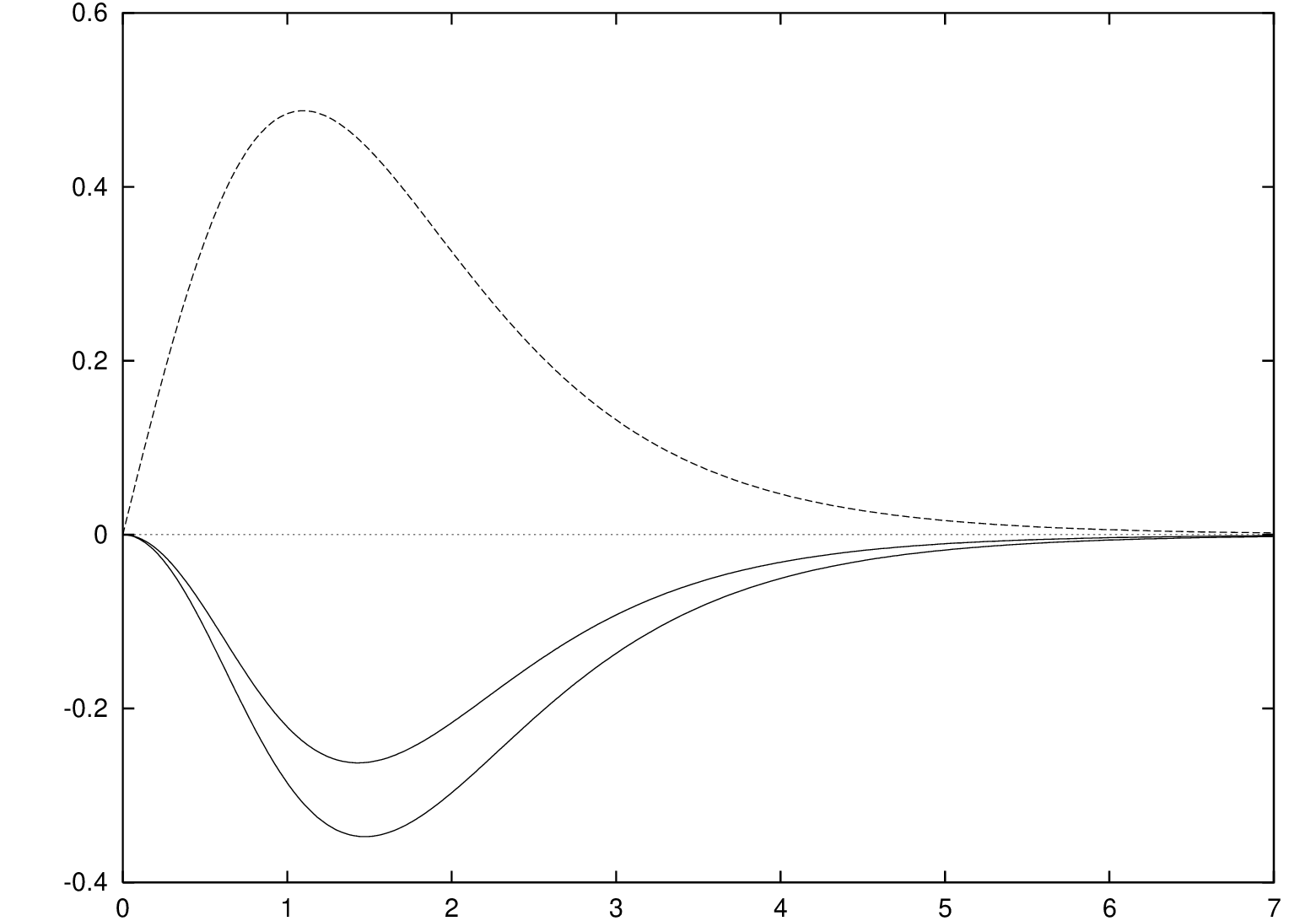,width=0.6\myfigwidth} \ }

{\small
\vskip -6.1 in
\hskip 1.8 in \parbox{1in}{$l=0$ \\ $w = 2.017$ \\ (real)}
\hskip 2.0 in \parbox{1in}{$l=0$ \\ $w = 2.017$ \\ (imag)}
\vskip 1.7 in
\hskip 1.8 in \parbox{1in}{$l=1$ \\ $w = 1.581$}
\vskip 0.6 in 
\hskip 4.8 in \parbox{1in}{$l=1$ \\ $w = 2.247$}
\vskip 1.0 in
\hskip 1.8 in \parbox{1in}{$l=1$ \\ $w = 2.265$}
\hskip 2.0 in \parbox{1in}{$l=2$ \\ $w = 2.098$}
\vskip 1.5 in}
\caption{Fermion bound state solutions in the abelian string model with
$\mass{f}/\mass{s} = 2.3$.}
\label{bssol fig}
\end{figure}

Plots of the solutions for $\mass{f}/\mass{s} = 2.3$ are shown in
figure~\ref{bssol fig}. As expected they decay outside the string
(which is approximately of radius 3 on the plots).

\renewcommand{\baselinestretch}{1}

\renewcommand{\baselinestretch}{\myblstr}

\section{Bound States in Other Theories}
\label{BS other}

We know that the right-handed neutrino zero modes on an $SO(10)$
abelian string do not survive the electroweak phase transition (see
section~\ref{Ind abstrZM}). Considering the form of the mass terms in
the theory will give some insight into what happens to them.

After the electroweak phase transition the left and right neutrino
fields ($\nu$ and $\nu^c$) receive masses from both $\ph{10}$ and
$\ph{126}$. In the absence of cosmic strings, $\ph{10}$ contributes a
Dirac mass term $\Me = \yuke \etau$ and $\ph{126}$ contributes a
Majorana mass term $\Mg = \yukg \etag$. Assuming $\Mg \gg \Me$ the
neutrino mass eigenvalues are
\bea
m_R &=& \frac{\sqrt{\Mg^2 + 4\Me^2} + \Mg}{2} \approx \Mg \nonumber \\
m_L &=& \frac{\sqrt{\Mg^2 + 4\Me^2} - \Mg}{2} \approx \frac{\Me^2}{\Mg} \ .
\eea
The mass eigenstates are then approximately $\nu^c + \eps \nu$ and
$\nu - \eps \nu^c$, with $\eps = \Me / \Mg$.

Since $|\ph{126}|$ and $|\ph{10}|$ vary inside a cosmic string, the
neutrino masses will too. In section~\ref{SO10 EWZsect} we showed that
there is a region around the string core in which $|\ph{10}|$ is reduced
but $|\ph{126}|$ takes its usual VEV. Thus $m_L$ will be lower in this
region. At the centre of the string $\ph{126}=0$ and 
$|\ph{10}| \sim \Me/\Mg$ (see figure~\ref{ewh0 fig}), thus $m_L=m_R
\sim \Me^2/\Mg$ there.

We see that there is a potential well inside the string, which
suggests that neutrino bound states will be present. As with the
abelian string, we can investigate the existence of bound states by
examining the approximate large and small $r$ solutions of \bref{BS Aeqn}, 
and then trying to match them at intermediate $r$. The 8 small
$r$ solutions have the same behaviour as the small $r$ zero mode
solutions given by (\ref{leftUsol},\ref{leftVsol}) and their right
moving equivalents. Four of them have acceptable small $r$
behaviour. Of the 8 large $r$ solutions 4 decay outside the string (as
required) if $w^2 - k^2 < m^2_L$, so bound states could exist. If
$w^2 - k^2 > m^2_L$ no more than 2 of the solutions are acceptable,
and so there are no bound states. Thus any fermion solutions which are
localised to the string will have $w^2 - k^2 < m^2_L$. The maximal
current in this case will be tiny.

At the phase transition the right-handed neutrino zero modes will mix
with the left-handed neutrinos to produce the bound states. Any currents
with rest mass greater than the left-handed neutrino mass will be
able to escape from the string. The remaining current will be far to
small to stabilise vortons, so they will certainly
collapse. Furthermore we expect the bound state to be spread over the region
of electroweak symmetry restoration. This is far larger than the size
of a vorton, which is a couple of orders of magnitude greater than the
GUT string radius~\cite{vorton}. The current
on one part of the string will then interact with current on the opposite
side of the loop, increasing the vorton's instability. Thus we
conjecture that vortons in $SO(10)$ will decay at the electroweak
phase transition, even though currents may continue to exist.

The electron and quark masses are also reduced inside the string, so
they may have bound state solutions too. Their off-string mass is far
greater than the neutrinos, so these currents may become large enough to
have detectable effects. Since they are charged they will produce a
wider range of effects than the neutrino currents. Even so, they will
still be far smaller than the GUT currents that were present before
the electroweak phase transition.

The situation is likely to be different in supersymmetric
theories. Here, SUSY-breaking destroys the zero modes by mixing them with other
zero modes, rather than massless fields. Again we expect the current
to be limited by the off-string mass of the particles it is composed
of. SUSY-breaking does not significantly alter this, so the current will
not fly off the string as it could in the $SO(10)$ model. Additionally,
extra symmetry restoration does not occur, so the width of the string will
not increase and the current on a vorton will not scatter
off itself. This suggests vortons arising in supersymmetric theories are more
stable than those in ordinary GUTs. Since the mass of the SUSY bound states
will be of order the SUSY breaking scale, it is still possible the
currents could decay into Standard Model particles. This will
depend on the couplings of the theory. Hence fermion currents may
provide constraints on SUSY models after all.

\section{Summary}
\label{BS sum}

The existence and form of fermion bound states and their corresponding
currents on cosmic strings were investigated in this chapter.
We have shown that only time-like and light-like currents can
occur. Using numerical methods, the discrete spectrum of states for the abelian
string model was determined. We found that it varied with the magnitude of the
Yukawa coupling of the theory. For very low values there is just one bound
state and a zero mode.

Since the bound states can carry angular momentum, they will
contribute to vorton formation. However, unless interactions between
the states and the electroweak sector are suppressed, they will decay.

We also speculated about the existence of massive fermion currents in models
whose fermion zero modes are destroyed by a phase transition or
supersymmetry breaking. We conjecture that when the zero mode mixes with
a massless field, the size of the current will be reduced. If the
energy scale at which this mixing occurs is much lower than that at
which the string formed (as in the GUT considered), the current will
be drastically reduced. Any vortons will then certainly decay. If
the zero mode mixes with another zero mode (as in the SUSY theories
considered), the current will not be significantly altered. Any such
vortons must decay by another mechanism.

\chapter{Summary and Future Work}
\label{Ch:Summary}
\section{Summary}
In this thesis we have investigated cosmic strings in realistic
particle physics theories. We have seen that these strings have a far
richer microstructure than their counterparts in the simple abelian
string model. This microstructure can lead to observational
consequences, and will change the physical predictions of the theory. 
The additional properties of these strings can help to provide explanations
for cosmological phenomena. They may also be used to constrain the
underlying theory.

We began by looking at cosmic strings in theories with several phase
transitions. In chapter~\ref{Ch:SO10} we showed that a string formed at one
phase transition could affect future phase transitions, and force
their Higgs fields to take string-like solutions. Even if string-like
solutions formed at such transitions would normally decay, the
presence of the GUT defect stabilises them. The existence of these
string-like solutions leads to extra symmetry restoration.

This effect was considered in detail for a realistic grand unified
theory with an $SO(10)$ symmetry group. After the electroweak phase
transition there are a total of five distinct types of cosmic string
in this model. Three of them restored electroweak symmetry in a region
around the string core. The size of this region is of the same order
as the electroweak scale, so it is far greater (about $10^{14}$ times
bigger) than the string width. Profiles of the electroweak fields were
found numerically.

Previous work suggests that if fermions couple to the Higgs field of
an abelian cosmic string, fermion zero modes (and hence massless
currents) will exist on
the strings~\cite{Jackiw}. In chapter~\ref{Ch:Index} the existence of
these currents in a general theory was investigated. We considered
theories with more exotic cosmic strings whose Higgs fields can have
more than one winding number. An index theorem giving the number and type of
massless fermion currents was derived. The theorem depends only on the type of
Yukawa terms present, and their angular dependence. It does not depend
on the size of the Yukawa couplings, or the radial dependence of the
string solution. This allows a wide range of theories to be examined
with relative ease. We showed that massless currents can be created or
destroyed at phase transitions. This provides a solution to the
potential vorton problem.

The index theorem was applied to the
strings discussed in chapter~\ref{Ch:SO10}. We found that for
topologically stable strings, only the abelian ones had massless currents at
high temperatures. In this case the current carriers were right handed
neutrinos. Following the electroweak phase transition, no
topologically stable strings admitted massless currents. Thus fermion
superconductivity is unlikely to occur in this model at low
temperatures. The model provides an example of massless current
destruction. Some of the strings with higher winding numbers did have
fermion currents, although such strings are unlikely to be stable.

The idea that particle physics theories are supersymmetric above the
electroweak scale has become increasingly favoured in recent years. In
chapter~\ref{Ch:SUSY} the implications of supersymmetry for abelian
cosmic strings were investigated. Two methods of gauge symmetry
breaking were considered, giving two types of string. The strings
broke (fully in one theory, partially in the other) supersymmetry in their
cores. Applying supersymmetry transformations did not therefore leave
the solution invariant. However, since the field equations were still
invariant under supersymmetry, the transformed solutions still solved
them. These new solutions corresponded to strings with fermion zero
modes on them. Furthermore these solutions were expressed in terms of
the background string fields, whose properties have been extensively
studied.

Supersymmetry is clearly broken in the present universe. We examined
the effects of soft supersymmetry breaking terms on cosmic strings in
chapter~\ref{Ch:SUSY2}. A theory with abelian and nonabelian cosmic
strings was considered. Soft terms are those which break
supersymmetry, but do not give quadratically divergent
contributions to the electroweak Higgs field mass. Some of them
destroyed the zero modes. There is some similarity between this
mechanism and the destruction of high temperature neutrino zero modes in
chapter~\ref{Ch:Index}. The effects of soft supersymmetry breaking on
the abelian strings in chapter~\ref{Ch:SUSY} were also considered. The
results were the same for the $F$-term symmetry breaking, but the zero
modes on strings formed by $D$-term symmetry breaking were unaffected.

Lastly, in chapter~\ref{Ch:Bound}, we investigated the spectrum of massive
fermion bound states and currents in an abelian string model. We found
that unlike the massless currents, the number of massive currents is
dependent on the strength of the couplings in the theory. We also
showed that space-like fermion currents do not exist on cosmic strings
in any model.

We have shown that cosmic strings in realistic
models can be significantly different to those in simple abelian
models. If quantum field theory is supersymmetric at high
temperatures cosmic strings are certain to have conserved
currents. This can radically alter the cosmological implications of
the strings. We have also found that the properties of realistic
strings are not constant, but change as the universe passes
through phase transitions (or as supersymmetry is broken). For
example, the stability of currents on the strings can be changed.

\section{Future Work}

There are clearly many extensions of the work on fermion bound states
started in chapter~\ref{Ch:Bound}. The destruction of massless fermion
currents at gauge and supersymmetry breakings has been seen in
chapters \ref{Ch:SO10} and \ref{Ch:SUSY2}. It seems likely that they
turn into bound states, which may then decay. I will examine this
conjecture in more detail. If such currents do decay, analysis of the
bound states will help determine the decay products. Such decays may
help provide a mechanism for baryogenesis.

Although the existence of conserved fermion currents on strings could
stabilise loops, there is still debate about this. The currents are
not topologically conserved, and there are possible mechanisms that
could destabilise them. One possibility is the interaction of the
current carriers with plasma particles or other current
carriers. These interactions can also create currents. The presence of
fermion bound states on a string may alter these processes. I will
investigate these ideas.

The only GUTs considered in chapter~\ref{Ch:SO10} involved
$SO(10)$ (or a subgroup), but many of the results can be generalised to
other theories, such as $E_6$. Chapter~\ref{Ch:Index} could be
extended in a similar way. I am currently investigating the zero modes
in an axion cosmic string theory~\cite{axionstring}, which can arise
naturally in superstring theory. It is possible that phase transitions
will also affect gauge boson currents, in a similar way to fermion
currents. Such currents are a common feature of nonabelian theories.

Another possible area of investigation is the effect of the other
symmetry restorations (particularly the $SU(5)$ restoration) on
monopoles. In the $SO(10)$ GUT considered, monopoles form when $SU(5)$ is
broken~\cite{su5 mono}. It is credible that it would be energetically
favourable for monopoles to sit on nonabelian strings, since this
would reduce the variation in $\ph{45}$, and so possibly the total
energy. If this does happen, the likelihood of monopole collision will
be greatly increased, as collisions in one dimension are far more
frequent than in three. Since they are $Z_2$ monopoles (and hence
monopoles are topologically equivalent to anti-monopoles), this results
in a higher annihilation rate. While monopoles occur in most GUTs, their
properties conflict with observations, so a mechanism is required to
get rid of them. Since cosmic strings could reduce the
number of monopoles they could solve, or help to solve, this
problem. I will see if this mechanism works, and then determine its
effectiveness. I will also see how the presence of a gauge boson
current on the cosmic string affects it. This mechanism has
similarities to the Langacker-Pi mechanism~\cite{LangackerPi}, in which
strings form that link the monopoles together. The monopoles are then
ends to the strings. These rapidly contract, pulling the monopoles
into each other, causing them to annihilate. This mechanism is
different because the strings are attached to the monopoles for
topological reasons rather than just dynamical ones.
While the Langacker-Pi mechanism is very efficient, it
requires the breaking and later restoration of electromagnetism, which
is hard to reconcile with experiment.

One more aspect of GUT strings that I intend to examine is the
interaction of several strings. Collisions between abelian strings
have been extensively studied, while those between the more exotic
string solutions have received less attention. I will look at these in
more detail, particularly in realistic GUTs. Previously, strings that
form in a  $G \longrightarrow H \times D$ symmetry breaking, where the discrete
group $D$ is nonabelian, have been considered. These strings cannot
intercommute for topological reasons~\cite{NAnoncomm}. This leads to a
very different evolution of cosmic string
networks~\cite{NAnetwork}. Unfortunately such strings are generally
Alice strings, and so are not physical. It is possible similar effects
will occur in more realistic theories, although for dynamical rather
than topological reasons. Even the simplest abelian strings do not
intercommute in some cases~\cite{abnoncomm}. If strings do not
intercommute, loop production will be suppressed. A different
mechanism for energy loss is then needed if the universe is not to
become string dominated. If there are significant
differences with GUT strings, they will have implications for the evolution of
string networks. This is particularly important given that recent
simulations with abelian strings~\cite{Turok} indicate that they
produce too little temperature anisotropy at small angular scales to
explain the observed data. The cosmology of more exotic cosmic strings
remains an open question.

The effects of soft SUSY breaking have only been partially explored in
chapter~\ref{Ch:SUSY2}, and I will be considering them further. It is
not clear what form the minimum of a general scalar potential takes when
soft SUSY breaking terms are present. It appears that it may have a
flat direction (in addition to the one arising from the gauge
symmetry). This could be of interest, since it has been suggested that
it could lead to a force between strings. Flat directions are not
uncommon in SUSY theories, and a more approachable theory which also
has strings has been considered in~\cite{Rubakov}. If a similar
mechanism can be found for monopoles, it could provide a possible
solution to the monopole problem. It is also possible the flat
direction will give some kind of inflation (rapid expansion of the
universe, which could solve many cosmological problems), as it does in
unbroken supersymmetry. I hope to extend this work to supergravity
theories.

Another interesting property of SUSY is that inflation and cosmic
strings frequently occur together. When cosmic strings are formed because
of $F$-terms in the potential, hybrid inflation also occurs. When
a $D$-term is used to produce inflation, the breaking of the $U(1)$
will result in string formation. I will look at the implications these
two theories have for each other. In particular, the cosmological
predictions are likely to be much richer than those resulting from
strings or inflation alone.

\newcommand{\Vs}{\frac{B'}{\sqrt{6}}}

\newcommand{\vd}[1]{\hat{d}_{{#1}}}
\newcommand{\vu}[1]{\hat{u}_{{#1}}}
\newcommand{\ve}{\hat{e}^-}
\newcommand{\vn}{\hat{\nu}}
\newcommand{\vuc}[1]{\hat{u}^c_{{#1}}}
\newcommand{\vdc}[1]{\hat{d}^c_{{#1}}}
\newcommand{\vep}{\hat{e}^+}
\newcommand{\vnc}{\hat{\nu}^c}
\newcommand{\vl}{\hat{\lambda}}
\newcommand{\vv}[1]{\hat{{#1}}}
\newcommand{\vvc}[1]{\hat{{#1}}^c}
\newcommand{\vvp}[1]{\hat{{#1}}'}
\newcommand{\vvcp}[1]{\hat{{#1}}'^c}
\newcommand{\smfrac}[2]{\mbox{$\frac{#1}{#2}$}}

\appendix
\chapter{An SO(10) Grand Unified Theory}
\label{Ch:app}

\section{Fermion and Gauge Fields}

Under $SO(10)$, all left-handed fermions transform under one
representation, and right-handed fermions transform under its
conjugate~\cite{Ross}. It is convienient to use just one
representation. This can be achieved by using the charge conjugates of
the fermions
($\psi_R^c = C\bar{\psi}_R^T$, $\psi_L^c = C\bar{\psi}_L^T$).
The charge conjugate of a right handed fermion transforms as a left handed
fermion, and vice versa. Thus $\fl = \psi_L + \psi^c_R$ is left
handed. For $SO(10)$ this definition is necessary, as well as
convienient. $\psi_L$ could be gauge transformed to $\psi_R^c$, so
any gauge invariant quantities will have to involve just $\fl$ and
$\fr$ (right-handed equivalent of $\fl$). However, $\fr$ is superfluous, 
since it is equal to $i\sig{2} \fl^\ast = \flc$, so the theory can be
described entirely in terms of $\fl$. For the electron family, it can
be written as
\be
\fl^{(e)}= \left(u_1, u_2, u_3, \nu_e, d_1, d_2, d_3, e^-, d^c_1,
	d^c_2, d^c_3, e^+, -u^c_1, -u^c_2, -u^c_3, -\nu^c_e \right)^T \ ,
\ee
where $d_i = d_{iL}$, $d^c_i= i\sig{2} d_{iR}^\ast$, etc.\ so all
the fields are left handed. The other two families of fermions can be
described similarly. Adapting work by Rajpoot~\cite{Rajpoot}, the
gauge fields can be expressed explicitly as $16\times 16$
matrices, which act on the fermion fields.
\be 
A^a\gena = \sqrt{2} \left( \begin{array}{cc}
\begin{array}{cc} H_{16} & \WL{+} \\ \WL{-} & H_{16} \end{array}
 & M_{16} \\ M_{16}^\dagger &
\begin{array}{cc} -H_{16}^\ast & \WR{+} \\ \WR{-} & -H_{16}^\ast \end{array}
 \end{array} \right) + \Lambda_{16} \ ,
\ee 
where
\be
M_{16} = \left( \begin{array}{cccccccc} 
0 & -\Yg{3}{+} & \Yg{2}{+} & -\Y{1}{-}  
				&  0 & \X{3}{+} & -\X{2}{+} & -\Xg{1}{+} \\
\Yg{3}{+} & 0 & -\Yg{1}{+} &-\Y{2}{-}
				&  -\X{3}{+} & 0 & \X{1}{+} & -\Xg{2}{+} \\
-\Yg{2}{+} & \Yg{1}{+} & 0 &-\Y{3}{-}
				&  \X{2}{+} & -\X{1}{+} & 0 & -\Xg{3}{+} \\ 
\Y{1}{-} & \Y{2}{-} & \Y{3}{-} & 0
				&  \Xg{1}{+} & \Xg{2}{+} & \Xg{3}{+} & 0 \\

0 & -\Xg{3}{-} & \Xg{2}{-} & \X{1}{-}  
				&  0 & \Y{3}{+} & -\Y{2}{+} & \Yg{1}{-} \\
\Xg{3}{-} & 0 & -\Xg{1}{-} &\X{2}{-}
				&  -\Y{3}{+} & 0 & \Y{1}{+} & \Yg{2}{-} \\
-\Xg{2}{-} & \Xg{1}{-} & 0 &\X{3}{-}
				&  \Y{2}{+} & -\Y{1}{+} & 0 & \Yg{3}{-} \\ 
-\X{1}{-} & -\X{2}{-} & -\X{3}{-} & 0
				&  -\Yg{1}{-} & -\Yg{2}{-} & -\Yg{3}{-} & 0
\end{array} \right) \ ,
\ee
and
\be
H_{16} = \left( \begin{array}{cccc}
 &&& \Xs{1}{+} \\ & G && \Xs{2}{+} \\ &&& \Xs{3}{+} \\
 \Xs{1}{-} & \Xs{2}{-} & \Xs{3}{-} & 0 \end{array} \right) \ .
\ee 
$G$ is a $3\times 3$ matrix of containing the gluon fields. It is
hermitian, and so $H$ is too. The other fields are contained in the
diagonal matrix $\Lambda_{16}$
\bearr{l@{}r@{}l}
\Lambda_{16} &\, {}=\, \mbox{diag}&\left(
\left(\Vs+\WL{3}\right)_3, -3\Vs+\WL{3}, 
	\left(\Vs-\WL{3}\right)_3, -3\Vs-\WL{3}, \right.\\
& &\left.\hspace{.2in}
\left(-\Vs+\WR{3}\right)_3, 3\Vs+\WR{3},
	\left(-\Vs-\WR{3}\right)_3, 3\Vs-\WR{3} \right)\\
 &\, {}=\, \mbox{diag}&\left(
(s+2a+2z)_3, -3s+4z, (s-3z-a)_3, -3s-3a-z, \right.\\
& &\left.\hspace{.2in}	   (-3s+a-z)_3, s+3a-3z, (s-2a+2z)_3, 5s\right)
\ .
\eearr
The subscripts indicate repeated values, and
\bea
s &=& \frac{1}{5}\left( -\WR{3} + \sqrt{\frac{3}{2}}B' \right) \ ,\\
B &=& \sqrt{\frac{3}{5}}\WR{3} + \sqrt{\frac{2}{5}}B' \ , \\
z &=& \frac{1}{4}\left(\WL{3} - \sqrt{\frac{3}{5}}B \right) \ , \\
a &=& \frac{1}{4}\left(\WL{3} + \sqrt{\frac{5}{3}}B \right) \ .
\eea
$Z=\sqrt{10}z$ and $A=\sqrt{6}a$ are the unrenormalised electroweak
$Z^{0}$ boson and photon respectively. $Z'=-\sqrt{10}s$ is a
high energy $SO(10)$ boson. The generator $P$ is obtained by putting
$s=1$ and $a=z=0$ in the expression for $\Lambda_{16}$. The substitutions
$s=z=0$ and $a=1/3$ give the charge operator.

\section{Higgs Fields}

The electroweak Higgs field transforms under the ${\bf 5}_{-2}$ and
$\bar{\bf 5}_2$ representations of $SU(5)$. They are contained in
the {\bf 10} of $SO(10)$, and so the components of $\ph{10}$ can be
expressed as symmetric products of spinors transforming under the {\bf
16} representation. Thus $\ph{10}$ can be expressed as
$\phi_\sbd^\alpha \Hd{\alpha} + \phi_\sbu^\alpha \Hu{\alpha}$, where
$\Hd{\alpha}$ ($\alpha = 0,\pm,1,2,3$), are the five components of 
${\bf 5}_{-2}$, and $\Hu{\alpha}$ are the corresponding 
components of $\bar{\bf 5}_2$. $(\Hd{0},\Hd{+})$ and $(\Hu{0},\Hu{-})$ form
$SU(2)_L$ doublets, while $\Hd{i}$ and $\Hu{i}$ ($i=1,2,3$) form an
$SU(3)_c$ triplet and anti-triplet.

Expressing these components of {\bf 10} in terms of symmetric products
of {\bf 16}s gives   
\bea
\Hu{0} &=& \smfrac{1}{4}[\sprod{\vu{j}}{\vuc{j}} + \sprod{\vn}{\vnc}] 
\ , \nonumber \\ 
\Hu{-} &=& -\smfrac{1}{4}[\sprod{\vuc{j}}{\vd{j}} + \sprod{\ve}{\vnc}]
\ , \nonumber \\ 
\Hu{i} &=& -\smfrac{1}{4}[\eps_{ijk}\sprod{\vu{j}}{\vd{k}}
	+ \sprod{\vdc{i}}{\vnc} - \sprod{\vuc{i}}{\vep}]
\ , \nonumber \\ 
\Hd{0} &=& \smfrac{1}{4}[\sprod{\vd{j}}{\vdc{j}} + \sprod{\vep}{\ve}]
\ , \nonumber \\ 
\Hd{+} &=& \smfrac{1}{4}[\sprod{\vu{j}}{\vdc{j}} + \sprod{\vep}{\vn}]
\ , \nonumber \\ 
\Hd{i} &=& \smfrac{1}{4}[\eps_{ijk}\sprod{\vuc{j}}{\vdc{k}} 
	+ \sprod{\vd{i}}{\vn} - \sprod{\vu{i}}{\ve}] \ ,
\eea
where $\vn$ is the basis vector corresponding the $\nu$ field, etc.
The gauge fields can also be expressed as $10\times 10$ matrices,
which will that act on $(\phi_\sbd^\alpha, \phi_\sbu^\alpha)$.
\be
A^a\gena = \sqrt{2} \left( \begin{array}{cc} 
H_{10} & M_{10} \\ M_{10}^\dagger & -H_{10}^\ast
\end{array} \right) + \Lambda_{10}
\ee
with
\be
H_{10} = \left( \begin{array}{ccccc} 
&&& \X{1}{-} & \Y{1}{-} \\
& G && \X{2}{-} & \Y{2}{-} \\
&&& \X{3}{-} & \Y{3}{-} \\
 \X{1}{+} & \X{2}{+} & \X{3}{+} & 0 & \WL{+} \\
 \Y{1}{+} & \Y{2}{+} & \Y{3}{+} & \WL{-} & 0
\end{array} \right) \ ,
\ee
\be 
M_{10} = \left( \begin{array}{ccccc}
0 & -\Xs{3}{-} & \Xs{2}{-} & \Xg{1}{+} & -\Yg{1}{-} \\
\Xs{3}{-} & 0 & -\Xs{1}{-} & \Xg{2}{+} & -\Yg{2}{-} \\
-\Xs{2}{-} & \Xs{1}{-} & 0 & \Xg{3}{+} & -\Yg{3}{-} \\
-\Xg{1}{+} & -\Xg{2}{+} & -\Xg{3}{+} & 0 & -\WR{+} \\
\Yg{1}{-} & \Yg{2}{-} & \Yg{3}{-} & \WR{+} & 0 
\end{array} \right) \ ,
\ee
and
\bea
\Lambda_{10} &\!\!=\!\!&
\mbox{diag}\left(
	\left(-2\Vs\right)_3, \WL{3}+\WR{3}, -\WL{3}+\WR{3},
	\left(2\Vs\right)_3, -\WL{3}-\WR{3}, \WL{3}-\WR{3} \right)
\nonumber \\
&\!\!=\!\!&
\mbox{diag}\left(
	(-2s-a+z)_3, -2s+3a+z, -2s-4z,
\right. \nonumber \\ && \hspace{2in} \left. 
	(2s+a-z)_3, 2s-3a-z, 2s+4z \right) \ .
\eea
The second Higgs field $\ph{45}$ is in the ${\bf 24}_0$ component of
the {\bf 45} representation, and its usual vacuum expectation value is
proportional to the generator of the $B$ field. In the absence of a
string, $\ph{126}$ is proportional to $\sprod{\Nr}{\Nr}$, with $\Nr = -\vnc$.

\section{Fermion Masses}

The masses of the fermions arise from Yukawa couplings to the Higgs
fields. These must of course be Lorentz and gauge invariant. Since the
theory contains only left handed spinors the only possible Lorentz
invariant mass terms are Majorana masses
($\bar{\psi}^c_L \psi_{L} + \bar{\psi}_L \psi^c_L +
(\mbox{right-hand terms})$). The Majorana masses transform as a
product of {\bf 16}s, and so can be coupled to similarly transforming
Higgs fields, $\ph{126}$ and $\ph{10}$, but not $\ph{45}$. This gives
the fermionic Lagrangian \bref{fermLag}. We have considered only one
family of fermions for simplicity. 

Of course, since (\ref{fermLag}) is invariant under $SO(10)$, it must
be invariant under $SU(5)$ as well. Thus $\phvac{126}$ can only couple to
$SU(5)$ singlets (i.e. products of the conjugate neutrino
field). The two components of $\phvac{10}$ couple to 
${\bf 5}_{-2}$ and $\bar{\bf 5}_2$ products of fermions. Under
$SU(5)$, the remaining fermions transform under ${\bf 10}_1$ and
$\bar{\bf 5}_{-3}$ representations. To find allowable mass terms,
products of these representations need to be expressed in terms of
irreducible representations. ${\bf 10}_1 \times {\bf 10}_1$ and 
${\bf 1}_5 \times \bar{\bf 5}_{-3}$ both contain $\bar{\bf 5}_2$s,
 and ${\bf 10}_1 \times \bar{\bf 5}_{-3}$ contains a ${\bf 5}_{-2}$. 
Thus, for the usual VEVs of the Higgs fields, the mass terms are
written in terms of particle fields as 
\bea
\flb \Hd{0} \flc &=& \smfrac{1}{8}\left[ 
d_i^{\dagger} i\sig{2} d_i^{c\ast} + e^{-\dagger} i\sig{2} e^{+\ast} 
+ d_i^{c\dagger} i\sig{2} d_i^\ast + e^{+\dagger} i\sig{2} e^{-\ast} 
\right] \nonumber \\ &=& 
 \smfrac{1}{4}\left[\bar{d}_{iL} d_{iR} + \bar{e}_L^- e_R^- \right] \ ,
\eea
and similarly for $\Hu{0}$, so 
\bea
\flb \phvac{10} \flc &\!\!=\!\!& 
\frac{\etad}{4} \left[
 \bar{d}_{iL} d_{iR} + \bar{e}^-_L e^-_R \right]
+ \frac{\etau}{4} \left[
 \bar{u}_{iL} u_{iR} + \bar{\nu}_L \nu_R \right] \ , \\
\flb \phvac{126} \flc &\!\!=\!\!& \etag \nu^{c\dagger} i\sig{2} \nu^{c\ast}
			= \etag \nu^T_R i\sig{2} \nu_R \ .
\eea
This model, unlike the standard model, has non-zero neutrino masses.
If the $\nu^T_R i\sig{2} \nu_R$ term is much larger than the
$\bar{\nu}_L \nu_R$ term, the mass eigenstates will be
approximately $\nu_L$ and $\nu_R$, and have very small and very
large mass eigenvalues respectively, giving an almost massless
left-handed neutrino.

\end{document}